\documentclass[11pt]{article}
\usepackage{geom}
\graphicspath{./images/}

\title{Near-Optimal Min-Sum Multi-Robot Motion Planning in a Planar Polygonal Environment}

\author{Pankaj K. Agarwal \thanks{Department of Computer Science, Duke University, USA} \and 
Benjamin Holmgren \footnotemark[2]
\and Alex Steiger \footnotemark[2]}

\begin{document}
\begin{titlepage}

\maketitle

\begin{abstract}  
    Let $\P \subset \R^2$ be a planar polygonal environment with $n$ vertices, and let $[k] = \{1,\ldots,k\}$
    denote $k$ unit-square robots translating in $\P$.
    Given source and target placements $s_1, t_1, \ldots, s_k, t_k \in \P$
    for each robot, we wish to compute a collision-free motion plan $\plan$, i.e.,
    a coordinated motion for each robot $\robot{i}$ along a continuous path from $s_i$ to $t_i$ so that robot $i$ does not leave $\P$ or
    collide with any other $\robot{j}$. Moreover, we additionally require that $\plan$
    minimizes the sum of the path lengths; this variant is known as \textit{min-sum motion planning}.

    Even computing a feasible motion plan for $k$ unit-square robots in a polygonal environment
	is {\textsf PSPACE}-hard.
    For $r > 0$, let $\opt(\bs,\bt, r)$ denote the cost of a min-sum motion plan for $k$ square
	robots of radius $r$ each from $\bs=(s_1,\ldots,s_k)$ to $\bt=(t_1,\ldots,t_k)$.
    Given a parameter $\epsilon > 0$, we present an algorithm for computing a coordinated motion plan 
	for $k$ unit radius square robots of cost at most $(1+\epsilon)\opt(\bs,\bt, 1+\eps)+\epsilon$,
    which improves to
	$(1+\epsilon)\opt(\bs,\bt, 1+\epsilon)$ if $\opt(\bs,\bt, 1+\eps)\geq 1$, that runs
	in time $f(k,\epsilon)n^{O(k)}$, where $f(k,\epsilon) = (k/\epsilon)^{O(k^2)}$.
    Our result is the first polynomial-time bicriteria $(1+\epsilon)$-approximation algorithm for any 
	optimal multi-robot motion planning problem amidst obstacles for a constant value of $k > 2$. 
	The algorithm also works even if robots are modeled as $k$ congruent disks.
\end{abstract}
\end{titlepage}

\newpage
\newpage

\section{Introduction}\label{sec:intro}

With the rise of autonomous systems in
widespread application areas such as search and rescue,
warehouse management, extraplanetary exploration, handling hazardous materials, and 
transportation~\cite{applications}, there has been extensive work on computing a coordinated motion 
plan for a team of robtots moving inside a 2D or 3D environment. In addition to avoiding 
collisions with obstacles, motion plans must also avoid collisions between robots, and one thus needs to consider the problem in a higher dimensional \emph{configuration space}. Furthermore, we wish to ensure a good quality of motion such as being short or having a small makespan.

In this paper, we consider the problem of coordinated motion planning of $k$ identical robots of simple shape
(e.g., squares, disks) in a polygonal work environment. For simplicity, we focus on the case in which each
robot is modeled as a unit square. Note that if $k$ is part of the input, even the problem of computing
a feasible plan is \pspacehard~(see e.g.~\cite{DBLP:conf/compgeom/AbrahamsenBBKLS25}). We seek an approximation 
algorithm that runs in roughly $n^{O(k)}$ time and returns a motion plan that aims to minimize the 
total length of the paths traversed by the robots. No such algorithm is known for $k > 2$.

\paragraph{Problem statement.}

Let $\poly := \{x \in \R^2 \mid ||x||_\infty \leq 1\}$ denote the 
axis-aligned square of unit radius---referred to as 
a \emph{unit square} for short---centered at the origin. 
Let $\BR = \{1, \ldots, k\}$ denote a system of $k$ robots, each modeled as a
unit square, that can translate inside the same closed 
polygonal environment (possibly with holes) $\P \subset \R^2$,
called the shared \emph{workspace}.
Let $n$ be the number of vertices of $\P$.
A placement of one robot of $\BR$ is represented by 
the position of its center in the workspace~$\P$. For such a placement 
to be free of collision with the boundary $\partial\P$ of $\P$,
the representing point should be at $\ell_\infty$-distance at least $1$ 
from $\partial \P$. (Note that the robot is allowed to touch
an obstacle, since we define $\F$ to be a closed set.)
We let $\F \subset \P$ denote
the \emph{free space} of a single robot, which is the subset of $\P$ 
consisting of all collision-free placements.  
A (joint) \emph{configuration} $\bp$ of $\BR$ 
is represented as a $k$-tuple $\bp = (p_1, \ldots, p_k) \in \P^k$, 
where $p_i$ is the placement of a robot $i \in \BR$.
The \emph{configuration space}, called \emph{C-space} for short, 
is the set of all configurations, 
and is thus represented as $\P^k \subset\R^{2k}$. 
A configuration $\bp = (p_1, \ldots, p_k) \in \R^{2k}$ is called \emph{free} 
if $p_i \in\F$ for all $i \in \BR$ 
and  $||p_i-p_j||_\infty \geq 2$ for every pair $i \not= j$.
Let $\BF := \BF(\P)$ 
denote the \emph{$2k$-dimensional free space}, comprising the set of all free configurations.
Clearly, $\BF \subseteq \F^k$.

Let $\bs = (s_1, \ldots, s_k)$ be a given \emph{source configuration} and $\bt = (t_1, \ldots, t_k)$ a given \emph{target configuration}.
An \emph{$(\bs,\bt)$-plan} is a continuous function $\pth:[0,T] \to \P^k$,
for some $T \in \RR_{\geq 0}$, with $\pth(0) = \bs$ and $\pth(T) = \bt$.
The image of $\plan$ is a (continuous) curve in the C-space, referred to 
as an \emph{$(\bs, \bt)$-path}. 
With a slight abuse of notation, we use $\plan$ to denote its image 
as well.
If $\pth \subset \BF$ we say that $\pth$ is \emph{feasible},
and if there exists a feasible $(\bs,\bt)$-plan we say that the 
pair $(\bs,\bt)$ is \emph{reachable}.
For a plan $\pth:[0,T] \to \P^k$, let
$\pi_i:[0,T] \to \P$ be the projection of $\pth$ onto the two-dimensional plane 
spanned by the $2i - 1$ and the $2i$ coordinates specifying the motion of robot $i$
that $\pth$ induces; that is,
$\pth(\lambda) = \left(\pi_1(\lambda), \ldots, \pi_k(\lambda)\right)$ for all $\lambda \in [0,T]$. 
Again, with a slight abuse of notation, we also use $\pi_i$
to denote the paths followed by robot $i$.

For a path $\gamma$ in $\P$, let $|\gamma|$ denote its
arc-length.
We define $\cost(\pth)$, the \emph{cost} of an $(\bs,\bt)$-plan $\pth$,
by $\cost(\pth) := \sum_{i \in \BR} |\pi_i|$, that is, 
$\cost(\pth)$ is the sum of the arc-lengths of the paths of the $k$ robots.
If $(\bs,\bt)$ is reachable, a minimum-cost feasible $(\bs,\bt)$-plan
is referred to as a \emph{min-sum} plan. In this paper, we investigate the problem of
deciding whether a given pair $(\bs,\bt) \in \BF^2$ is reachable, and, if so, computing a
min-sum $(\bs,\bt)$-plan.
We will also study the min-sum problem when each robot is a
unit disk.
The above definitions extend verbatim to this settings, though of course the regions $\F$
and $\BF$ depend on the shape of the robots.

\paragraph{Related work.}

Algorithmic motion planning has been extensively studied in the computational geometry
and robotics communities for more than four decades: see the books and reviews \cite{halperin2018handbook, sharir2018handbook, lavalle2006} for a high-level overview of the most relevant results.
Multi-robot motion planning is computationally intractable even in simple settings. If the number of robots is part of the input, the feasibility problem is already \pspacehard \cite{DBLP:conf/compgeom/AbrahamsenBBKLS25,geft2019,brocken2020,solovey2016,hopcroft1984}.

Notwithstanding a rich literature on multi-robot motion planning 
in both continuous and discrete settings (robots moving on a graph in the latter setting),~see, e.g.,~\cite{DBLP:journals/trob/DayanSPH23,
DBLP:journals/ijrr/KaramanF11,
KavSveLatOve96,
DBLP:journals/cacm/Salzman19,
DBLP:journals/arobots/ShomeSDHB20,
stern2019multiagent,
DBLP:journals/arobots/TurpinMMK14},
relatively little is known about algorithms producing paths with provable quality guarantees.
Approximation algorithms for minimizing the total path-length 
are given in~\cite{geft2019,DBLP:conf/wafr/SolomonH18,solovey2015,DBLP:conf/compgeom/AbrahamsenBBKLS25}
for a set of unit-disk robots assuming a certain separation between the start and goal positions, as well as from the obstacles. 
The separation assumption makes the problem considerably easier.
An $O(1)$-approximation algorithm was proposed in~\cite{demaine2019} for computing a plan that minimizes the makespan for a set of unit discs (or squares) in the plane without obstacles, again assuming some separation.

Computing the min-sum motion plan for two unit squares/disks even in the absence of 
obstacles is non-trivial \cite{esteban2023,liu2016}.
There is some recent progress on min-sum motion planning amid obstacles~\cite{steiger2024, struijs}.
Agarwal et~al. \cite{struijs} give an $O(n^4\log n)$-time algorithm for computing a min-sum plan for two unit-square robots in a rectilinear environment when the path length is measured in the $\ell_1$-metric.
However, their approach neither extends to arbitrary polygonal environment nor to the $\ell_2$-metric,
and it is an open question whether the min-sum problem for $2$ robots (under the $\ell_2$-metric) is in \polynomial.
For arbitrary polygonal environments, Agarwal et~al. \cite{steiger2024} give an $(1+\epsilon)$-approximation for two unit-square robots.
To accomplish the FPTAS, the algorithm in \cite{steiger2024} strongly relies on using motion in which one robot moves at a time.
For $k> 2$ robots, it is easy to construct examples where the optimal motion must move many robots simultaneously, which adds considerable complexity to
the problem and the approach in \cite{steiger2024} does not extend.
It remains an open question
whether a polynomial-time approximation algorithm exists for min-sum motion planning if $k>2$ is a constant. 
Note that we cannot hope for a PTAS for the min-sum problem if $k$ is part of the input since even the feasibility problem
is \pspacehard.

\paragraph{Our results.}
Our main result is a polynomial-time bicriteria approximation algorithm (PTAS) for
min-sum motion-planning with $k$ congruent axis-aligned square
robots translating in a planar polygonal environment, where $k$ is a constant.
For a real parameter $r \geq 1$, and for $\bs,\bt \in \BF$, let $\opt(\bs,\bt,r)$
denote the cost of the min-sum $(\bs,\bt)$-plan for $k$ square robots
each of radius $r$. If $(\bs,\bt)$ is not reachable with respect to $k$
squares of radius $r$, then $\opt(\bs,\bt,r)$ is undefined.
If a plan $\pth \subset \BF$ is also a feasible plan with respect to $k$
squares of radius $1 + \rho$ each, we say that $\pth$ is \emph{$\rho$-robust}.
We say that $(\bs,\bt)$ is \emph{$\rho$-reachable} if there exists a $\rho$-robust
plan.
We note that our use of the parameter $\rho > 0$ can be interpreted
as a typical robustness parameter in the sampling-based motion planning literature;
see, e.g., \cite{sampling}.
%
%
The following theorem states our main result:
\begin{theorem}\label{thm:result}
    Let $\P$ be a closed polygonal environment in $\R^2$ with $n$ vertices.
    Let $\BR = \{1,\ldots,k\}$ be $k$ axis-aligned unit-square robots translating in $\P$,
    let $\bs, \bt \in \BF$ be a pair of free
    configurations of $\BR$, and let $\epsilon \in (0,1)$
    be a parameter. If there exists a feasible $\epsilon$-robust $(\bs,\bt)$-plan,
    then a feasible $(\bs,\bt)$-plan of $\BR$ such that
    $\cost(\pth) \leq (1+\epsilon)\opt(\bs,\bt,1+\epsilon)+\epsilon$ can be
	computed in $f(k,\epsilon)n^{O(k)}$ time, where $f(k,\epsilon) = \left(k/\epsilon\right)^{O(k^2)}$.
    If $\opt(\bs,\bt,1+\epsilon) > 1$, then $\cost(\pth) \leq (1+\epsilon)\opt(\bs,\bt,1+\epsilon)$.
\end{theorem}

Our algorithm extends to regular convex polygons in a straightforward manner. Since a disk can be approximated by regular convex polygon, we obtain the following result. See \secref{ext}.


\begin{theorem}\label{thm:result-disk}
    Let $\P$ be a closed polygonal environment in $\R^2$ with $n$ vertices.
    Let $\BR$ be $k$ congruent disk robots translating in $\P$.
    Let $\bs, \bt \in \BF$ be 
    a pair of free configurations of $\BR$, and let $\epsilon \in (0,1)$
    be a parameter. If there exists a feasible $\epsilon$-robust $(\bs,\bt)$-plan,
    then a feasible $(\bs,\bt)$-plan of $\BR$ such that
    $\cost(\pth) \leq (1+\epsilon)\opt(\bs,\bt,1+\epsilon)+\epsilon$
	can be computed in $f(k,\epsilon)(n)^{O(k)}$ time, where $f(k,\epsilon) = \left(k/\epsilon\right)^{O(k^2)}$.
    If $\opt(\bs,\bt,1+\epsilon) > 1$, then $\cost(\pth) \leq (1+\epsilon)\opt(\bs,\bt,1+\epsilon)$.
\end{theorem}

As a shorthand, we call a plan of the form in Theorems~\ref{thm:result}--\ref{thm:result-disk}
i.e., an $(\bs,\bt)$-plan $\pth$ such that $\cost(\pth) \leq \opt(\bs,\bt, 1 + \epsilon)+\epsilon$,
an \emph{$\epsilon$-optimal} plan.
Our main technical contribution is \thmref{result}, which, to our knowledge, is
the first $n^{O(k)}$-time algorithm for computing an $\epsilon$-optimal plan a constant number of 
$k > 2$ robots in a planar polygonal environment.

At a very high level, our algorithm (Section 2) is sampling-based and similar to 
many approximation algorithms for optimal motion planning \cite{steiger2024}\todo{more refs},
namely we construct a finite graph $\PG = (\V,\E)$ in the $2k$-dimensional free space $\BF$ such that 
$\bs,\bt \in \BF$, and argue that a shortest path from $\bs$ to $\bt$ in $\PG$ is a feasible 
$(\bs,\bt)$-plan of cost at most $(1+\varepsilon)\opt(\bs,\bt,1+\epsilon)+\epsilon$. 
We prove two key properties to bound the size of $\PG$ and the cost of the shortest path in $\PG$: (1) the
existence of a near-optimal \emph{tame} plan $\widetilde{\pth} = (\widetilde{\pi}_1,\ldots, \widetilde{\pi}_k)$
in which breakpoints of each $\widetilde{\pi_i}$ lie in the $O(k^2)$-sized neighborhood of a fixed set of $O(n)$ \emph{landmarks} (Section 3).
(2) There exists a near-optimal plan $\widetilde{\pth}$ in which the number of breakpoints is proportional to
$\cost(\widetilde{\pth})$ (Section 4). Similar properties of a near-optimal plan were proved in \cite{steiger2024},
but the situation was considerably simpler, since altering the path of a robot could only impact the motion of
one other robot; this was core to the argument there. We need a completely different approach here, and this is the main technical contribution
of the paper. To prove the existence of a near-optimal tame plan, we start with an arbitrary optimal plan $\pth$ and perform a two-level
surgery on $\pth$ to eliminate all breakpoints that are far from the landmarks. To this end, we perform a \emph{global surgery} that
reroutes a robot whenever its path contains a far-away breakpoint. However, doing so may interfere with 
the motion of other robots, so we park the others in appropriately chosen parking areas, by performing a \emph{local surgery}, while the rerouted robot
passes, and then move the others one by one. How exactly is this accomplished without incurring too much cost lies at the core of the paper.

We show that $\PG$ contains a path in a small neighborhood of a tame plan $\pth$. We prove this by arguing that there is a vertex $v_i$ of $\V$ in an $\epsilon$-neighborhood of
each breakpoint $\bp_i$ of $\pth$ and that the segment $v_iv_{i+1}$ connecting two consecutive vertices is in $\F$ and thus $(v_i,v_{i+1}) \in \PE$.
We thus retract $\pth$ to a path $\widetilde{\pth}$ in $\PG$.
This is where we use the robustness assumption.
Retracting $\bp_i$ to $v_i$ for each breakpoint adds roughly $||\bp_i-v_i||$
to the cost of the plan. We bound the number of breakpoints to bound this additive error,
which determines how densely we have to sample points in the neighborhood of each landmark, and which in turn determines the size of $\PG$.
In order to bound the number of breakpoints, roughly speaking, we exploit the $\rho$-robustness of a given
min-sum plan\footnote{We believe we can bound the number of breakpoints even without a robustness assumption, but the argument becomes even more complicated.
Since we already need robustness for the retraction, we exploit it here as well and simplify the argument.}
and define, for any fixed configuration $C \in \F$, we partition an $O(1)$-size neighborhood of $\C$ into $f(k,\rho)$ equivalence classes.
We show that an optimal straight-line plan exists between any two equivalent configurations.
Using this observation we greedily "shortcut" the plan between any two equivalent configurations, and argue that the number of breakpoints in the
resulting plan is proportional to $f(k,\rho)$, which is exponential in $k$ and quadratic in $\rho^{-1}$ (cf. \lemref{bound-full}).

The algorithm itself is fairly general and works for robots of other shapes. The main challenge is to prove the existence of near-optimal tame plans. Our argument for squares extends to regular convex polygons (Section 5).
Since a disk can be approximated by a regular convex polygon, the robustness assumption enables our argument for convex polygons to extend to disks.

\section{Overview of the Algorithm}\label{sec:algo}

Given a polygonal environment $\P$, $k$ unit squares $\BR$ that can
translate inside a polygonal workspace $\P$, and a 
pair of start and target configurations $\bs$
and $\bt$ respectively, we describe an algorithm for computing an $\epsilon$-optimal
$(\bs,\bt)$-plan for $\BR$. We first state the existence of a near-optimal
plan with some desirable properties (\secref{well-behaved}),
which are proved in later sections, 
then describe the algorithm (\secref{algo-description}). A formal analysis of its performance is deferred to the appendix
(\secref{correct}).
As in \cite{steiger2024}, we choose a set $\X$ of sample points in $\F$,
construct a graph $\PG = (\PV,\PE)$ in the $2k$-dimensional free space
$\BF$ such that $\PV \subset (\F \cap X)^k$ and $\bs,\bt \in \PV$,
and compute a shortest path from $\bs$ to $\bt$ in $\PG$.
Finally, we analyze the performance of our algorithm.

\subsection{Well-behaved near-optimal plans}\label{sec:well-behaved}

We describe two key properties and claim the existence of a near-optimal
plan with these properties. Since $\P$ is a polygonal environment
and the robots are squares, $\BF$ is polyhedral.
Hence, for a reachable pair of configurations $(\bs,\bt) \in \BF^2$,
any min-sum $(\bs,\bt)$-plan $\pth$ is piecewise linear \cite{struijs,steiger2024}.
For a piecewise-linear plan $\pth$, we call its vertices \emph{breakpoints}.

\paragraph{Landmarks and tame plans.}
We define a set $\Lambda \subset \F$ of $O(n)$ points, which we refer to as \emph{landmarks}.
For a value $r > 0$, let $\F^{(r)} = \P \ominus r\square$, i.e., $\F^{(r)}$
is the 2D free space with respect to a single square robot of radius $r$; $\F = \F^{(1)}$.
Let $V^{(r)}$ be the set of vertices of $\F^{(r)}$.
It is known that $|V^{(r)}| = O(n)$ \cite{klps86}.
We define
\begin{equation}
    \Lambda = V^{(1)} \cup V^{(2)} \cup \{s_i, t_i \mid i \in \BR\}.
\end{equation}

For a value $r > 0$ and a point $q \in \R^2$, let $B(q,r) = q + r\square$ denote the square of radius $r$
centered at $q$.
For a value $D \geq 0$, a point $p \in \R^2$ is called \emph{$D$-close} if there is a landmark in $\Lambda$ within
$\ell_\infty$-distance $D$ from $p$, i.e., $\min_{u \in V}||p - u||_\infty \leq D$,
or equivalently $B(p,D) \cap \Lambda \not= \emptyset$, and called \emph{$D$-far}
otherwise. The union of $D$-radius squares centered at landmarks,
$U(\Lambda,D) = \bigcup_{u \in \Lambda} B(u,D)$, is the set of $D$-close
points. We refer to $\partial U(\Lambda,D)$, the boundary of $U(\Lambda,D)$, as the
\emph{$D$-frontier}. A configuration $\bp = (p_1, \ldots, p_k)$ is called \emph{$D$-close}
if $p_i$ is $D$-close for all $i \in \BR$ and \emph{$D$-far} otherwise.
A plan $\pth$ is called \emph{$D$-tame} if all of its breakpoints are $D$-close.
The following lemma states the first property of near-optimal plans.

\begin{lemma}\label{lem:far}
    Let $\Delta = C\cdot k^2\epsilon^{-1}$ for some sufficiently large constant $C>0$. For any reachable pair
    $(\bs,\bt) \in \BF^2$, there is a $\Delta$-tame $(\bs,\bt)$-plan $\pth$ such that $\cost(\pth) \leq (1+\epsilon)\opt(\bs,\bt, 1)$.
\end{lemma}

Note that \lemref{far} does not require any robustness assumption. The lemma suggests that we can restrict our search to $\Delta$-tame plans
and that it suffices to sample points in the $\Delta$-neighborhoods of landmarks in $\Lambda$, so $\X \subset U(\Lambda,\Delta)$.
The proof of \lemref{far}, presented in \secref{far}, is one of the main technical contributions of the paper.
Given a min-sum plan $\pth$ that is not $\Delta$-tame, we perform surgery on $\pth$ in order to make it $\Delta$-tame at a slight
increase in its cost.

\paragraph{Plans with few breakpoints.}

Next, we show the existence of an $(\bs,\bt)$-plan with few breakpoints, whose cost is close to that of an $\epsilon$-robust $(\bs,\bt)$-plan.
The following lemma formally states the property.

\begin{lemma}\label{lem:bound}
    Given a parameter $\rho > 0$ and a $\rho$-reachable pair $(\bs,\bt) \in \BF^2$, there exists a $\rho$-robust $(\bs,\bt)$-plan $\pth$
    of $\BR$ such that $\cost(\pth) \leq \opt(\bs,\bt,1+\rho)$, with at most $f(k, \rho) \cdot \left(\cost(\pth) + 1\right)$
    breakpoints, where $f(k,\rho) = \left(k/\rho\right)^{O(k^2)}$.
\end{lemma}
 
\lemref{bound} guides how densely we sample inside the $\Delta$-neighborhood of each landmark.
When we retract a $\Delta$-tame plan $\pth$ to the vertices of $\PG$, we pay additive cost per vertex, say, $\varphi$,
which is inversely proportional to the sampling density. If $\beta$ is the number of breakpoints in $\pth$ then we need to ensure that
$\beta\varphi \leq \epsilon \cost(\pth)$ or $\varphi \leq (\epsilon/\beta)\cost(\pth)$, so the sampling density needs to be $\omega(varphi^{-2})$. We prove \lemref{bound} in \secref{nice},
and it is the second main technical result of the paper.

\subsection{Computing a near-optimal min-sum plan}\label{sec:algo-description}

We now describe the algorithm for computing a near-optimal $(\bs,\bt)$-plan for
$\BR$ inside $\P$. The algorithm consists of the following steps:

\begin{enumerate}
    \item \textbf{Computing landmarks.} Compute $\F^{(1)}$ and $\F^{(2)}$ using the algorithm described in \cite{klps86}.
    Set $\Lambda = V^{(1)} \cup V^{(2)} \cup \{s_i,t_i \mid i \in \BR\}$.
    \item \textbf{Sampling.} Let $\bar{\epsilon} = \frac{\epsilon}{f(k,\epsilon) \cdot \sqrt{2} k}$, where $f(k,\epsilon)$ is the
    function defined in \lemref{bound-full}. Let $\G$ be the set of vertices of the uniform grid
    in $\R^2$ with cells of width $\bar{\epsilon}$. That is, $\G = \{(i \bar{\epsilon}, j \bar{\epsilon}) \mid i,j \in \ZZ\}$.
    Compute 
    \begin{equation}
        \X = \{u \in \G \mid u \in \F \text{ and $u$ is $\Delta$-close}\}.
    \end{equation}
    $|\X| = O(n / \bar{\epsilon}^2) = f'(k,\epsilon)n$, where $f'(k,\epsilon) = O(f^2(k,\epsilon)k^2\epsilon^{-2}) = \left(k/\epsilon\right)^{O(k^2)}$.
    \item \textbf{Configuration graph.} We now construct a weighted graph $\PG = (\PV,\PE), w: \PE \to \R_{\geq 0}$ in $\BF$.
    Set $\PV = \{\bp = (p_1, \ldots, p_k) \in \X^k \mid \bp \in \BF\}.$
    Set $\PE = \{(\bp,\bq) \in \PV \times \PV \mid \bp\bq \subset \BF\}.$ For an edge $(\bp,\bq) \in \PE$, we set its weight to be
    $w(p,q) = \sum_{i \in \BR}||p_i - q_i||.$
    \item \textbf{Path computation.} If $\bs,\bt$ lie in different connected components of $\PG$, return $(\bs,\bt)$ is not reachable.
    Otherwise, compute a shortest path $\tilde{\pth}$ from $\bs$ to $\bt$ in $\PG$ using Dijkstra's algorithm, and return $\tilde{\pth}$ as the desired $(\bs,\bt)$-plan.
\end{enumerate}

This completes the description of the algorithm. 
We now defer the analysis of its performance using Lemmas \ref{lem:far} and \ref{lem:bound}
to \secref{append}. We show that $\PG$ contains a path from $\bs$ to $\bt$ of cost at most
$(1+\epsilon)\opt(\bs,\bt,1+\epsilon) + O(\epsilon)$ and that the time spent constructing $\PG$
and computing $\tilde{\pth}$ is $f(k,\epsilon) n^{2k}$, where $f(k,\epsilon) = (k/\epsilon)^{O(k^2)}$.
This implies \thmref{result}.

\section{Near-Optimal Motion Far from Vertices}\label{sec:far}

In this section we give an overview of the proof of \lemref{far}, which shows the existence of an $\epsilon$-optimal
$\Delta$-tame $(\bs,\bt)$-plan $\tilde\plan$, in which all breakpoints are $\Delta$-close
for $\Delta = O(k^2/\epsilon)$. 
We prove the lemma by showing that if a plan $\pth$ has a breakpoint at time $\lambda$
such that $\pth(\lambda)$ is $\Delta$-far, we can remove the 
breakpoint $\pth(\lambda)$ at a slight increase in the cost of $\plan$. 
The exact surgery to remove a $\Delta$-far breakpoint
$\pth(\lambda)$ requires careful consideration of the position of any robot $i$ which is $\Delta$-far at a placement $\pi_i(\lambda)$; removing this breakpoint
depends on the topology of $\F$ in a neighborhood
of $\pi_i(\lambda)$ of size $O(k)$. Roughly speaking, $\pi_i(\lambda)$ either lies in a ``narrow''
corridor, and motion inside a corridor is relatively constrained, or there is sufficient space
to park all robots that come sufficiently close to $\pi_i(\lambda)$.
We first characterize the neighborhood of a $\Delta$-far point (\subsecref{ab-topology}), then describe the surgery for
$\pi_i(\lambda)$ lying in a corridor (\subsecref{ab-corridor}), followed by the surgery 
procedure when $\pi_i(\lambda)$ is in a wide area  (\subsecref{ab-fat}).  
We present the main ideas here, and refer to the full version \ben{\secref{far-full}} for all the details.

Recall that for a value $r > 0$ and a point $q \in \R^2$,  $B(q,r) = q + r\square$
denotes the square of radius $r$
centered at $q$. Let $K(q,r) \subseteq B(q,r) \cap \F$ be the connected component of $B(q,r) \cap \F$
that contains $q$ (assuming $q \in \F$; otherwise it is undefined). 
For a set $P \subset \F$ of points, $|P| < k$, we define $\F[P] =  \F \setminus \bigcup_{p \in P}B(p,2)$, which
is the \emph{conditional free space} $\F$ for one robot when the $|P|$ robots are placed at points
in $P$. $\F[P]$ can be computed in $O(n + |P| \log^2{n})$ time \cite[Chapter~13]{debergbook}.

\subsection{Topology far from vertices}
\label{subsec:ab-topology}

The following lemma characterizes the topology  of $\F$ in a neighborhood sufficiently far from the 
vertices of $\F$.

\begin{figure}
\centering
        \includegraphics[width=0.55\textwidth]{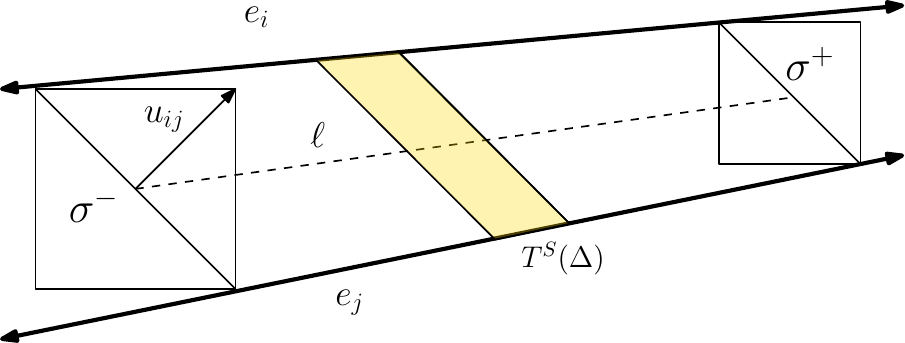}\hspace{2cm}
	\includegraphics[width=0.25\textwidth]{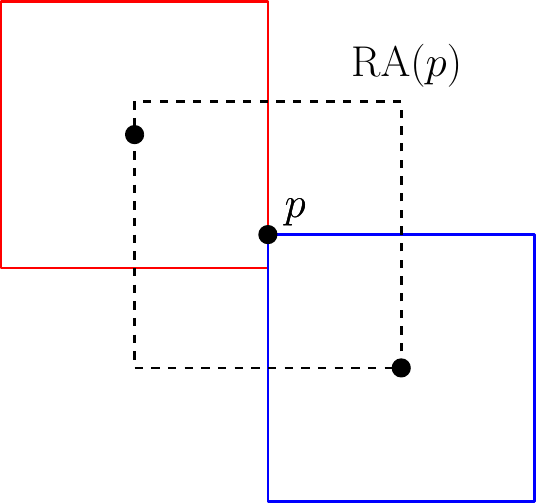}
	\caption{(left) An example corridor $\corr$ containing a $\Delta$-far region, $\corr^S(\Delta)$. (right) A revolving area.}
        \label{fig:RA+corr}
\end{figure}

\begin{lemma}\label{lem:ab-fat-square}
    For any point $q \in \F$ that is $D$-far, for some value $D > 0$, the boundary of $K(q, D/3)$
    intersects at most two edges of $\F$.
\end{lemma}

\lemref{ab-fat-square} allows us to limit attention to regions $K(q,r)$ whose boundaries intersect 
with at most two edges of $\F$. 
This motivates the definition of \emph{corridors}, which are relatively narrow regions within $\F$, and are handled in our surgery separately.

\paragraph{Corridors.}
Intuitively, \emph{corridors} are regions in $\F$ bounded by two of its edges that are too narrow for robots to freely
pass one another. 
Let $e_i, e_j$ be a pair of edges in $\F$ supporting a square (of any radius) in $\F$;
for simplicity assume that neither of the edges is axis-aligned (the definition of corridors is simpler in that case).
That is, there exists some square $B(q,r) \subset \F$ with radius $r > 0$ contained in $\F$,
such that $e_i$ (resp. $e_j$)
touches $B(q,r)$ at a point $v_i$ (resp. $v_j$), but does not intersect $\interior(\poly)$.
Let $u_{ij} \in [0,\pi)$ be a direction normal to the segment $v_i v_j$.
Consider a trapezoid $\corr$ bounded by $e_i, e_j$
such that
(i) two of its boundary edges are contained in $e_i, e_j$, which are called \emph{blockers},
(ii) no landmarks of $\Lambda$ are in the interior of $\corr$, and
(iii) the edges of $\corr$ that are not blockers, denoted by $\sigma^-,\sigma^+$ and called \emph{portals},
are normal to $u_{ij}$. See \figref{corridor}.

$\corr$ is called a \emph{corridor} if the $\ell_\infty$-distance between the endpoints of each of $\sigma^-$ and $\sigma^+$
is at most 2, i.e., if $\sigma^- = a^-b^-$ and $\sigma^+ = a^+b^+$ then the \emph{width} of $\corr$, defined as
$\width(\corr) := \max\{||a^- - b^-||_\infty, ||a^+ - b^+||_\infty\}$, is at most 2. 
Let $\ell_\corr$ be the line
segment bisecting its portals. Then for any point $p \in \ell_\corr$, the $\ell_\infty$-distance to $e_i$ or $e_j$
is at most 1. We set $u_\corr := u_{ij}$, refer to it as the \emph{portal direction} of $\corr$,
and assume that $\sigma_T^-$ appears before $\sigma_T^+$ in direction $u_T$.
Similar to the width of $\corr$, we refer to the $\ell_\infty$ distance between
the endpoints of $\ell_\corr$ as the \emph{depth} of $\corr$, denoted $\depth(\corr)$.
We say that a corridor $\corr$ is \emph{$D$-deep} if the depth of $\corr$ is at least $D \geq 0$, and its
is a \emph{maximal} corridor if no other corridor contains $\corr$. If $\corr$ is maximal,
then there is a landmark of $\Lambda$ on at least one of the portals;
if both portals of $\corr$ have the same length then there are landmarks on each portal.
These conditions imply that there can only be $O(n)$ maximal corridors in $\F$.

For a value $r \geq 0$, let $\sigma^-(r)$ (resp. $\sigma^+(r)$) be the line segment
parallel to portals at $\ell_\infty$-distance $r$ from $\sigma^-$ (resp. $\sigma^+$) inside $\corr$,
i.e., in direction $u_\corr$ (resp. $-u_\corr$). If $\corr$ is not $r$-deep, $\sigma^-(r)$ and $\sigma^+(r)$ are undefined.
Let $\corr^-(r)$ (resp. $\corr^+(r)$) denote the trapezoid formed by $\sigma^-, \sigma^-(r)$ (resp. $\sigma^+, \sigma^+(r)$)
and the portions of the blockers $e_i, e_j$ between them, and let $\corr^{S}(r) \subset \corr$ denote the corridor formed
by $e_i, e_j, \sigma^-(r),$ and $\sigma^+(r)$, which we refer to as the $r$-sanctum of $\corr$.

\paragraph{Revolving areas.}

We next define regions in $\F$ in which robots can move around one another
in a relatively unconstrained manner.
For a placement $p \in \F$ of a unit square robot $i$, we say that $i$ has a
\emph{revolving area} at $p$ if $p + \poly \subset \F$, and denote it by $\RA(p)$.
It will be useful to consider
\emph{well-separated revolving areas} (WSRAs), which are configurations of revolving areas with no conflicting placements.
Specifically, for a set of conflict-free placements $P = \{p_1, ..., p_\ell\}$ of $l \leq k$ robots,
(where $\interior(p_i + \poly) \cap \interior(p_j + \poly) = \emptyset$),
the corresponding set of revolving areas is denoted $\RA(P) = (\RA(p_1), ..., \RA(p_\ell))$. We say that $\RA(P)$
is \emph{well-separated} if $\|p_i-p_j\|_\infty \ge 4$ for every $i \not=j$. See \figref{RA}.
The following lemma, also observed in \cite{steiger2024}, summarizes the relationship between
revolving areas and corridors.

\begin{lemma}\label{lem:ab-ra-in-corr}
    Suppose $p \in \F$ is a $1$-far point that does not lie in any corridor.
    Then $p$ lies in a revolving area $\RA(q)$ for some point $q \in \F$.
\end{lemma}

Our surgery to remove $\Delta$-far breakpoints not in a corridor will heavily exploit RAs,
and in particular, WSRAs. We refer to the $O(k)$-far points of $\F$ not lying in a corridor as \emph{wide areas}.

\subsection{Surgery inside a corridor}
\label{subsec:ab-corridor}

The high-level idea of the surgery inside a corridor is to park any robots that ever travel 
$\Delta$-deep in the corridor, for $\Delta=\Omega(k)$,
at one of the  well-behaved positions we call ``parking places,'' which are $O(k)$-deep in 
the corridor, and 
keep the robot at one of these places until it exits $\corr$. 
We begin by stating a key property of corridors, proved in~\cite{steiger2024},
which shows that the ordering of the robots inside a corridor $\corr$, with respect to its
direction vector $u$, remains the same for the duration of a continuous time interval that
the robots are in $\corr$.

\begin{lemma}[Lemma 2.2 of \cite{steiger2024}]\label{lem:ab-inner-products}
    Let $i,j \in \BR$ s.t. $i \not= j$, let $\corr$ be a maximal corridor, and let $u$ be its direction vector.
    Let $[\lambda_1, \lambda_2] \subseteq [0,1]$ be a time interval in a plan $\pth$
    such that $\pi_i[\lambda_1, \lambda_2], \pi_j[\lambda_1, \lambda_2] \subset \corr$.
    Then the sign of
    $\langle \pi_i(\lambda) - \pi_j(\lambda), u\rangle$ is the same for every $\lambda \in [\lambda_1, \lambda_2]$.
\end{lemma}

Using this lemma, we devise a \emph{local surgery}
that modifies an arbitrary min-sum plan into one only
parking the robots at well-behaved placements inside a corridor.
That is, whenever a robot enters a corridor, our surgery parks robots only
within $O(k)$ distance from its portals,
at a small increase in the cost of the plan. First, we need some technical lemmas
allowing us to move robots individually to these well-behaved parking places near the portals,
and to prove the existence of sufficiently many such parking places.

\begin{lemma}\label{lem:ab-portal-cases}
    Let $\corr$ be an $8k$-deep corridor, let $\ell$ be the bisecting segment of its portals, and let $\sigma$ be
    one of its portals, say, $\sigma^-$. 
    Let $\bp = (p_1, \ldots, p_k) \in \BF$ be a feasible configuration such that $p_i \in \sigma(4k)$ for some $i \in \BR$,
    and for every other $p_j \not= p_i$, $p_j \not \in \corr^-(4k, 4k+4)$. Let $q = \sigma(4k + 2) \cap \ell$.
    Let $\bq = (p_1, \ldots, q, \ldots, p_k)$ denote the configuration with robot $i$ placed at $q$ and every other robot $j \not= i$
    remain at $p_j$. There exists a feasible $(\bp,\bq)$-plan $\pth$, with $\cost(\pth) \leq k (4\sqrt{2} + 2)$.
\end{lemma}

\begin{lemma}\label{lem:ab-corridor-vertices}
    Let $\corr$ be a $20k$-deep corridor.
    Then there exists a sequence $P = \langle p_1, \ldots, p_k \rangle$ of at least $k$
    placements in $\corr^-(10k)$ (resp. $\corr^+(10k)$) such that $p_i$ lies on the bisecting
    line $\ell_\corr$ of $\corr$,
    for every $i \in \BR$, and $||p_i - p_j||_\infty \geq 4$ for every pair $i \not= j$.
\end{lemma}

\begin{figure}
    \centering
    \includegraphics[width=0.75\textwidth]{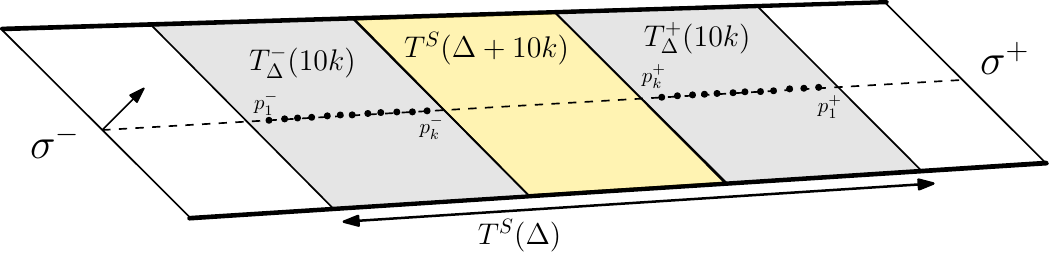}
    \caption{A corridor $\corr$, with designated parking places defined by \lemref{thin-corridor-vertices}
    at the sequences of points $P^- = \langle p_1^-, \ldots, p_k^- \rangle$
    and $P^+ = \langle p_1^+, \ldots, p_k^+ \rangle$.
    We use $P^-$ and $P^+$ to move robots across $\corr^S(\Delta + 10k)$ without creating breakpoints.}
    \label{fig:parking-spots-full}
\end{figure}

\paragraph{Surgery.} 
Let $\pth$ be a piecewise-linear $(\bs,\bt)$-plan. Let $\Delta = Ck$ for a sufficiently large constant $C\geq 10$.
We now describe the surgery on $\pth$ to remove all $\Delta+10k$-far breakpoints that lie inside
corridors. Let $\corr$ be a maximal corridor that contains a $\Delta+10k$-far breakpoint of $\pth$;
$\Delta$ is at least $2\Delta+20k$ deep.
Let $\ell_\corr$, $u_\corr$ be its portal bisecting segment and portal direction respectively, and let $\corr_\Delta := \corr^S(\Delta)$
be the $\Delta$-sanctum of $\corr$. $\corr_\Delta$ is also a corridor.
Let $P_\corr^- = \langle p_1^-, \ldots, p_k^-\rangle$ (resp. $P_\corr^+ = \langle p_1^+, \ldots, p_k^+\rangle$) be the sequence of points in $\corr_\Delta^-(10k)$
(resp. $\corr_\Delta^+(10k)$) as defined in \lemref{ab-corridor-vertices},
sorted in increasing (resp. decreasing) order in direction $u_\corr$ (i.e., $\langle p_i^-, u_\corr \rangle < \langle p_{i+1}^-, u_\corr \rangle$
and $\langle p_i^+, u_\corr \rangle > \langle p_{i+1}, u_\corr\rangle$; see \figref{parking-spots-full}.).
$P^-, P^+$ act as parking places for robots entering $\corr_\Delta$.

Suppose a robot $i \in \BR$ enters the $\Delta$-sanctum $\corr^S(\Delta)$ of $\corr$
at time $\lambda_0$, and let $[\lambda_0^-, \lambda_0^+]$ be the maximal time interval containing
$\lambda_0$ during which $i$ lies in $\corr$. We create two events for $i$ during $[\lambda_0^-, \lambda_0^+]$
as follows: Let
\[\bar{\lambda}_0 = \inf\{\lambda \in [\lambda_0^-, \lambda_0^+] \mid \pi_i(\lambda) \in \corr^S(\Delta)\},\]
and
\[\bar{\lambda}_0^+ = \sup\{\lambda \in [\lambda_0^-, \lambda_0^+] \mid \pi_i(\lambda) \in \corr^S(4k)\}.\]
We create events at times $\bar{\lambda}_0$ and $\bar{\lambda}_0^+$ for robot $i$ during the 
interval $[\lambda_0^-, \lambda_0^+]$, corresponding to the time $\bar{\lambda}_0$ that $i$ enters the 
sanctum $\corr^S(\Delta)$ for the first time in $[\lambda_0^-, \lambda_0^+]$, and the time $\bar{\lambda}_0^+$ that it 
leaves $\corr^S(4k)$.
We collect all events associated with $\corr$ and process them in increasing order
of time. At any given time, some parking places of $P_\corr^-$ and $P_\corr^+$ will be occupied.
We maintain the invariant that at any given time, the deepest parking places are occupied, namely
if $p_j^-$ (resp. $p_j^+$) is occupied and $j < k$, so is $p_{j+1}^-$ (resp. $p_{j+1}^+$).
We process each event as follows:

\begin{enumerate}[(i)]
    \item \textit{$\bar{\lambda}_0$ event for robot $i$}: If $i$ enters $\corr$ through the portal $\sigma_\corr^-$ (resp. $\corr_\sigma^+$)
    then move $i$ to the deepest unoccupied parking place of $P_\corr^-$ (resp. $P_\corr^+$)
		while keeping all other robots at their current positions, using a local surgery (cf.\ \lemref{ab-portal-cases}).
		It remains parked at one of the points of $P_\corr^-$ (resp.\ $P_\corr^+$) 
		until the time instance
    $\lambda_0^+$, though it may be moved to a deeper parking place of the set
    to maintain the above invariant.
    \item \textit{$\bar{\lambda}_0^+$ event for robot $i$}: Without loss of generality assume $i$ exits $\corr^S(4k)$ at time $\bar{\lambda}_0^+$ through the
    portal $\sigma_\corr(4k)^+$. Suppose $i$ had entered $\corr$ through $\sigma_\corr^+$, then it is parked at a point
    $p_j^+$ of $P_\corr^+$ and $p_1^+, \ldots p_{j-1}^+$ are unoccupied (by \lemref{ab-inner-products}).
		We move $i$ from $p_j^+$ to $\pi_i(\bar{\lambda}_0^+)$ using a local surgery 
		(cf.\ \lemref{ab-portal-cases}) while
    keeping all other robots at their current positions. If $i$ entered $\corr$ 
		through $\sigma_\corr^-$ then
    \lemref{ab-inner-products} implies that $i$ is parked at $p_k^-$ of $P_\corr^-$. Again, we move $i$ from $p_k^-$ to $\pi_i(\bar{\lambda}_0^+)$ using a local surgery as above.
     Finally, we move all robots parked at points of $P_\corr^-$ from their current positions to the next deeper position of $P_\corr^-$ 
    to maintain the invariant that the deepest parking placements of $P_\corr^-$ are occupied.
\end{enumerate}

    We show in \secref{far-full}  that the modified plan is feasible, does not contain
 any $(\Delta+10k)$-far breakpoints inside $\corr$, and bound the cost of the new plan.
Applying this procedure repeatedly for all corridors, we obtain the following:

\begin{lemma}\label{lem:ab-corr}
    Let $\pth$ be a min-sum $(\bs,\bt)$-plan.  There exists a plan $\pth'$ with 
no $(\Delta + 10k)$-far breakpoints inside a corridor whose cost is
$\cost(\pth') \leq (1 + \frac{C_\corr k}{\Delta})\cost(\pth)$,
    where $C_\corr$ is a constant.
\end{lemma}

\subsection{Near-optimal motion in wide  areas}
\label{subsec:ab-fat}

We now describe the surgery 
to remove $\Delta$-far breakpoints from wide areas.
Here we set $\Delta = Ck^2$, for some sufficiently large constant $C>0$.
Unlike corridors, the relative ordering of robots in a wide area may change and the robots may 
cross each other,
which makes the surgery considerably more involved. We heavily rely on revolving areas, particularly WSRAs.
The following two lemmas lay the groundwork for performing the surgery and are used repeatedly. 
The first lemma states that one can carve out sufficient number of WSRAs in a neighborhood of sufficiently far point, and the second lemma states that two robots can move freely within an RA. 

\begin{lemma}\label{lem:ab-wsras-count}
    Let $j \geq 0$ be an integer. For any $3(24)j$-far point $q \in \F$ that lies in a wide area, 
    there are $j$ well-separated revolving areas (WSRAs) in $K(q,24j)$.
\end{lemma}

\begin{lemma}\label{lem:ab-jiggle}
    Let $p \in \F$ be a point such that $\RA(p) \subseteq \F$, and let $a,b \in \partial B(p,2)$ such that $||a-b||_\infty \geq 2$.
	Then there exists a feasible
    $((p,a),(p,b))$-plan $\tilde{\pth} = (\pi_1,\pi_2)$ for two robots (i.e., one robot moves from $a$ to $b$ while the other
    begins and ends at $p$) such that $\pi_1, \pi_2 \subset B(p,2)$, and $\cost(\tilde{\pth}) < 20$.
\end{lemma}

With these two lemmas in hand, we show that if some of the robots are placed inside 
$K(q,24k)$, for a $90k$-far point $q\in\F$, at any given time in $\plan$, they 
can be \emph{dispersed} to WSRAs within $K(q,24k)$ by performing a \emph{local surgery} on the 
plan $\plan$ at a cost of $O(k^2)$. 
To accomplish this dispersion, we heavily rely on RAs, which 
exist in wide areas by \lemref{ra-in-corr}, and in order to ensure that moving in and out of RAs does not cause conflicts, it is convenient to work with WSRAs.
Skipping the details of the local surgery and its correctness from here, 
which can be found in \secref{far-full}, we summarize the result:

\begin{lemma}\label{lem:ab-dispersion}
    Let $q \in \F$ be a $90k$-far point that lies in a wide area, let $R = \{r_1, \ldots, r_k\}$ be a set of centers of WSRAs
    in $K(q,24k)$, let $\BS \subseteq [k]$ be a subset of $j \leq k$ robots, and let 
$\ba = (a_1, \ldots, a_k)$ and $\bb = (b_1, \ldots, b_k)$
    be two feasible configurations such that $a_j \in K(q,24k)$ and $b_j \in R$ for all $j \in \BS$. Then there
    exists a feasible $(\ba,\bb)$-plan of cost at most $C_1k^2$ for some constant $C_1 > 0$.
\end{lemma}

\paragraph{Surgery.}

We fix four parameters: $\delta := Ck$ where $C$ is a sufficiently large constant,
$\Delta := \delta k = Ck^2$, $\Delta^- := \Delta - 90k$ and $\Delta^+ := \Delta + 90k$. 
For $90k$-far point $q$, let $K(q) := K(q,24k)$. We now describe the \emph{global} surgery
to remove all $\Delta^+$-far breakpoints from $\plan$ that lie in  wide areas.
Roughly speaking, if the path $\pi_i$ of a robot $i$ in $\plan$ contains a $\Delta^+$-far breakpoint,
robot $i$ must reach the $\Delta$-frontier at some time $\lambda$.  Let 
$[\lambda^-,\lambda^+]$ be the maximal interval containing $\lambda$ such that $\lambda^-<\lambda$ 
is the last time when $i$ is $\Delta$-close before time $\lambda$, 
and $\lambda^+>\lambda$ is the first time after time $\lambda$ when $i$ is $\delta$-close.
We park $i$ at a point $q$, the center of a WSRA of $K(\pi_i[\lambda^-])$.
We also park all other robots that pass
through $K(\pi_i[\lambda^-])$ in individual WSRAs of $K(\pi_i[\lambda^-])$, using the local surgery. 
We keep each robot $j$ that has parked in a WSRA ``frozen'' there until we reach a time
$\lambda'$ at which $\pi_j(\lambda')$ becomes $\delta$-close again.
Since $t_j \in \Lambda$, for all $j \in [k]$, 
robot $j$ eventually becomes $\delta$-close.
At time instance $\lambda'$, we freeze all other robots at their current positions
and move $j$ from its current position to $\pi_j(\lambda')$, using Lemmas~\ref{lem:ab-dispersion} 
and~\ref{lem:ab-jiggle} repeatedly. We now describe this global surgery in more detail.

Let $\pth:[0,1] \to \BF$ be a feasible (piecewise-linear) $(\bs,\bt)$-plan. We maintain the following information:
  (i) a set $Z \subseteq \BR$ of frozen robots;
  (ii) a set $Q = \{q_1, \ldots, q_s\}$, where $s \leq k$, of $\Delta$-frontier points such that 
the robots of $Z$ are parked in WSRAs of $K(q_i) := K(q,24k)$ for some $q_i \in Q$;
  (iii) for $i \le s$, $R_i = \{r_{i1}, \ldots, r_{ik}\} \subset K(q_i)$, the set of centers of $k$ 
WSRAs according to \lemref{wsras-count};
  (iv) $R_i^{O} \subseteq R_i$ and $R_i^{U} \subseteq R_i$: the set of \emph{occupied} and \emph{unoccupied} (centers of) WSRAs of $R_i$;
  (v) $Z_i \subset Z$, the set of robots parked in the occupied WSRAs of $K(q_i)$; note that
  $|Z_i| = |R_i^O|$ and $Z_1, \ldots, Z_s$ is a partition of $Z$ (a frozen robot may lie in the neighborhood
  of multiple points in $Q$, but it is assigned to only one of them); and (vi) 
  a matching $M_i \subseteq Z_i \times R_i^O$ where $(j,p)$ implies that robot $j$ is parked 
at $p \in R_i^O$ (only one robot is parked at $p$).
Initially, $Z = \emptyset$, $Q = \emptyset$, and thus the rest of the sets are undefined.

We construct an $(\bs,\bt)$-plan $\pth':[0,1] \to \BF$
in which at any time $\lambda \in [0,1]$, any frozen robots in $Z$ are parked at a $\left(\Delta\right)$-far
(but $\Delta^+$-close) WSRA,
and for any unfrozen robot $j \in \BR \setminus Z$, $\pi_j'(\lambda) = \pi_j(\lambda)$. Any robot $i \in Z$ remains frozen while $\pi_i(\lambda)$
is $\delta$-far, and any robot $j \in \BR \setminus Z$ remains thawed while $\pi_j(\lambda)$ is both $\Delta$-close and outside of any neighborhoods $K(q_i)$, where
$q_i \in Q$.
Thus, we have one of three types of \emph{critical events} at a time
$\lambda \in [0,1]$ that cause a robot to either become frozen or thawed:

{
  \renewcommand{\labelenumi}{\textbf{(T\arabic{enumi})}:}
  \setlength{\leftmargini}{3em}  
\begin{enumerate}
    \item A thawed robot $i \in \BR \setminus Z$ reaches the $\Delta$-frontier at time $\lambda$, i.e.,
    $\pi_i(\lambda)$ lies at the $\Delta$ frontier.
    \item A thawed robot $j \in \BR \setminus Z$ enters $K(q_i)$ for some $q_i \in Q$ at time $\lambda$, i.e., $\pi_j(\lambda)\in \partial K(q_i)$.
    \item A frozen robot $i \in Z$ becomes $\delta$-close at time $\lambda$, i.e., $\pi_i(\lambda)$ lies at the $\delta$-frontier.
\end{enumerate}
}

By traversing $\pth$ in increasing order of time and maintaining $Z$ and $Q$, we can determine all critical events in
a straightforward manner. Let $\lambda_1 = 0 < \lambda_2 < \ldots < \lambda_m = 1$ be the sequence
of critical events. For simplicity, we assume that only one event occurs at each critical event,
though it is easy to adapt the procedure to handle multiple simultaneous events.
Suppose we have processed $\lambda_1, \ldots, \lambda_{i-1}$ and constructed $\pth'[0,\lambda_i)$. We process the event $\lambda_i$
and construct $\pth'[\lambda_i, \lambda_{i+1})$.
The surgery proceeds by case analysis
depending on the type of critical event for $\lambda_i$:

    \paragraph{(T1)} Suppose robot $j \in \BR \setminus Z$ reaches the $\Delta$-frontier. Add $q_{s+1} := \pi_j(\lambda_i)$
    to $Q$. Compute WSRAs $R_{s+1}$ of $K(q_{s+1})$ using \lemref{ab-wsras-count}. Let $Z_{s+1} \subseteq \BR \setminus Z$
    be the set of unfrozen robots including $j$, that lie in $K(q_{s+1})$, i.e., the set of all $j \in \BR \setminus Z$
    such that $\pi_j(\lambda_i) \in K(q_{s+1})$. Assign $|Z_{s+1}|$ WSRAs of $R_{s+1}$ to $Z_{s+1}$, set them to be
    $R^{O}_{s+1}$, and set $R^{U}_{s+1} = R_{s+1} \setminus R_{s+1}^{O}$.
    Let $M_{s+1}$ be the resulting assignment. Using \lemref{ab-dispersion},
    move all robots in $Z_{s+1}$
    to their assigned parking placements, namely, centers of the assigned WSRAs. Let $\bphi$ be the plan computed by the local surgery.
    We define $\bpsi:(\lambda_i, \lambda_{i+1}) \to \BF$ as follows:
    For $j \in Z$, $\psi_{j}(\lambda) = \varphi_j$ for all $\lambda \in (\lambda_i, \lambda_{i+1})$
    where $\varphi_j$ is the assigned parking place of $j$. For $j \in \BR \setminus Z$,
    $\psi_{j}(\lambda) = \pi_j(\lambda)$ for all $\lambda \in (\lambda_i, \lambda_{i+1})$. We set $\pth'[\lambda_i, \lambda_{i+1}) = \bphi \circ \bpsi$.

    \paragraph{(T2)} Suppose robot $j \in \BR \setminus Z$ enters $\partial K(q_u, 80k)$ for some $q_u \in Q$. We choose an unoccupied point
    $\xi \in R_u^U$ and assign $\RA(\xi)$ to $j$. Add $j$ to $Z_u$ and $Z$, move $\xi$ from $R_u^U$ to $R_u^O$, and add $(j, \xi)$ to $M_u$.
    Using the local surgery we compute a feasible plan $Q$ that moves $j$ from $\pi_j(\lambda_i)$ to $\xi$ while the other robots
    begin and end at their current position. Let $\bphi$ be the plan computed by the local surgery, and define $\bpsi$ as in a (T1) event.
    Set $\pth'[\lambda_i, \lambda_{i+1}) = \bphi \circ \bpsi$.

  \paragraph{(T3)} Suppose $j \in Z_u$ for some $u \leq s$. 
    Remove $j$ from $Z$ and $Z_u$, update the sets $R_u^O, R_u^U, M_u$, and remove $q_u$
    from $Q$ if $R_u^O = \emptyset$ (all the sets associated with $q_u$ are also removed).
    Let $P_j$ be the shortest $\delta$-far path from the WSRA of $j$ to its next $\delta$-close
    position, $\pi_j(\lambda_i)$ in $\pth$. If there exists a point $\xi \in P_i$
    and an unfrozen robot $h \in \BR \setminus Z$ such that 
    $\|\pi_h(\lambda_i)-\xi\|_\infty < 2$,
    i.e., $h$ conflicts with $P_i$, using the local surgery for the unfrozen robots in $K(\xi)$,
    we temporarily move them to WSRAs in $K(\xi, 80k)$ and freeze them there. We repeat this for 
all robots that interfere with $P_j$.
    Let $Y_i \subseteq \BR \setminus (Z \cup \{j\})$ be the set of robots that are moved to 
    WSRAs in this step.
    Move $j$ to $\pi_j(\lambda_i)$ along $P_j$ using \lemref{ab-jiggle} if $j$ ever enters a WSRA 
    of some frozen robot in $Z\cup Y_i$. Once $j$ reaches $\pi_i(\lambda_i)$, undo the 
surgery performed above to move all robots $h \in Y_i$ back to their original positions 
$\pi_h(\lambda_i)$. Let $\bphi$ be the plan
    describing the above motion and $\bpsi$ as in (T1),
    then we set $\pth'[\lambda_i, \lambda_{i+1}) = \bphi \circ \bpsi$.

Let $\pth'$ be the $(\bs,\bt)$-plan obtained after processing all events. We prove in \secref{far-full}
that $\pth'$ is feasible and does not contain any $\Delta^+$-far breakpoints, and bound 
$\cost(\pth')$. Omitting the details from this section, we conclude the following:

\begin{lemma}\label{lem:ab-wide}
  Let $\pth$ be a min-sum $(\bs,\bt)$-plan, and let $\Delta = C_1k^2$ for some constant $C_1>0$.
  There exists a feasible, piecewise-linear $(\bs,\bt)$-plan $\pth'$
  that does not contain any $\Delta$-far breakpoints and has cost
  $\cost(\pth') \leq \left(1 + \frac{C_W k^2}{\Delta}\right)\cost(\pth)$, where $C_W$ is a constant independent of $k$ and $\Delta$.
\end{lemma}

Combining Lemmas~\ref{lem:ab-corr} and~\ref{lem:ab-wide} and choosing an appropriate value 
of $\Delta$, we obtain \lemref{far}.

\section{Optimal, Robust Plans with Few Breakpoints}\label{sec:nice}




In this section, we show that there exists a min-sum plan with a number of breakpoints proportional to its length.
Recall that, for a piecewise-linear $(\bs,\bt)$-plan $\pth$, we denote the number of breakpoints of $\pth$ by $\beta(\pth)$.
First, we begin by establishing some auxiliary lemmas, starting with the following elementary observation:

\begin{lemma}\label{lem:segment}
    Let $s$ denote an axis-aligned segment with length at most $2$.
    Then $s \cap \F$ is connected (and possibly empty) in $\F$.
\end{lemma}

We say that a path $\pi$ is $xy$-monotone if $\pi$ is monotone in both of the $x$- and $y$- coordinates.
Then it is easy to see from \lemref{segment} that any shortest path in a
connected component of $p + \square \cap \F$ is $xy$-monotone.
Further, it is shown in \cite{steiger2024} that \lemref{segment}
gives directly the following lemma.

\begin{lemma}\label{lem:shortest-local}
    Let $p \in \F$. Then,
    \begin{enumerate}[(i)]
        \item $\partial(p + \square) \cap \F$ has at most two connected components,
        \item $p + \square \cap \F$ is composed of $xy$-monotone components (without holes), and
        \item if $a,b \in \F$ lie in the same connected component of $p + \square \cap \F$,
        then the shortest path $\pi^*$ from $a$ to $b$ is $xy$-monotone and is contained in $p + \square$.
    \end{enumerate}
\end{lemma}



\begin{figure}
    \centering
    \includegraphics[width=0.45\textwidth]{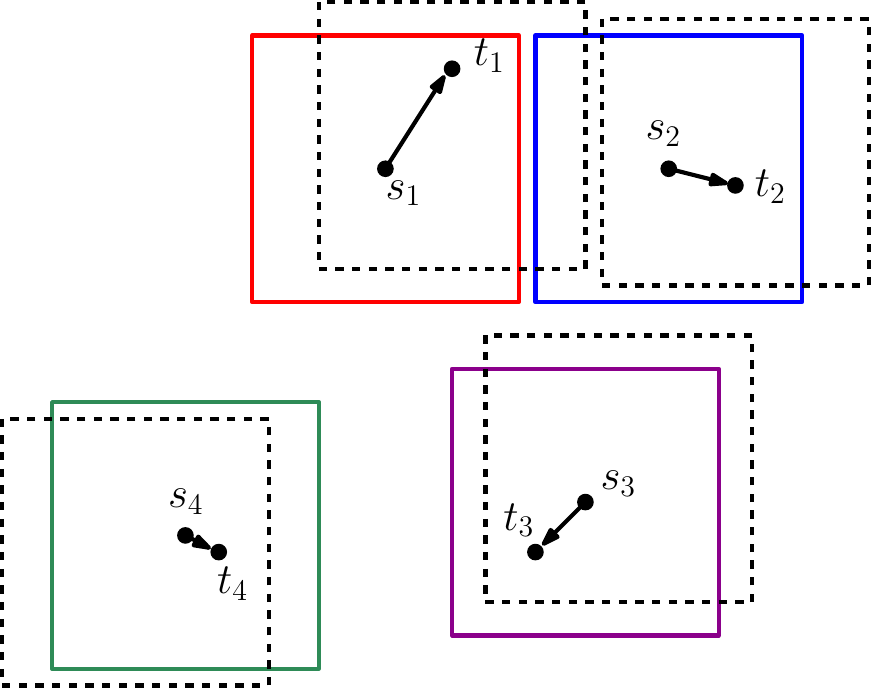} \hspace{1cm} \includegraphics[width=0.44\textwidth]{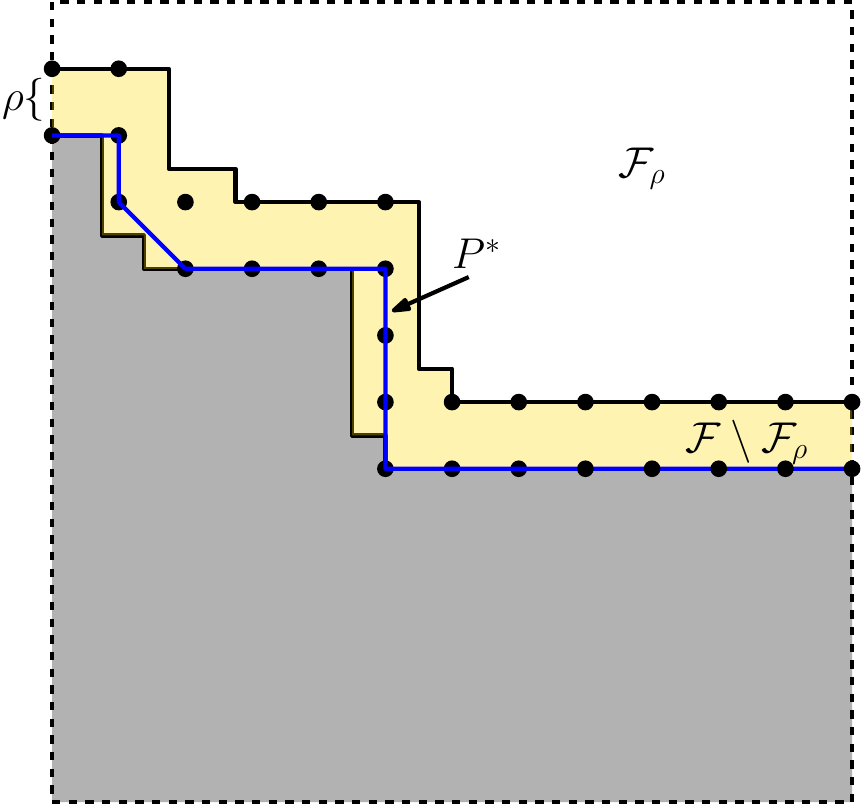}
    \caption{(Left) an example configuration satisfying \lemref{geodesic-plan}. Note that the order type
    is defined by $\theta_{ij} \in \{x,y\}$ for each pair $(i,j) \in [4]^2$, where $\theta_{12} = x, \theta_{13} = y,
    \theta_{14}=y, \theta_{23}=y, \theta_{24}=x \text{ or } y, \text{ and } \theta_{34} = x$. (Right) Reducing the number of vertices in $\square_i \cap \F_\rho$ using $\rho$-robustness, by snapping vertices of $\F$ to
    the nearest $\rho/2$-width grid vertices contained in $\F_\rho$. The blue polygonal curve $P^*$ after snapping is used in place of $\partial(\F)$.
    The resulting free space $\F^*$ now has $|V(\F^*_i)| \leq 2/\rho$, and still contains $\F_\rho$.}
    \label{fig:f-rho-gridsnap}
\end{figure}

Let $\rho > 0$ be a parameter. We next describe a simplification of $\F$ in terms of $\rho$-robustness. Let $\F_\rho \subset \F$ define the 2D free space for a $(1+\rho)$-radius square and let $\square_i$ be a unit square placed in $\P$ for some robot $\robot{i} \in \BR$. Examine $\F \cap \square_i$ in comparison with $\F_\rho \cap \square_i$. We claim that it is straightforward to
simplify $\partial(\F) \cap \square_i$ into another polygonal curve $P^*$ for which
$P^* \subset \cl(\F \setminus \F_\rho)$ and $P^*$ has only $2\rho^{-1}$ vertices:
In a connected component $C_\rho$ of $\cl(\F \setminus \F_\rho)\cap \square_i$, consider the overlay of the axis-aligned grid
$\G_\rho$ with $\rho/2$-width grid cells on $C_\rho$. Snap every vertex of $\partial(\F) \cap C_\rho$ to its nearest vertex
in the overlay of $\G_\rho$ that is contained in $C_\rho$. Denote by $P^*$ the resulting polygonal curve; by construction
$P^*$ has $2\rho^{-1}$ vertices since $C_\rho$ is $xy$-monotone by \lemref{segment}.    
Let $\F^i_*$ denote the polygonal free space within $\square_i$ that results by substituting $P^*$ as the boundary of $\F_\rho \cap \square_i$. By construction, 
$(\F_\rho \cap \square_i) \subseteq \F^* \subseteq (\F \cap \square_i)$. See \figref{f-rho-gridsnap}. Since $\partial(\F^i_*)$ has $2\rho^{-1}$
vertices, the vertical decomposition $\VD(\F^i_*)$ also has complexity $2\rho^{-1}$. This is summarized in the following lemma:

\begin{lemma}\label{lem:robust-freespace}
    Let $\rho > 0$ be a parameter, let $\square_1, \square_2, \ldots, \square_k$ denote $k$ axis-aligned, unit-radius squares, let $\F_\rho \subset \F$ define the 2D free space for a $(1+\rho)$-radius square robot, and let $\F^i_* \supseteq \F \cap \square_i$ be as defined above for each robot $\robot{i} \in \BR$. Then the vertical decomposition $\VD(\F^i_*)$ of $\F^i_*$ has complexity $2\rho^{-1}$ for each $\robot{i} \in \BR$.
\end{lemma}

Suppose for $\bp = (p_1, \ldots, p_k),\bq = (q_1, \ldots, q_k) \in \BF$
that, for each $i \in \BR$, the shortest path from $p_i$ to $q_i$ (in the absence of any other robots) is
$\pi_i^*$. We call a min-sum plan $\pth^*$ a \emph{geodesic plan} if $\pi_i = \pi^*_i$ for every robot $\robot{i}\in \BR$.
In the following, we observe a situation in which a geodesic plan is indeed feasible, and hence it is optimal since, for any $(\bs,\bt)$-plan $\pth$, by definition we have $\cost(\pth^*) \leq \cost(\pth)$. In particular, this occurs when the geodesic paths between each pair $p_i,q_i$ trace a line segment in $\F$, which is the case when $p_i,q_i$ lie in some convex region in $\F$ (such as a cell of a vertical decomposition).
In the end, this leads to a bound on the maximum number of breakpoints in a min-sum plan by a packing argument, formalized in \lemref{bound}.

For any configuration $\bp \in \BF$ and any pair of robots $\robot{i},\robot{j} \in \BR$, there exists a direction $\theta_{ij} \in \{x,y\}$ such that $\bp_i,\bp_j$ are $\theta_{ij}$-separated.
We call the corresponding tuple of $k \choose 2$ directions
$\mathbf{\Theta} = (\theta_{12}, \theta_{13}, \ldots \theta_{ij}, \ldots, \theta_{(k-1)k})$
the \emph{order type} with respect to the configuration $\bp$ (if a pair of placements is both $x$-separated and $y$-separated, the configuration has multiple order types).

\begin{lemma}\label{lem:geodesic-plan}
    Let $\square_1, \square_2, \ldots, \square_k$ denote $k$ axis-aligned, unit-radius squares
    placed in $\P$ so that their interiors do not intersect.
    Suppose $\bs = (s_1, s_2, \ldots, s_k), \bt = (t_1, t_2, \ldots, t_k) \in \BF$ are such that the segment $s_it_i$ is contained in $\F$ for every $i \in [k]$ and $\bs,\bt$ have the same order type.
    Then there exists a geodesic $(\bs,\bt)$-plan that has no breakpoints in its interior.
\end{lemma}

\begin{proof}
    For each robot $\robot{i} \in \BR$, let $\pi_i^*$ be the line segment from $s_i$ to $t_i$.
    Let $\pth^* = (\pi_1^*, \ldots, \pi_k^*)$ be the $(\bs,\bt)$-plan obtained by linearly interpolating each robot $\robot{i}$ between the endpoints of segment $\pi_i^*$. That is, for any $\gamma \in [0,1]$ and any robot $\robot{i} \in \BR$, define $\pi_i^*(\gamma)$ by:
    \[\pi_i^*(\gamma) = \gamma t_i + (1-\gamma) s_i. \]
    
    We next argue that the resulting plan $\pth^*$, parameterizing every path in this manner, is feasible. Let $\robot{i},\robot{j}$ be any two robots in $\BR$.
    By assumption, there is some $\theta \in \{x,y\}$
    for which both configurations $(s_i, s_j)$ and $(t_i, t_j)$ are $\theta$-separated.
    Without loss of generality, assume $\theta = x$ so the configurations are both $x$-separated.
    Additionally, using standard transformations as needed, assume that $s_i$ lies to the left of $s_j$, i.e., $x(s_i) \leq x(s_j) - 2$. Then either $x(t_i) \leq x(t_j)-2$ or $x(t_i) \geq x(t_j)+2$ since $(t_i,t_j)$ is $x$-separated.
    Furthermore, we have $x(t_i) \leq x(s_i)+1$ and $x(t_j) \geq x(s_j)-1$ and thus $$x(t_i) \leq x(s_i)+1 \leq x(s_j)-1 \leq x(t_j)$$ since $\square_i,\square_j$ have unit radii. Hence it cannot be that $t_i$ is right of $t_j$ (otherwise $x(t_i) \geq x(t_j)+2$, a contradiction), so $x(t_i) \leq x(t_j)-2$.
    
    We next show that $x(\pi_i^*(\gamma)) \leq x(\pi_i^*(\gamma)) - 2$ for any $\gamma \in [0,1]$,
    which implies the feasibility of the plan $(\pi_1^*, \pi_2^*)$ between any pair of robots.
    Expanding definitions,
    \begin{align*}
        x(\pi_i^*(\gamma)) - x(\pi_j^*(\gamma)) &= \gamma \cdot x(t_i) + (1-\gamma)x(s_i) - \left(\gamma \cdot x(t_j) + (1-\gamma)x(s_j)\right) \\
        &= \gamma\left(x(t_i) - x(t_j)\right) + (1-\gamma)\left(x(s_i) - x(s_j)\right)\\
        &\leq \gamma \cdot (-2) + (1-\gamma) \cdot (-2)\\
        &= -2.
    \end{align*}
    Hence, $x(\pi_i^*(\gamma)) \leq x(\pi_j^*(\gamma)) - 2$, for any time $\gamma \in [0,1]$. It follows that the parameterization given by the specified linear interpolation is feasible.
\end{proof}

\subsection{Min-sum plans with few breakpoints}\label{sec:breakpoints}
We next bound the number of breakpoints in any min-sum plan in terms of its cost by applying the previous lemma and a packing argument.

\begin{lemma}\label{lem:bound-full}
    Given $\rho$-reachable configurations $\bs,\bt \in \BF$, there exists a $(\bs, \bt)$-plan
    $\pth' = (\pi_1', \ldots, \pi_k')$, for which $\cost(\pth') \leq \opt(\bs,\bt, 1+\rho)$,
    with $\beta(\pth') \leq f(k, \rho) (\cost(\pth') + 1)$ for a function $f(k, \rho)$
    that is exponential in $k$.
\end{lemma}

\begin{proof}
    Let $\pth$ be any $\rho$-robust $(\bs,\bt)$-plan.
    We construct a plan $\pth'$ (that lies in $\F$) from $\pth$ such that $\cost(\pth') \leq \cost(\pth)$,
    so that $\beta(\pth')$ is proportional to $\cost(\pth')$,
    using \lemref{geodesic-plan} and the fact that $\pth$ is $\rho$-robust.

    As defined above, let $\F_\rho \subset \F$ define the 2D free space for a $(1+\rho)$-radius square robot.
    The high-level idea is that we can simplify $\pth$ into a new plan $\pth'$ by shortcutting between close configurations of $\pth$ with the same order type (out of the $2^k$ possible order types) using \lemref{geodesic-plan}, while introducing at most one new configuration for each such pair, then charge the breakpoints of $\pth$ uniquely to (i) the length of the subplan between the breakpoints or (ii) the number of combinations of order types and vertices of (a simplification of) $\F$ within the neighborhoods of the breakpoints. The details are as follows.
    
    The construction of our $(\bs,\bt)$-plan $\pth'$ is described in an iterative fashion, which maintains a time variable $\lambda' \in [0,1]$ that increases monotonically from each iteration to the next. 
    Each iteration is within an \emph{epoch}, and each epoch may consist of multiple iterations. Each epoch is defined by a time $\lambda_0' \in [0,1]$.
    
    We initialize the first epoch at time $\lambda_0' = 0$. In any epoch, let $\bp = \pth(\lambda_0')$ and let $\square_1, \ldots, \square_k$ be the unit squares centered at the respective points of $\bp$; that is, $\square_i = p_i + \square$ for each robot $i \in \BR$. Additionally, in any fixed iteration (of some epoch), let $\Theta$ be an order type of $\bp$, let $\tau_i$ be the cell of the vertical decomposition $\VD(\F^i_*)$ containing\footnote{If a robot $\robot{i}$ lies on an edge of $\VD(\square_i \cap \F^*)$, we either choose $\tau_i$ as the cell next intersected by $\pth_i[\lambda',1]$, if any, and otherwise pick either arbitrarily.} $\pth_i(\lambda')$, where $\F^i_*$ is the simplification of $\square_i \cap \F$ as defined earlier in this section. Note that the $\square_i$'s do not change within an epoch but the $\tau_i$'s may change.
    
    We next describe an iteration of our construction.
    We choose $\lambda_\Theta \in (\lambda,1]$ to be maximum time such that $\pth(\lambda_\Theta)$ has order type $\Theta$ and $\pi_i(\lambda_\Theta) \in \tau_i$ for all robots $\robot{i} \in \BR$, if it exists. There are two cases to consider:
    \begin{itemize}
        \item $\lambda_\Theta$ does not exist. Let $\bq$ be the next breakpoint in $\pth$ after $\bp$, and let $\lambda_q > \lambda'$ be the (first) time after $\lambda'$ such that $\pth(\lambda_q) = \bq$. We simply concatenate subplan $\pth(\lambda',\lambda_q)$ to the end of our partial plan $\pth'$. If $\bq = \bt$ then we are done, so suppose $\bq \neq \bt$. If $q_i \in \square_i$ for each robot $\robot{i} \in \BR$ then we update $\lambda'$ to $\lambda_q$ then proceed to the next iteration (in the same epoch). Otherwise, we start a new epoch at $\lambda_q$. Note that the last subcase is the only instance in which a new epoch starts.
        
        \item Otherwise, $\lambda_\Theta$ exists. Let $\bq = \pth(\lambda_\Theta)$. At time $\lambda_\Theta$, either $\lambda_\Theta = 1$ or at least one robot crosses the boundary of its current cell $\tau_i$ into the interior of another cell in $\square_i$ immediately after time $\lambda_\Theta$.
        We have that $\pth(\lambda'),\pth(\lambda_\Theta)$ have order type $\Theta$ and $\pi_i(\lambda'),\pi_i(\lambda_\Theta) \in \tau_i$ for all robots $\robot{i} \in \BR$. By \lemref{geodesic-plan}, there exists a geodesic $(\bp,\bq)$-plan $\pth''$. So, we concatenate $\pth''$ (reparameterized from $\lambda'$ to $\lambda_\Theta$) to the end of our partial plan $\pth'$. Since $(\F_{\rho} \cap \square_i) \subseteq \F^*_i$ and $\pi''_i$ is simply a line segment for all robots $i \in \BR$, we have $\cost(\pi_i[\lambda',\lambda_{\Theta}]) \geq \cost(\pi_i'[\lambda',\lambda_{\Theta}])$.
    If $\lambda_\Theta = 1$ then we are done, and otherwise we update $\lambda'$ to be $\lambda_\Theta$ then proceed to the next iteration (in the same epoch).
    \end{itemize}
    

    Clearly, the iterative process ends with $\pth'$ as a $(\bs,\bt)$-plan with cost $\cost(\pth') \leq \cost(\pth)$. It remains to show that the maximum number of breakpoints of $\pth'$, $\beta(\pth')$, is upper-bounded by $f(k,\rho)\cost(\pth')$, where $f(k,\rho) = 2^{k}\cdot(2/\rho^k)$.
    
    Consider the partition of $\pth'$ into subplans where the first breakpoint in each subplan is the start of an epoch. By definition, an epoch ends because the subplan of $\pth$ between the penultimate and last breakpoints of the epoch involves a robot $\robot{i} \in \BR$ leaving its square $\square_i$. Since such a robot was placed at the center of its square $\square_i$ at the start of the epoch, it must move at least distance $1$ from the center to the boundary of $\square_i$ during the epoch. It follows that there are at most $1+\cost(\pth)$ epochs, as the sum of lengths of such subpaths of $\pth$ that cause epochs to end can be at most the entire cost of the plan. Furthermore, since $|\VD(\F^i_*)| = 2\rho^{-1}$ by \lemref{robust-freespace}, we have $(2/\rho)^{k}$ possible distinct tuples of cells among the vertical decompositions of all $k$ robots in their respective fixed unit squares during an epoch, and there are $2^k$ distinct order types. Therefore, there are at most $f(k,\rho) = 2^{k}\cdot(2/\rho^k)$ breakpoints per epoch (excluding the its starting configuration). Putting everything together, we have that $\pth'$ has $\beta(\pth') \leq f(k,\rho)\cdot(\cost(\pth)+1)$ breakpoints.
    
    
\end{proof}

We remark that the bound in \lemref{bound} only depends on the complexity of the $\F_i^*$'s
if there are potentially very many vertices of $\F$ lying in any one unit-radius square in $\P$.
Assuming that there are no edges of $\F$ with length less than,
say, $1/c$ for a constant $c \geq 1$, we obtain the following easy corollary, and remove any dependence on the complexity
from the number of breakpoints. That is, the $\beta(\pth')$ is upper bounded by a function $f$ multiplied with the
cost of the plan, where $f$ depends only on $k$.

\begin{corollary}\label{cor:assume-breakpoints}
    Suppose the minimum length of an edge in $\F$ is at least $1/c$, for a constant $c \geq 1$.
    Given $\rho$-reachable configurations $\bs,\bt \in \BF$, there exists a $(\bs, \bt)$-plan
    $\pth' = (\pi_1', \ldots, \pi_k')$, for which $\cost(\pth') \leq \charge(\bs,\bt, 1+\rho)$,
    with $\beta(\pth') \leq f(k, \rho) (\cost(\pth') + 1)$ for a function $f(k)$
    that is exponential in $k$ (and has no dependence on $\rho$). 
\end{corollary}
\section{Extensions}\label{sec:ext}

In this section we briefly sketch how we extend the algorithm to centrally-symmetric
regular convex polygons and disks.

\begin{figure}
    \includegraphics[width=0.45\textwidth]{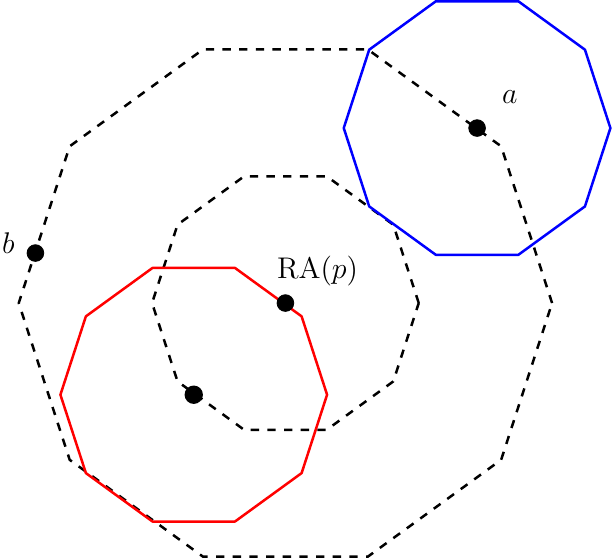} \hspace{1cm} \includegraphics[width=0.45\textwidth]{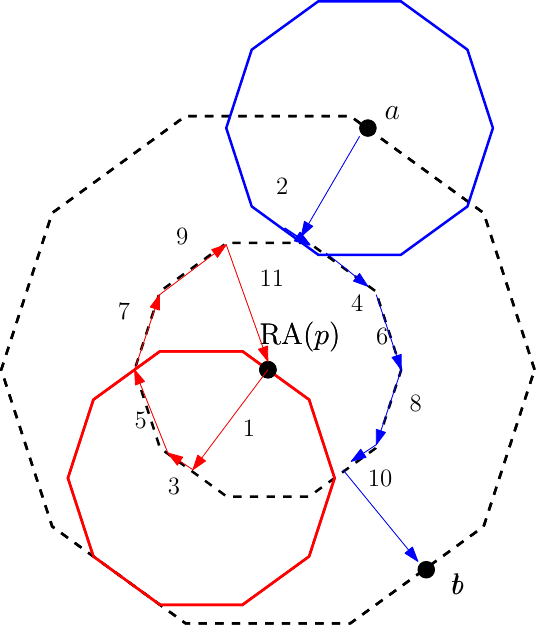}
    \caption{An example decoupled $((p,a), (p,b))$-plan $\pth$, which works identically to the one for squares in \lemref{jiggle}.}
    \label{fig:jiggle-poly}
\end{figure}

\paragraph{Case of regular $m$-gons.}
Let $m$ be an even integer, and let $\regpoly$ be a regular convex $m$-gon of unit apothem
centered at the origin, i.e., $\regpoly$ inscribes the unit circle. We now assume that each of the
robots of $\BR$ is modeled as a copy of $\regpoly$. The algorithm extends to this setting in a straightforward
manner--we have to redefine some of the concepts more generally. First, $\F$ is now defined
as $\F := \{x \in \P \mid x \oplus \regpoly \subseteq \P\}$. We define the \emph{Minkowski metric}
with respect to $\regpoly$, $\dpoly : \R^2 \times \R^2 \to \R_{\geq 0}$ as
\[\dpoly(p,q) = \inf \{\lambda \mid q \in p + \lambda \regpoly\}.\]

Since $\regpoly$ is centrally symmetric, $\dpoly(p,q) = \dpoly(q,p)$ and $\dpoly$ is a metric.
We replace the $\ell_\infty$-metric with the $\dpoly$-metric whenever we wish to measure distance between
two points, so $\bp = (p_1, \ldots, p_k)$ is a feasible configuration if $\dpoly(p_i, p_j) \geq 2$
for all $i \not= j$. However, we can continue to define a point being near or far
with respect to the $\ell_\infty$-metric for the sake of simplicity, and $K(q,r)$ is the connected
component of $\F \cap (q + r\poly)$ that contains $q$, where $\poly$ as above is the unit square centered
at the origin. We now define the revolving area centered at a point $p \in \F$ to be $\RA(p) = p + \regpoly$
if $p + \regpoly \subset \F$, and otherwise $\RA(p)$ is undefined.

There are two main modifications needed in our construction. First, portals of a corridor are now
defined by which pair of vertices of a homothet of $\regpoly$ are supported by the blockers, so there are
$O(m^2)$ different orientations of portals and $\dpoly$-distance between the endpoints
of a portal have to be less than 2 for a corridor.
It can be verified that \lemref{ab-ra-in-corr}
and \lemref{ab-inner-products} hold for $\regpoly$ with the modified definitions of corridors and revolving area. We state and prove the extension of \lemref{ab-inner-products} to a regular $m$-gon here.

The rest of the lemmas in \secref{thin-corridor} basically use packing and separation arguments
and thus hold verbatim for a centrally-symmetric regular polygon. As such, the surgery For
corridors works as described above. The surgery for wide areas relies on WSRAs and repeatedly
uses packing arguments, which also extend to centrally-symmetric regular $m$-gons without modification.

\begin{figure}
    \centering
    \includegraphics[height=2in]{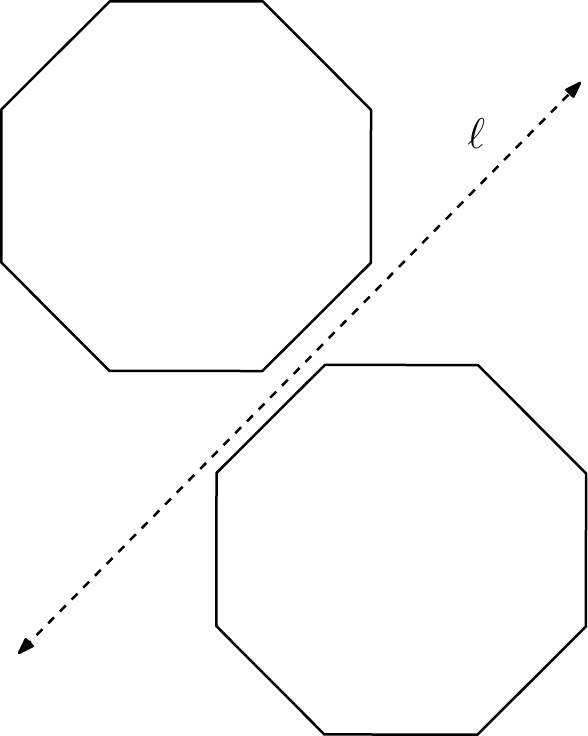}
    \caption{A pair of octagonal robots that are $\theta = \pi/4$-separated.}
    \label{fig:separated}
\end{figure}

The second major change is in bounding the number of breakpoints. Here we used the notion of two squares being $x$-separated
or $y$-separated and defined the order type of a configuration. We extend these notions to $\regpoly$
by defining $\theta$-separation. For $\theta \in [0,\pi)$, we say that two placements $p,q \in \F$ 
of $\regpoly$ are called \emph{$\theta$-separated} if a line in direction $\theta$
separated $\regpoly + p$ and $\regpoly + q$, i.e., they are contained in different
halfplanes bounded by $\ell$. See \figref{separated}.

Let $\bp = (p_1, \ldots, p_k)$ be a feasible configuration. It can be verified that for any pair $i \not= j$,
$p_i, p_j$ are $\theta$-separated for some $\theta \in \{\frac{h\pi}{m} \mid 0 \leq h < m\}$.
As earlier, we define the \emph{order type} of $\bp$ to be a vector of length $k \choose 2$ that
extends the separation type of each pair of robots. Note that the order type is not unique.

With this definition of order type at hand,
we get the analogous result to \lemref{geodesic-plan} for polygons,
Thus, following the same argument as in \secref{breakpoints}
shows that there exists an $\epsilon$-reachable, $\epsilon$-optimal plan
with $O(f(km,\epsilon))$ breakpoints.
Putting everything together, we obtain the following.

\begin{theorem}\label{thm:result-poly}
    Let $\P$ be a closed polygonal environment with $n$ vertices.
    Let $\BR$ be $k$ identical robots translating in $\P$, each
    modeled as a centrally-symmetric convex $m$-gon.
    Let $\bs, \bt$ be 
    source and target configurations of $\BR$, and let $\epsilon \in (0,1)$
    be a parameter. If there exists a feasible $\epsilon$-robust $(\bs,\bt)$-plan,
    then a feasible $(\bs,\bt)$-plan $\pth$ of $\BR$ with
    $\cost(\pth) \leq (1+\epsilon)\charge(\bs,\bt,1+\epsilon)+\epsilon$ can be
    computed in $f(km,\epsilon)(n)^{O(k)}$ time,  where $f(k,\epsilon) = \left(km/\epsilon\right)^{O(k^2)}$.
    If $\charge(\bs,\bt,1+\epsilon) > 1$, then $\cost(\pth) \leq (1+\epsilon)\charge(\bs,\bt,1+\epsilon)$.
\end{theorem}

\paragraph{Case of disks.}
It is well known that for any $\epsilon \in (0,1)$ a unit-radius disk $D$ can be $(\epsilon/3)$-approximated
by a centrally-symmetric $m$-gon $\regpoly_{\epsilon}$ with $m = O(\frac{1}{\sqrt{\epsilon}})$
in the sense that $D \subseteq \regpoly_{\epsilon} \subseteq (1+\epsilon/3)D$. Using \thmref{result-poly}
above, we compute an $(\epsilon/3)$-reachable $(1+\epsilon)$-approximate $(\bs,\bt)$-plan
$\pth$ for $\regpoly_\epsilon$.
Since $(1+\epsilon/3)\regpoly_\epsilon \subseteq (1+\epsilon/3)^2D \subseteq (1+\epsilon)D$,
$\cost(\pth) \leq (1+\epsilon/3) \charge(\bs,\bt, 1+\epsilon) + \epsilon/3$, where
$\charge(\bs,\bt, r)$ is the cost of an optimal $(\bs,\bt)$-plan for $k$ disks of radius $r$.
Hence, we obtain \thmref{result-disk}.

\section{Near-Optimal Motion Far from Vertices: A Complete Proof}\label{sec:far-full}

In this section we give the full proof of \lemref{far}, namely the existence of an $\epsilon$-optimal
$\Delta$-tame plan, in which all breakpoints of the paths of all robots lie in $\Delta$-sized
neighborhoods of landmarks, where $\Delta = O(k^2/\epsilon)$. As mentioned above, we do not need any robustness assumption here.
We prove the lemma by showing that if a plan $\pth$ has a breakpoint at time $\lambda$
such that $\pth(\lambda)$ is $\Delta$-far, we perform a surgery on $\pth$ that removes $\pth(\lambda)$
at a slight increase in the cost. At a very high level, suppose $\pi_i(\lambda)$
is $\Delta$-far for some $i \in \BR$. Let $(\lambda^-, \lambda^+)$ be the maximal interval such that
$\pi_i(\lambda)$ is $\Delta$-far for every $\lambda \in [\lambda^-, \lambda^+]$. The idea is to
\emph{freeze} the robot $i$ at time $\lambda^-$ by parking it at a location near $\pi_i(\lambda)$,
and keeping it at this position until time $\lambda^+$. We \emph{thaw} (unfreeze)
robot $i$ at time $\lambda^+$, freeze all other robots, and move $i$ along the shortest path in $\F$ from
$\pi_i(\lambda^-)$ to $\pi_i(\lambda^+)$. When $i$ reaches $\pi_i(\lambda^+)$,
we thaw all other robots and follow the original plan $\pth$. There are several challenges
in implementing this strategy. The frozen robots may conflict with the motion of other
thawed robots that are traversing $\pth$ as normal, so we also freeze any robots entering the
neighborhood of a frozen robot. Without care, this may cause a cascading effect. Similarly,
when we move robot $i$ from $\pi_i(\lambda^-)$ to $\pi_i(\lambda^+)$, $i$ may interfere
with other robots and we park them at an appropriate location as well. How exactly freezing,
thawing, and parking is performed is highly nontrivial and depends on the topology of $\F$ in a neighborhood
of $\pi_i(\lambda)$ of size $O(k)$. Roughly speaking, $\pi_i(\lambda)$ either lies in a ``narrow''
corridor, and motion inside a corridor is relatively constrained, or there is sufficient space
to park all robots that come sufficiently close to $\pi_i(\lambda)$.
We first characterize the neighborhood of a $\Delta$-far point (\secref{topology}), then describe the surgery for
$\pi_i(\lambda)$ lying in a corridor (\secref{thin-corridor}), followed by some useful surgery procedures that we
will use as subroutines of our global surgery
when $\pi_i(\lambda)$ is in a wide area  (\secref{fat}, local surgery). Finally,
we prove the feasibility of the modified plan and bound its cost (\secref{fat}, global surgery). First, we need
a few notations:

Recall that for a value $r > 0$ and a point $q \in \R^2$, we write $B(q,r) = q + r\square$ to
denote the square of radius $r$
centered at $q$, and let $K(q,r) \subseteq B(q,r) \cap \F$ be the connected component of $B(q,r) \cap \F$
that contains $q$ (assuming $q \in \F$; otherwise it is undefined). For simplicity we use $B(q)$ and $K(q)$
to denote $B(q,24k)$ and $K(q,24k)$, respectively, which will become important in \secref{fat}.
For a set $P \subset \F$ of points, we define $\F[P] =  \F \setminus \bigcup_{p \in P}B(p,2)$, which
is the conditional free space $\F$ for one robot when the $|P|$ robots are parked at points
in $P$. $\F[P]$ can be computed in $O(n + |P| \log^2{n})$ time \cite[Chapter~13]{debergbook}.

Similarly, for a pair of placements $(p_i, p_j) \in \F^2$, we say that $(p_i,p_j)$ is $x$-separated
(resp. $y$-separated) if the $x$ (resp. $y$) coordinates of $p_i$ and $p_j$ differ by at most two, i.e.,
$|x(p_i) - x(p_j)| \geq 2$ (resp. $|y(p_i) - y(p_j)| \geq 2$). Abusing notation slightly,
for a connected set of points $S \subset \F$, we may denote by $x(S)$ (resp. $y(S)$) the range in $x$ (resp. $y$)
coordinate values of $S$. That is, $x(S) = [\min_{p \in S}x(p),\max_{q \in S}x(q)]$ (resp. $y(S) = [\min_{p \in S}y(p),\max_{q \in S}y(q)]$).


\subsection{Topology far from vertices}\label{sec:topology}

In this subsection, we show that the topology of $\F$ in a neighborhood sufficiently far from the vertices
of $\F$ is simple, and then characterize these regions rigorously. Recall that
for a value $D \geq 0$, a point $p \in \F$ is called $D$-far (resp. $D$-close)
if the nearest landmark in $\Lambda$ is at least (resp. most) distance $D$ from $p$.

\begin{figure}
    \centering
    \begin{subfigure}[b]{0.47\textwidth}
        \centering
        \includegraphics[width=\textwidth]{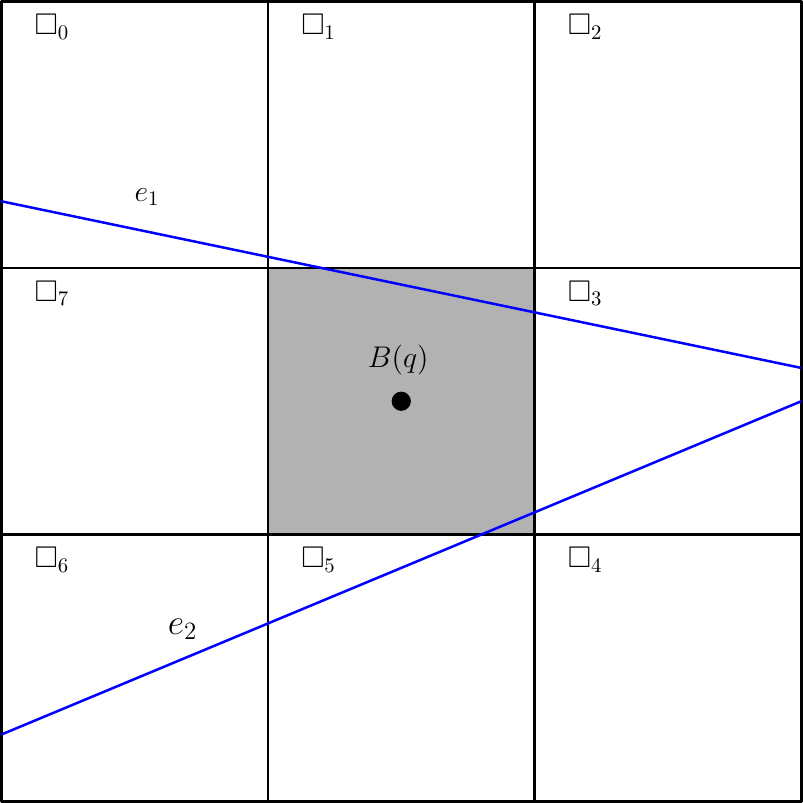}
        \caption{An example $B(q,D)$ for a value $D > 0$, containing $B(q, D/3)$ in gray, with center $q$ that is $D$-far.
            No landmarks of $\Lambda$ may lie inside $\interior(B(q, 3r))$, so at most two edges of $\F$ intersect $B(q, r)$.}
        \label{fig:fat-annulus}
    \end{subfigure}
    \hfill
    \begin{subfigure}[b]{0.47\textwidth}
        \centering
        \includegraphics[width=\textwidth]{images/corridor}
        \caption{An example corridor $\corr$.
            The left and right squares each have radius $\alpha < 2$, with its blockers $e_i, e_j$
            and portals $\sigma_L, \sigma_R$. The direction of the portals is $u_{ij}$, and
            the portal bisector is the line segment $\ell$. Since $\alpha < 2$, there are no revolving areas inside $\corr$.}
        \label{fig:corridor}
    \end{subfigure}
\end{figure}

\begin{lemma}\label{lem:fat-square}
    For any point $q \in \F$ that is $D$-far for some value $D > 0$, the boundary of $K(q, D/3)$
    intersects at most two edges of $\F$.
\end{lemma}
\begin{proof}
    By definition, no landmark of $\Lambda$ lies in $\interior B(q,D)$; it is possible that
    one or more points of $\Lambda$ lie on $\partial B(q,D)$.
    Observe that the annulus defining $\cl(B(q,D)\setminus B(q,D/3))$ can be subdivided into
    eight radius $D/3$ squares $(\square_0, ..., \square_7)$, written in clockwise order,
    surrounding $B(q, D/3)$. We label the portion of $\partial B(q,D)$ that overlaps with
    $\partial \square_i$ as $i$, for $0 \leq i \leq 7$.

    So, suppose for contradiction $\partial K(q,D/3)$ intersects three edges $e_1, e_2, e_3$
    of $\partial \F$. The endpoints of these edges do not lie in $\interior B(q,D)$ and their relative
    interiors are pairwise disjoint. Let $u_iv_i = e_i \cap B(q,D)$, for $1 \leq i \leq 3$.
    For each $i$, $u_iv_i$ appear in consecutive order along $\partial B(q,D)$ since the
    relative interiors of these
    segments are pairwise disjoint. See \figref{fat-annulus}.
    Assume their ordering along $\partial B(q,D)$ is $u_1, v_1, u_2, v_2, u_3, v_3$. A simple
    calculation shows that if $u_i$ appears in $i$ of $\partial B(q,D)$ then $v_i$ cannot
    appear before part $i+3\left(\textrm{mod\ } 8\right)$ of the boundary.
    
    Now, a simple packing argument shows that we cannot pack the endpoint of all three edges
    on $\partial B(q, D)$. Indeed, suppose $u_1$
    appears on part 1 of $\partial B(q,D)$. Then $v_1,u_2$ may not appear before part 4 of
    $\partial B(q,D)$, which implies $v_2,u_3$ cannot appear before
    part 7 of $\partial B(q,D)$. 
    But then $v_3$ has to appear in part 2 or later, contradicting the fact that $v_3$ appears before
    $u_1$. See \figref{fat-annulus}.
\end{proof}

\lemref{fat-square} allows us to limit attention to regions $K(q,r)$ whose boundaries intersect with at most two edges of $\F$. 
This motivates the definition of \emph{corridors}, which are relatively narrow regions within $\F$, and are handled in our surgery separately.

\paragraph{Corridors.}
Intuitively, \emph{corridors} are regions in $\F$ bounded by two of its edges that are too narrow for robots to freely
pass one another. Our definition is analogous to that in \cite{steiger2024} and requires careful consideration
in order to show that no robots ever need to have a breakpoint deep inside a corridor,
which is crucial for obtaining an efficient FPTAS. We now define a corridor formally.

Let $e_i, e_j$ be a pair of edges in $\F$ supporting a square (of any radius) in $\F$;
for simplicity assume that neither of them is axis-aligned (the definition of corridors is simpler in that case).
That is, there exists some square $B(q,r) \subset \F$ with radius $r > 0$ contained in $\F$,
such that $e_i$ (resp. $e_j$)
touches $B(q,r)$ at a point $v_i$ (resp. $v_j$), but does not intersect $\interior(\poly)$.
Let $u_{ij} \in [0,\pi)$ be a direction normal to the segment $v_i v_j$.
Consider a trapezoid $\corr$ bounded by $e_i, e_j$
such that
(i) two of its boundary edges are contained in $e_i, e_j$, which are called \emph{blockers},
(ii) no landmarks of $\Lambda$ are in the interior of $\corr$, and
(iii) the edges of $\corr$ that are not blockers, denoted by $\sigma^-,\sigma^+$ and called \emph{portals},
are normal to $u_{ij}$. See \figref{corridor}.

$\corr$ is called a \emph{corridor} if the $\ell_\infty$-distance between the endpoints of each of $\sigma^-$ and $\sigma^+$
is at most 2, i.e., if $\sigma^- = a^-b^-$ and $\sigma^+ = a^+b^+$ then the \emph{width} of $\corr$, defined as
$\width(\corr) := \max\{||a^- - b^-||_\infty, ||a^+ - b^+||_\infty\}$, is at most 2. 
Let $\ell_\corr$ be the line
segment bisecting its portals. Then for any point $p \in \ell_\corr$, the $\ell_\infty$-distance to $e_i$ or $e_j$
is at most 1. Similar to the width of $\corr$, we refer to the $\ell_\infty$ distance between
the endpoints of $\ell_\corr$ as the \emph{depth} of $\corr$, denoted $\depth(\corr)$.
We say that $\corr$ is a \emph{maximal} corridor if no other corridor contains $\corr$. If $\corr$ is maximal,
then there is a landmark of $\Lambda$ on at least one of the portals;
if both portals of $\corr$ have the same length then there are landmarks on each portal.
These conditions imply that there can only be $O(n)$ maximal corridors in $\F$.

Finally, we say that a corridor $\corr$ is \emph{$D$-deep} if the depth of $\corr$ is at least $D \geq 0$.
We will later show in \secref{thin-corridor} that if $\corr$ is $D$-deep for a sufficiently large value $D$
(dependent on $k$), we can perform a surgery
resulting in a plan with no breakpoints $D$-deep within $\corr$.

\paragraph{Revolving areas.}

\begin{figure}
    \centering
    \includegraphics[width=0.3\textwidth]{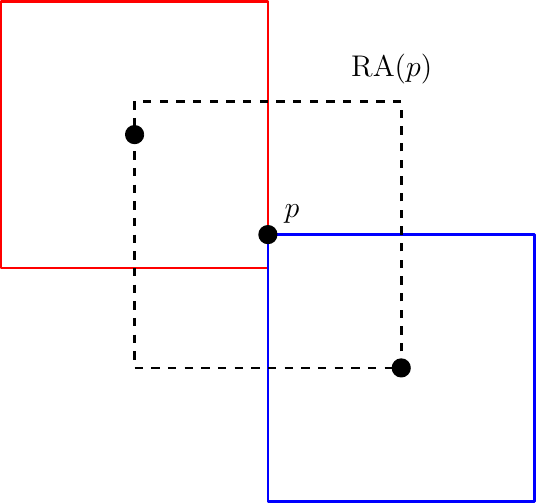} \hspace{1cm} \includegraphics[width=0.5\textwidth]{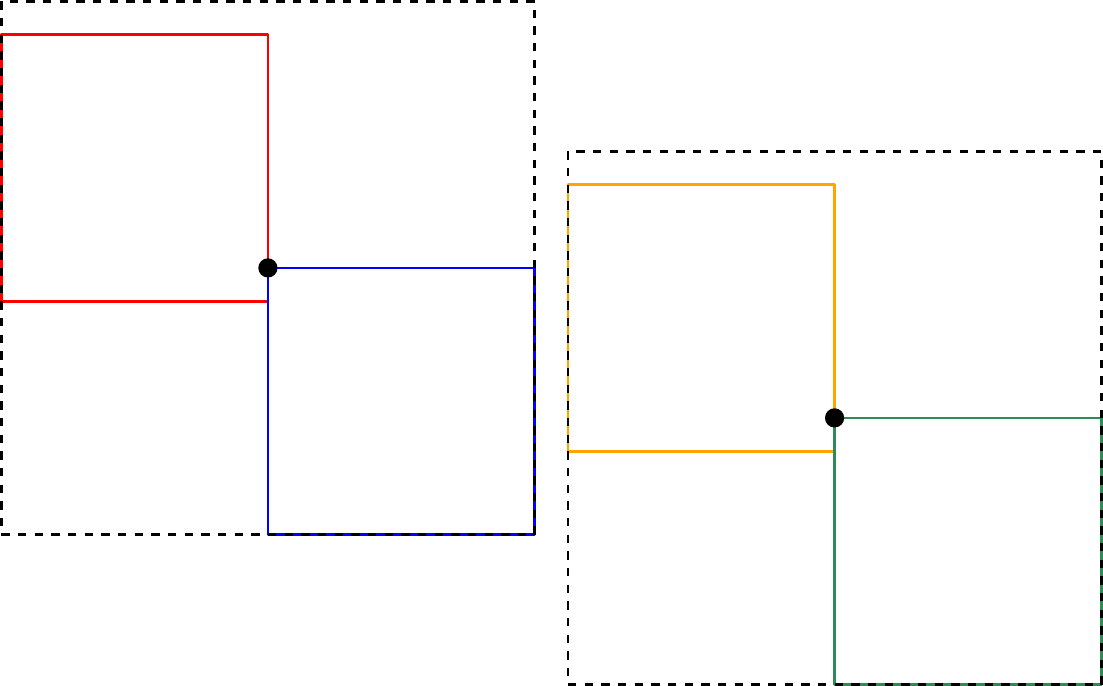}
    \caption{A revolving area $\RA(p)$, which allows two robots to move freely around one another (left), and a pair
    of WSRAs, where motion in $\RA(p_i)$ does not effect $\RA(p_j)$ (right).}
    \label{fig:RA}
\end{figure}

We next define regions in $\F$ in which robots can move around one another
in a relatively unconstrained manner.
For a placement $p \in \F$ of a unit square robot $i$, we say that $i$ has a
\emph{revolving area} at $p$ if $p + \poly \subset \F$, and denote it by $\RA(p)$.
It will be useful to consider
\emph{well-separated revolving areas} (WSRAs), which are configurations of revolving areas with no conflicting placements.
Specifically, for a set of conflict-free placements $P = \{p_1, ..., p_l\}$ of $l \leq k$ robots,
(where $\interior(p_i + \poly) \cap \interior(p_j + \poly) = \emptyset$),
the corresponding set of revolving areas is denoted $\RA(P) = (\RA(p_1), ..., \RA(p_l))$. We say that $\RA(P)$
is \emph{well-separated} if $p_i + 2\poly \cap p_j + 2\poly = \emptyset$ for every $i \not=j$. See \figref{RA}.
The following lemma, also observed in \cite{steiger2024}, summarizes the relationship between
revolving areas and corridors.

\begin{lemma}\label{lem:ra-in-corr}
    Suppose $p \in \F$ is a $1$-far point that does not lie in any corridor.
    Then there exists a revolving area $\RA(q)$ for some point $q \in \F$
    such that $p \in \RA(q)$.
\end{lemma}

Our surgery to remove $\Delta$-far breakpoints not in a corridor will heavily exploit RAs,
and in particular, WSRAs. Unfortunately, since WSRAs are not present inside corridors,
they require slightly different treatment than the rest of $\F \setminus U(\Lambda, \Delta)$.

\subsection{Near-optimal plans inside a corridor}\label{sec:thin-corridor}


Given a piecewise-linear $(\bs,\bt)$-plan $\pth$ that contains a $\Delta$-far breakpoint inside a corridor $\corr$,
we describe a surgery procedure that modifies $\pth$ into another $(\bs,\bt)$-plan $\pth'$ such that $\pth'$ does not have a $\Delta$-far
breakpoint, for $\Delta = \Omega(k)$, inside $\corr$ and $\cost(\pth') \leq \cost(\pth) + O(k)$.
The high-level idea of the surgery is to park any robots that ever travel $\Delta$-deep in the corridor
at a well-behaved position inside a ``parking area'', which is $O(k)$-deep in the corridor, and keep the robot
in this region until the next time that it reaches a portal of $\corr$. We argue that this surgery is feasible,
and we bound the cost we pay for the surgery.

\begin{figure}
    \centering
    \includegraphics[width=0.75\textwidth]{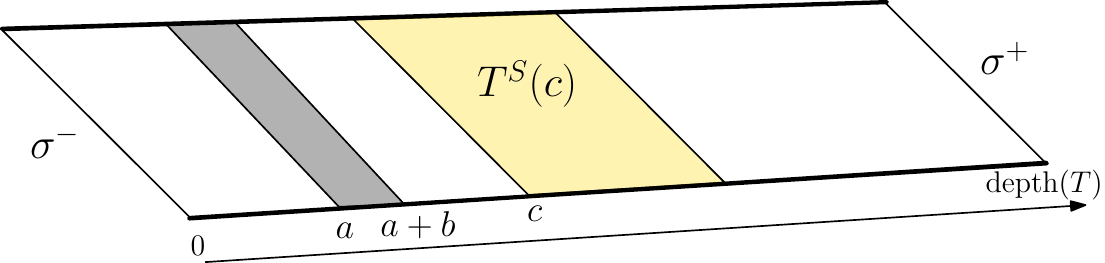}
    \caption{A corridor $\corr$, where $\corr^-(a, a+b)$ is shown in grey, and $\corr^S(c)$ is in yellow.
    Note that the left boundary of $\corr(a, a+b)$ is $\sigma^-(a)$, the right boundary is $\sigma^-(b)$,
    and $\corr^S(c)$ has left boundary $\sigma^-(c)$ and right boundary $\sigma^+(c)$.
    The bottom arrow plots $\depth(\corr)$ from the perspective of $\sigma^-$.}
    \label{fig:sanctum}
\end{figure}

We begin by describing a few notations and some key properties of a plan inside a corridor. Let $\corr$ be a
corridor with blockers $e_i, e_j$, portals $\sigma^-, \sigma^+$, and direction vector $u_\corr$,
as defined above. For a value $r \geq 0$, let $\sigma^-(r)$ (resp. $\sigma^+(r)$) be the line segment
parallel to portals at $\ell_\infty$-distance $r$ from $\sigma^-$ (resp. $\sigma^+$) inside $\corr$,
i.e., in direction $u_\corr$ (resp. $-u_\corr$). If $\corr$ is not $r$-deep, $\sigma^-(r)$ and $\sigma^+(r)$ are undefined.
Let $\corr^-(r)$ (resp. $\corr^+(r)$) denote the trapezoid formed by $\sigma^-, \sigma^-(r)$ (resp. $\sigma^+, \sigma^+(r)$)
and the portions of the blockers $e_i, e_j$ between them, and let $\corr^{S}(r) \subset \corr$ denote the corridor formed
by $e_i, e_j, \sigma^-(r),$ and $\sigma^+(r)$, which we refer to as the $r$-sanctum of $\corr$.
Similarly, for two values $r^- < r^+$, let $\corr^-(r^-, r^+)$ (resp. $\corr^-(r^-, r^+)$)
denote the trapezoid formed by $\sigma^-(r^-), \sigma^-(r^+),$ (resp. $\sigma^+(r^-), \sigma^+(r^+)$)
and the portions of the blockers $e_i,e_j$ between them.
Note, the sanctum is equivalently written
$\corr^-(\Delta, \depth(\corr) - \Delta) = \corr^+(\Delta, \depth(\corr)-\Delta) = \corr^{S}(\Delta)$.
See \figref{sanctum}.
The depth of $\corr^{S}(r)$ is $\depth(\corr) - 2r$ and thus undefined if $\corr$ is not $2r$-deep.
Note, since $e_i,e_j$ are line segments, the width of $\corr$ varies monotonically between the portals.
That is, the length of the segment $||\sigma^-(r)||_2$ as $r$ ranges from $0$ to $\depth(\corr)$
either grows or shrinks monotonically,
and either $\sigma^-(0)$ (i.e., $\sigma^-$) or $\sigma^-(\depth(\corr))$ (i.e., $\sigma^+$) attains the value
$\max_{0 \leq l \leq \depth(\corr)} ||\sigma^-(l)||_2$. If $||\sigma^-||_2 \not= ||\sigma^+||_2$ and $\sigma^+$ (resp. $\sigma^-$) attains this maximum, then we say that $\corr$ is
\emph{widening} (resp. \emph{narrowing}) in the direction $u$. 
If $||\sigma^-||_2 = ||\sigma^+||_2$ then the blockers are parallel.

We now state a few key properties of corridors. First, we inherit a useful property directly from \cite{steiger2024},
which shows that the ordering of the robots inside a corridor $\corr$, with respect to its
direction vector $u$, remains the same for the duration of a time interval that the robots are in $\corr$.

\begin{lemma}[Lemma 2.2 of \cite{steiger2024}]\label{lem:inner-products}
    Let $i,j \in \BR$ s.t. $i \not= j$, let $\corr$ be a maximal corridor, and let $u$ be its direction vector.
    Let $[\lambda_1, \lambda_2] \subseteq [0,1]$ be a time interval in a plan $\pth$
    such that $\pi_i[\lambda_1, \lambda_2], \pi_j[\lambda_1, \lambda_2] \subset \corr$.
    Then the sign of
    $\langle \pi_i(\lambda) - \pi_j(\lambda), u\rangle$ is the same for every $\lambda \in [\lambda_1, \lambda_2]$.
\end{lemma}


The next lemma gives a plan of bounded cost to move a robot into a corridor,
which we will call when assigning robots a parking area.



\begin{figure}
    \centering
    \includegraphics[width=0.45\textwidth]{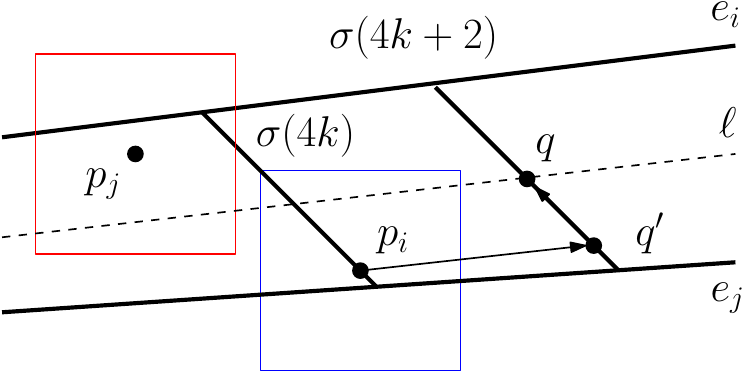} \hspace{1cm} \includegraphics[width=0.45\textwidth]{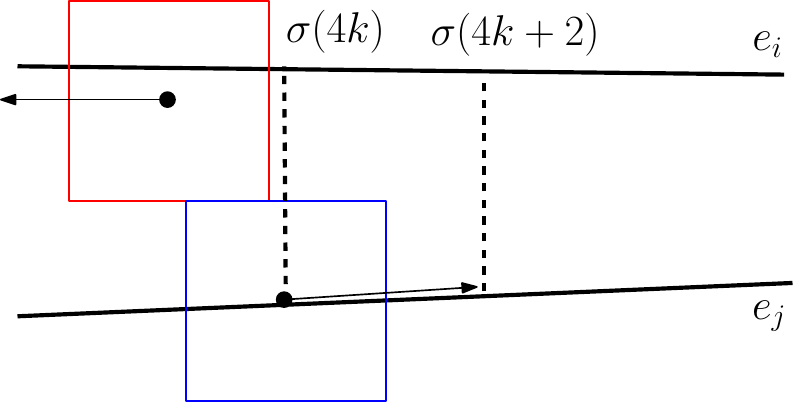}
    \caption{(Left) The motion to a parking area when a portal $\sigma$ with slope $-1$ as described in Case 1 (ii) of \lemref{portal-cases}.
    (Right) the motion described in Case 2, where it may be infeasible to move the red robot deeper into the corridor without first
    moving the blue robot in the $-u$ direction. This can cascade if there are $\Omega(k)$ robots which then conflict with the blue robot.}
    \label{fig:portal-cases}
\end{figure}

\begin{lemma}\label{lem:portal-cases}
    Let $\corr$ be an $8k$-deep corridor, let $\ell$ be the bisecting segment of its portals, and let $\sigma$ be
    one of its portals, say, $\sigma^-$. 
    Let $\bp = (p_1, \ldots, p_k) \in \BF$ be a feasible configuration such that $p_i \in \sigma(4k)$ for some $i \in \BR$,
    and for every other $p_j \not= p_i$, $p_j \not \in \corr^-(4k, 4k+4)$. Let $q = \sigma(4k + 2) \cap \ell$.
    Let $\bq = (p_1, \ldots, q, \ldots, p_k)$ denote the configuration with robot $i$ placed at $q$ and every other robot $j \not= i$
    remain at $p_j$. There exists a feasible $(\bp,\bq)$-plan $\pth$, with $\cost(\pth) \leq k (4\sqrt{2} + 2)$.
\end{lemma}
\begin{proof}
    See \figref{portal-cases}.
    Let $u$ be the direction of $\corr$, oriented toward $\sigma^+$.
    For concreteness, suppose the portals are either vertical or have slope $-1$, and that $u \in \{0,\pi/4\}$;
    the symmetric cases are easily handled after standard rotations of $\P$ by $\pi/2$ as needed.
    The slopes of the blockers of $\corr$ are strictly positive if its portals have slope $-1$, and if the portals
    are vertical one blocker has non-negative slope while the other has non-positive slope.
    There are two cases, depending on if $\corr$ is narrowing or widening with respect to $u$.
    (If $e_i$ and $e_j$ are parallel, either case suffices; we describe a feasible plan for this situation in Case 1.)
   
    \noindent\textbf{Case 1:} $||\sigma^-||_2 \leq ||\sigma^+||_2$ ($\corr$ is widening or parallel) :

    We observe that it is feasible to simply to move robot $i$ to $q$ directly,
    without moving any of the other robots. First, we argue that there is at most one $p_j$ such that $p_j \in p_i + 2\square$:
    Since $\corr$ is $8k$-deep, $p_i + 2\square \subset \corr$. On assumption that there can be no $p_j \in \corr^-(4k,4k+4)$,
    it is only possible to have $p_j \in p_i + 2\square \cap \corr^-(4k)$. Suppose, say, that $(p_i,p_j)$ are $x$-separated,
    and that there were some other point $p_l \in p_i + 2\square \cap \corr^-(4k)$, where $l \not \in \{j,i\}$. If $(p_l, p_j)$
    were $y$-separated, then there would be a vertical segment of length two contained in $\F \cap \corr^-(4k)$, contradicting that
    $\width(\corr) < 2$. On the other hand, if $(p_l,p_j)$ were $x$-separated, either $p_l \not \in p_i + 2\square \cap \corr^-(4k)$,
    or $(p_l, p_i)$ is not $x$-separated, and thus it is $y$-separated. But again, this implies a vertical segment of length at least two contained
    in $\F \cap \corr^-(4k)$, a contradiction. The case where $(p_i,p_j)$ are $y$-separated is similar.
    We conclude that there is at most one $p_j \in p_i + 2\square$. So, as long as we produce a plan
    such that $\pi_i \subset \corr^-(4k,4k+4)$, it suffices only to argue that $\pi_i \subset \F[p_j]$.

    There are two subcases, either (i) the portals are vertical, or (ii) the portals have slope $+1$.

    Suppose (i) that the portals are vertical. This implies $(p_i, p_j)$ is $x$-separated:
    On the assumption that $\corr$ is widening (in the $+x$ direction),
    the vertical segment $s = (x(\sigma(4k)) - h) \cap \corr$ for all $0 \leq h \leq 4k$ must satisfy
    $y(s) \subseteq y(\sigma)$, and $y(\sigma) \subset \interior(y(p_i + 2\square))$. Since
    $y(p_j) \in y(s)$, then $y(p_j) \in \interior(y(p_i) + 2\square)$, implying $||p_i - p_j||_\infty < 2$, contradicting that $\bp \in \BF$.
    So, $(p_i,p_j)$ are $x$-separated.

    Let $q'=y(p_i) \cap \sigma(4k+4)$ be the intersection point between $\sigma(4k+4)$ and
    the horizontal line with the same $y$-coordinate as $p_i$. Since $(p_i,p_j)$ are $x$-separated, this implies that
    $p_iq' \subset \F[p_j]$. The segment $q'q$ is vacuously in $\F[p_j]$, since $q'q \subseteq \sigma(4k+2)$, which has
    $\ell_\infty$-distance at least $2$ from any point of $\corr^-(4k)$ by definition.
    Hence, we define $\pi_i = p_iq' \cup q'q$, and conclude that $\pi_i \subset \F[p_j]$, and hence $\pi_i \subset \F[\{p_j \not= p_i\}]$. 
    
    Now suppose (ii) that the portals have slope $-1$. Since $\corr$ is widening, the segment $s=p_iq'$ parallel to $\ell$
    intersecting $\sigma(4k+2)$ at the point $q'$ is contained in $\F$, since for all points $l \in \ell$,
    there is a square $l + \alpha'\square \subset \F$, where $\alpha'\geq \alpha$. 
    Moreover, since the portals have slope $-1$, in whichever direction $\theta \in \{x,y\}$
    that $(p_i,p_j)$ are $\theta$-separated, $\theta(p_i) - \theta(p_j) \geq 2$. Hence, since $p_iq'$
    is $xy$-monotone, and in particular $\theta$-monotone,
    (as $\ell$ under our assumptions is $xy$-monotone, since the blockers each have positive slope)
    it follows that $p_iq' \cap p_j + 2\square = \emptyset$. As before, definitionally
    $\sigma(4k+2) \subset \F[p_j]$. Hence, we define $\pi_i = p_iq' \cup q'q$,
    and obtain by construction that
    $\pi_i \subset \F[p_j]$, and from the above argument, that $\pi_i \subset \F[\{p_j \not= p_i\}]$.
    
    
    We conclude that when $\corr$ is widening, there exists a feasible $(\bp,\bq)$-plan in which robot $i$ moves along the $(p_i,q)$-path
    $\pi_i$ as defined, while every other robot $j$ remains stationary at $p_j$. In either case,
    we have $||p_iq'||_2 \leq 4\sqrt{2}$, and $||q'q||_2 < 2$. So, $\cost(\pi_i) < 4\sqrt{2} + 2$,
    and thus $\cost(\pth) \leq 4\sqrt{2} + 2$.

    \noindent\textbf{Case 2:} $||\sigma^-||_2 > ||\sigma^+||_2$ ($\corr$ is narrowing):
    
    As above, we would like to simply move $p_i$ via the shortest path to $q$, but it could be the case that
    the segment $p_iq \not \subset \F[\{p_j \not= p_i\}]$. However, $\corr$ is widening in the $-u$ direction.
    So, suppose for at least one robot $j$, $p_iq \not \subset \F[p_j]$. We use the plan described in Case 1 to move
    robot $j$ to a point on $\sigma(4k-2)$. If this motion conflicts with some other robot $l$ (from the above, $l\not= i$)
    we move it to a point on $\sigma(4k-4)$. The process continues, where we park the $m$th conflicting robot at some
    point on $\sigma(4k-2m)$, until there are no more conflicts and we can move robot $i$ directly along $p_iq$ without conflict.
    Since $m \leq k$, we note that all robots remain in $\corr$ throughout the procedure, where the nearest any robot parks to $\sigma^-$
    is at a point on $\sigma(4k-2k)$ which is at least $2k$-deep.
    We then conduct the inverse plan to the one in Case 1 to move each robot $j$ to its initial position $p_j$, and denote by $\pth$ the resulting
    $(\bp,\bq)$-plan. Clearly, $\pth$ is feasible, and has $\cost(\pth) \leq (k-1) (4\sqrt{2} + 2) + 4\sqrt{2} < k (2\sqrt{2} + 4)$.
    
    We conclude the existence of a $(\bp,\bq)$-plan $\pth$, such that $\cost(\pth) \leq k (4\sqrt{2} + 2)$.

\end{proof}

%


We now establish the existence of parking areas for the robots in a corridor.


\begin{lemma}\label{lem:thin-corridor-vertices}
    Let $\corr$ be a $20k$-deep corridor.
    Then there exists a sequence $P = \langle p_1, \ldots, p_k \rangle$ of at least $k$
    placements in $\corr^-(10k)$ (resp. $\corr^+(10k)$) such that $p_i$ lies on the bisecting
    line $\ell_\corr$ of $\corr$,
    for every $i \in \BR$, and $||p_i - p_j||_\infty \geq 4$ for every pair $i \not= j$.
\end{lemma}
\begin{proof}
    Let $\G_4 = 4\ZZ^2$ be the grid with square cells of width four,
    and let $\F^{\#}$ be the overlay of $\G_4$
    on $\F$. Since $\width(\corr) < 2$,
    the faces of $\F^{\#}$ each correspond to a connected component of $\corr \cap g$ for some grid cell $g \in \G$,
    and $g$ cannot be contained entirely in $\F^{\#}$. Let $\F^{\verteq}$ (resp. $\F^{=}$) denote the overlay of
    the vertical (resp. horizontal) grid lines of $\G_4$ on $\F$.
    Consider the line $\ell_\corr$ bisecting the portals of $\corr$, and observe that
    the $L_\infty$-length of $\ell_\corr \cap \corr^-(8k)$ (resp. $\ell_\corr \cap \corr^-(8k)^+$) is at least $8k - \sqrt{2}$ and at most
    $8k + \sqrt{2}$, and for at least one of $\F^{\verteq}$ or $\F^{=}$,
    every line in either $\F^{\verteq} \cap \corr$ or $\F^{=} \cap \corr$ intersects $\ell_\corr$
    at a point. Let $\F^{\sim} \in \{\F^{\verteq}, \F^{=}\}$ be the overlay satisfying
    this property. So, there are at least $\frac{8k - \sqrt{2}}{4} > k$ vertices of $\F^{\sim} \cap \ell_\corr$ contained in
    $\corr^-(8k)$ (resp. $\corr^-(8k)^+$). 
    Every pair of vertices $v, u \in \F^{\sim} \cap \ell_\corr$ has $\|v - u\|_\infty = 4$,
    so the placements at $u$ and $v$ are $\theta$-separated for at least one value of $\theta \in \{0,\pi/2\}$, and thus they are feasible.
    Additionally, by construction for any pair $v, u$, we have that $v \not \in \interior(u + 4\poly)$ and vice versa.
    Now, it is possible that for at most one vertex $v$ of $\F^{\sim} \cap \ell_\corr$ in $\corr^-(8k)$
    (resp. $\corr^-(8k)^+$), $v + 4\poly \not \subset \interior(\corr)$,
    so there are at least $\frac{8k - \sqrt{2}}{4} - 1$
    placements satisfying the lemma statement in $\corr^-(8k)$ and $\corr^-(8k)^+$. Hence,
    in $\corr^-(10k)$ and $\corr^-(10k)^+$, there are at least $\frac{10k - \sqrt{2}}{4} - 1 > k$
    vertices $v$ of $\F^{\#} \cap \ell_\corr$, such that $v + 4\poly \subset \interior(\corr)$.
    
    Setting $P$ to be the vertices of
    $\F^{\#} \cap \ell_\corr$ inside $\corr^-(10k)$ (resp. $\corr^-(10k)^+$),
    there exist feasible configurations, with at least $k$ placements,
    in $\corr^-(10k)$ and $\corr^-(10k)^+$, satisfying the lemma.
\end{proof}

\begin{corollary}
    Let $\corr$ be a $2(r + 10k)$-deep corridor, for a value $r > 0$. Then there exists a sequence $P = \langle p_1, \ldots, p_k \rangle$
    of at least $k$ placements in $\corr^-(r, r+10k)$ (resp. $\corr^+(r,r+10k)$)
    such that $p_i$ lies on the bisecting line $\ell_\corr$ of $\corr$,
    for every $i \in \BR$, and $||p_i - p_j||_\infty \geq 4$ for every pair $i \not= j$.
\end{corollary}

\begin{figure}
    \centering
    \includegraphics[width=0.75\textwidth]{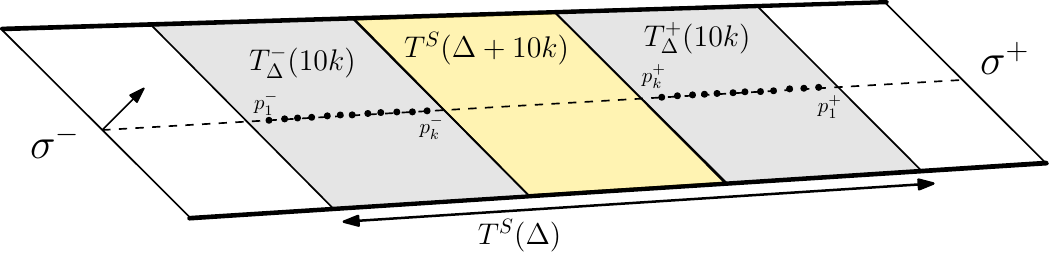}
    \caption{A corridor $\corr$, with designated parking places defined by \lemref{thin-corridor-vertices}
    at the sequences of points $P^- = \langle p_1^-, \ldots, p_k^- \rangle$
    and $P^+ = \langle p_1^+, \ldots, p_k^+ \rangle$.
    We use $P^-$ and $P^+$ to move robots across $\corr^S(\Delta + 10k)$ without creating breakpoints.}
    \label{fig:parking-spots}
\end{figure}

Let $\pth$ be a piecewise-linear $(\bs,\bt)$-plan. Let $\Delta = Ck$ for a sufficiently large constant $C$.
We now describe the surgery on $\pth$ to remove all $\Delta+10k$-far breakpoints that lie inside
corridors; so, suppose $\Delta \geq 10k$. Let $\corr$ be a maximal corridor that is $4\Delta$-deep, and contains a $\Delta$-far breakpoint of $\pth$.
Let $\ell_\corr$, $u_\corr$ be its portal bisecting segment and portal direction respectively, and let $\corr_\Delta := \corr^S(\Delta)$
be the $\Delta$-sanctum of $\corr$. $\corr_\Delta$ is also a corridor.
Let $P_\corr^- = \langle p_1^-, \ldots, p_k^-\rangle$ (resp. $P_\corr^+ = \langle p_1^+, \ldots, p_k^+\rangle$) be the sequence of points in $\corr_\Delta^-(10k)$
(resp. $\corr_\Delta^+(10k)$) as defined in \lemref{thin-corridor-vertices},
sorted in increasing (resp. decreasing) order in direction $u_\corr$ (i.e., $\langle p_i^-, u_\corr \rangle < \langle p_{i+1}^-, u_\corr \rangle$
and $\langle p_i^+, u_\corr \rangle > \langle p_{i+1}, u_\corr\rangle$; see \figref{parking-spots}.).
$P^-, P^+$ act as parking spots for robots entering $\corr_\Delta$.

Suppose a robot $i \in \BR$ enters the $\Delta$-sanctum $\corr^S(\Delta)$ of $\corr$
at time $\lambda_0$, and let $[\lambda_0^-, \lambda_0^+]$ be the maximal time interval containing
$\lambda_0$ during which $i$ lies in $\corr$. We create two events for $i$ during $[\lambda_0^-, \lambda_0^+]$
as follows: Let
\[\bar{\lambda}_0 = \inf\{\lambda \in [\lambda_0^-, \lambda_0^+] \mid \pi_i(\lambda) \in \corr^S(\Delta)\}.\]
Then, we create events at times $\bar{\lambda}_0^-$ and $\bar{\lambda}_0^+$ for robot $i$ during the interval
$[\lambda_0^-, \lambda_0^+]$, intuitively corresponding to the first time that $i$ enters the sanctum $\corr^S(\Delta)$ in $[\lambda_0^-, \lambda_0^+]$, and the time it leaves $\corr$.
We collect all events corresponding to $\corr$ and process them in increasing order
of time. At any given time, some parking places of $P_\corr^-$ and $P_\corr^+$ will be occupied.
We maintain the invariant that at any given time, the deepest parking places are occupied, namely
if $p_j^-$ (resp. $p_j^+$) is occupied and $j < k$, so is $p_{j+1}^-$ (resp. $p_{j+1}^+$).
We process each event as follows:

\begin{enumerate}[(i)]
    \item \textit{$\bar{\lambda}_0$ event for robot $i$}: If $i$ enters $\corr$ through the portal $\sigma_\corr^-$ (resp. $\corr_\sigma^+$)
    then move $i$ to the deepest unoccupied parking place of $P_\corr^-$ (resp. $P_\corr^+$)
    while freezing all other robots at their current positions. It remains parked until the time instance
    $\lambda_0^+$, though it may be moved to a deeper parking spot of $P_\corr^-$ (resp. $P_\corr^+$)
    to maintain the above invariant.
    \item \textit{$\lambda_0^+$ event for robot $i$}: Without loss of generality assume $i$ exits $\corr$ at time $\lambda_0^+$ through the
    portal $\sigma_\corr^+$. Suppose $i$ had entered $\corr$ through $\sigma_\corr^+$, then it is parked at a point
    $p_j^+$ of $P_\corr^+$ and $p_1^+, \ldots p_{j-1}^+$ are unoccupied (by \lemref{inner-products}).
    We move $i$ from $p_j^+$ to $\pi_i(\lambda_0^+)$ along the shortest feasible path while
    freezing all other robots at their current positions. If $i$ entered through $\sigma_\corr^-$ then
    \lemref{inner-products} implies that $i$ is parked at $p_k^-$ of $P_\corr^-$. Again, we move
    $i$ from $p_k^-$ to $\pi_i(\lambda_0^+)$ along the shortest feasible path while freezing other robots at their
    current positions. Finally, we move all robots parked at points of $P_\corr^-$ to the position incremented one index
    to maintain the invariant that the deepest parking placements of $P_\corr^-$ are occupied.
\end{enumerate}


    Apply the above procedure repeatedly, for every robot entering a $\Delta$-deep corridor during $[0,1]$,
    and denote the resulting plan by $\pth'$.
    We argue that $\pth'$ is feasible, has removed any $\Delta+10k$-far breakpoints in the sanctum,
    and bound the cost of $\pth'$.
    \begin{lemma}\label{lem:corridor-feasible}
        The modified plan $\pth'$ is both feasible and does not contain a $(\Delta+10k)$-far breakpoint inside any corridor.
    \end{lemma}
    \begin{proof}
    
    Fix any $\Delta$-deep corridor $\corr$, and let $\ell$ be the line bisecting the midpoints of the
    portals of both $\corr$ and $\corr^{S}(\Delta)$.
    Observe that the prefix $\pi_i'[\lambda_0^-, \bar{\lambda}_0]$
    of the path specified in both Case (i) and Case (ii) can be implemented by first moving
    $i$ to a point $q \in \ell$ using \lemref{portal-cases},
    and then by moving robot $i$ on $\ell$ directly to the deepest non-occupied point $p \in P^-$. Doing so is feasible:
    since $||\sigma^-||_\infty <2$,
    $i$ can be the only robot for which $\pi_i(\bar{\lambda}_0) \in \sigma_\Delta^-$, and by construction
    any robots placed in $\corr$ lie on some point in either $P^-$ or $P^+$. More specifically, if there are $l$ robots placed in
    $\corr_{\Delta}(10k)$ at time $\bar{\lambda}_0$, they lie on the points $p_k^-, p_{k-1}^-, \ldots, p_{k-l}^-$, so $p = p_{k-l-1}^-$ and
    the line segment $pq$ is contained in the free space, $\F[\{\pi_{j}(\bar{\lambda}_0) \mid j \not= i\}]$.
    Hence, the prefix $\pi_i'[\lambda_0^-, \bar{\lambda}_0]$
    is feasible by \lemref{portal-cases}.
    Moreover, since the prefix is feasible, by \lemref{inner-products} the total order induced by $\langle \pi_j'(\bar{\lambda}_0), u \rangle$,
    for the path $\pi_j'$ of every $j$, is the same as the total order induced by $\langle \pi_j(\bar{\lambda}_0), u \rangle$ was for the original plan. 
    The same statement is true for any additional critical events prior to $\lambda_0^+$.

    The suffix $\pi_i'[\bar{\lambda}_0, \lambda_0^+]$ is also feasible: suppose first that $i$ falls into Case (i).
    In this case, if $i$ is parked at $p_l^- \in P^-$, we may need to move $i$ to $p_{l+1}^-$ if some robot $j$ leaves
    $\corr$ at a time $\tilde{\lambda}_0^+ < \lambda_0^+$. This can be done in decoupled fashion directly along $\ell$,
    moving the robots at the deepest parking positions first, in the same manner as for the prefix. Clearly the procedure in
    this case never parks $i$ beyond $(\Delta + 10k)$-deep in $\corr$, since $i$ only increments its parking index when another
    robot $j$ parked in $P^-$ exits via $\sigma^+$; and if this is the case since $\pth'$ maintains the same total order in $\corr$
    as $\pth$ then $j$ must have been parked at a point $p$ with index exceeding $p_l^-$. Upon reaching the time $\lambda_0^+$,
    we move $i$ in decoupled fashion directly, while keeping the other robots parked at their current position,
    along $\ell$ until reaching the point $q = \sigma^+(2) \cap \ell$. 
    We then use the inverse plan to the one in \lemref{portal-cases} to move from $q$ to $\pi_i(\lambda_0^+)$.
    By construction, at time $\lambda_0^+$ it must be that $i$ is parked at $p_k^-$,
    and since $\pth'$ maintains the same total order in $\corr$ as $\pth$, it must be that
    $\langle \pi_i'(\lambda_0^+), u \rangle < \langle \pi_j'(\lambda_0^+), u \rangle$,
    for every $j \not= i$ lying in $\corr$. Hence, the only other robots parked in
    $\corr$ at time $\lambda_0^+$ must lie in $P^+$, and the suffix described is feasible.
    On the other hand, suppose $i$ falls into Case (ii). As before, we may need to move $i$
    along $\ell$ as other robots parked at deeper points in $P^-$ leave $\corr$,
    and we use an identical argument. Moreover, at time $\lambda_0^+$, again since $\pth'$
    preserves the total order in $\corr$ inherited from $\pth$, it must be that
    $\langle \pi_i'(\lambda_0^+), u \rangle > \langle \pi_j'(\lambda_0^+), u \rangle$,
    for every $j \not= i$ lying in $\corr$. Thus, the symmetric path to the one
    in Case (i) is feasible, traversing $\ell$ until reaching the boundary of the
    polygon defining $\sigma^+$ and then again using the inverse plan from \lemref{portal-cases}.
    In Case (i) $i$ only parks in $\corr(\Delta,\Delta + 10k)$,
    so the produced plan in this case vacuously
    has no breakpoints inside the sanctum $\corr^S(\Delta + 10(k))$,
    as claimed. In Case (ii) $i$ again only parks in $\corr(\Delta, \Delta+10k)$,
    and only enters $\corr^S(\Delta + 10(k))$ when traversing $\ell$.
    Hence, linearly interpolating $i$ along $\ell$ easily avoids creating
    any breakpoints inside $\corr^S(\Delta + 10(k+2))$ as well.

    Finally, we conclude that the plan $\pth'$ we produced has no breakpoints in $\corr^S(\Delta + 10k)$,
    since any robots entering $\corr^S(\Delta)$ have by construction only traveled along straight
    segments in $\corr^S(\Delta + 10k)$, and otherwise any robots that never enter the sanctum in $\pth$
    remain following their original paths. After repeating the argument for every maximal corridor in $\F$,
    the claim follows.
    \end{proof}

    \begin{lemma}\label{lem:corridor-cost}
        Let $\pth'$ denote the $(\bs,\bt)$-plan resulting from the above procedure. Then
        $\cost(\pth') \leq (1 + \frac{C_\corr k}{\Delta})\cost(\pth)$, where $C_\corr$ is a constant.
    \end{lemma}
    \begin{proof}
        Observe that in Case (i),
        $\cost(\pi_i[\bar{\lambda}_0, \lambda_0^+]) \geq \Delta$, since on assumption $i$ enters
        $\corr^S(\Delta)$, and in Case (ii), $\cost(\pi_i[\bar{\lambda}_0, \lambda_0^+]) \geq 4\Delta$ since
        $\corr$ is $4\Delta$-deep and $\pi_i(\bar{\lambda}_0)$ is
        $\Delta$-deep inside $\corr$. In Case (i), 
        the plan we construct incurs cost at most $k\left(4\sqrt{2} + 2\right)$
        for each call to \lemref{portal-cases}, and travels at most distance
        $\sqrt{2}(10k)$ along $\ell$ inside $\corr^S(\Delta)$ to and from
        its parking positions. Hence, the cost is bounded by

        \begin{align*}
            \cost(\pi_i'[\bar{\lambda}_0, \lambda_0^+]) &\leq k\left(4\sqrt{2} + 2\right) + 3\left(\sqrt{2} \cdot 10k\right) \\
            &<  11k + 40k \\
            &\leq \left(1 + \frac{11k}{4\Delta}\right)4\Delta \\
            &\leq \left(1 + \frac{11k/4}{\Delta}\right) \cost(\pi_i)\\
            &=\left(1 + \frac{C_\corr k}{\Delta}\right) \cost(\pi_i)
        \end{align*}

        Since $\pi_i[\lambda_0^-, \bar{\lambda}_0] = \pi_i'[\lambda_0^-, \bar{\lambda}_0]$,
        we conclude that $\cost(\pi_i') \leq \left(1 + \frac{C_\corr k}{\Delta}\right) \cost(\pi_i)$,
        for a constant $C_\corr$.
        Now consider Case (ii). As before, we call \lemref{portal-cases} twice,
        and we travel at most distance $\sqrt{2} \cdot 4\Delta$ along $\ell$ to reach
        $\sigma^+$. Since the cost of $\pi_i$ in this case exceeds that of Case (i),
        this only improves the bound above, and we conclude that
        $\cost(\pi_i') \leq \left(1 + \frac{C_\corr k}{\Delta}\right)\cost(\pi_i)$ in Case (ii) as well.
        Vacuously, in Case 2 $\cost(\pi_i) = \cost(\pi_i')$. Applying the bound
        over every $i \in \BR$ which enters $\corr$, we have that
        $\cost(\pth') \leq \left(1 + \frac{C_\corr k}{\Delta}\right)\cost(\pth)$.
    \end{proof}

Together, \lemref{corridor-feasible} and \lemref{corridor-cost}
prove the existence of a plan with no breakpoints beyond $\Delta + 10k$-deep
in the sanctum of a corridor, and we obtain the following lemma.

\begin{lemma}\label{lem:corr-bkpt}
    Let $\pth$ be a min-sum $(\bs,\bt)$-plan.
    Then, there exists a plan $\pth'$, with no $\Delta + 10k$-far breakpoints inside a corridor, such that $\cost(\pth') \leq (1 + \frac{C_\corr k}{\Delta})\cost(\pth)$,
    where $C_\corr$ is a constant.
\end{lemma}

\subsection{Near-optimal motion in wide  areas}\label{sec:fat}

By \lemref{corr-bkpt} we can assume that there are no $\Delta$-far breakpoints inside a corridor.
Throughout this section let $\Delta = Ck^2$;
the involved constant $C$ will be specified at the end of the section.
We perform a surgery on a given plan, at a small additional cost, to remove all $\Delta$-far breakpoints from wide areas.
However, unlike corridors, the relative ordering of robots in a wide area might change and the robots may cross each other,
which makes the surgery considerably more involved. We heavily rely on revolving areas, particularly WSRAs.
We first present a \textbf{local surgery} to reconfigure robots into WSRAs in a controlled manner,
before coordinating the motion fully over $[0,1]$ in a \textbf{global surgery}.
The following lemma lays the groundwork for performing the surgery.


\begin{lemma}\label{lem:wsras-count}
    Let $j \geq 0$ be an integer. For any $3(24j)$-far point $q \in \F$ that lies in a wide area, there are $j$ well-separated revolving areas
    (WSRAs) in $K(q,24j)$.
\end{lemma}
\begin{proof}
    \ben{This can be simplified a lot, and the constant should improve, using the analogous argument to the one for thin corridors.}
    \todo{simplify when rest is stable}
    Consider the 2D axis-aligned grid $\G$ whose cells have width four, $\G = 4\ZZ^2$.
    Let $\tilde{\F} = \P \oplus (-2\poly)$ be the set of all placements centering a revolving area
    in $\P$, let $0 < \alpha \leq 1$ and let $\FG$ be the overlay of $\G$ on $\tilde{\F} \cap K(q, \alpha 24k)$.
    Each face of $\FG$ is a connected component of $\left(\tilde{\F} \cap K(q, \alpha 24k) \right) \cap g$
    for some grid cell $g$ of $\G$. Let $V$ denote the set of vertices of $\FG$.
    Consider the vertices of $\FG$ lying on $\partial(\tilde{\F}) \cap \partial K(q, \alpha 24k)$. Abusing notation
    we denote it $\partial(V)$, and write $\interior(V) = V \setminus \partial(V)$.
    It is possible that a vertex $v \in \partial(V)$ can lie within
    $\ell_\infty$ distance four of other vertices of $V$, denote these neighbors by $N(v) = \{u \in V: \|u - v\|_\infty < 4, u \not= v\}$.
    Clearly $N(v)$ has constant size; there are two cases. If $v \in \interior(e)$ for some edge $e$ of a grid cell $g \in \G$,
    is possible to have two vertices in $N(v)$ that are collinear with $v$ on an edge of $\partial(\tilde{\F})$, and one vertex of $g$
    that remains free. The other edge of $\partial(\tilde{\F})$ could interesect $g$, adding an additional two vertices to $N(v)$.
    two vertices of $\G$ on either side of $v$. Otherwise, if $v$ is a vertex of $\G$, then either (i) $v$ must lie on $\partial(K(q, \alpha 24k))$ or (ii) an edge
    of $\partial(\tilde{\F})$ is axis-parallel, and in either case $|N(v)| \leq 2$ since in (i) at worst both edges of $\partial(\tilde{\F})$ can intersect
    a single edge of $g$ adjacent to $v$, and in (ii) the other edge of $\partial(\tilde{F})$ could intersect at most two edges of $g$.
    We conclude that $N(v) \leq 5$. 

    So, considering the vertices of $V$ in arbitrary order, we refine $V$ into a set $V' \subseteq V$
    where every $v \in V'$ has a WSRA simply by assigning $v \in V'$, and removing all $u \in N(v)$
    from further consideration. Then, by definition, any $v \in V'$ has no other vertices of
    $V'$ within $L_\infty$ distance four; either $N(v) = 0$ originally or $N(v)$ was removed in the refinement,
    so there is a WSRA centered at every $v \in \tilde{V}'$.

    We now argue that there are sufficiently many WSRAs (at least $\theta(k)$) in $ K(q, \alpha 24k)$ after the refinement
    process. Since $q$ is not in a thin corridor,
    for a point $p \in \partial(K(q, \alpha 24k))$, there is a line segment $pq \subset \tilde{F}$, with length at least $\alpha ck$.
    Then, $pq$ intersects at least $\lfloor\frac{\alpha ck}{4}\rfloor$ grid cells of $\G$, and since $pq \subset \tilde{F}$,
    there is at least one vertex of $\partial(V)$ that can be charged to each one of these grid cells, giving at least
    $\lfloor\frac{\alpha ck}{4}\rfloor$ vertices in $\tilde{V}$ before the refinement. After the refinement, since $N(v) \leq 5$
    for every vertex, and all vertices of $N(v)$ are excluded from $V'$ (so at least every 6th vertex is kept),
    we have $|V'|\leq \lfloor\frac{\alpha ck}{24}\rfloor$. Each vertex of $V'$ is the center for a WSRA by definition, and the claim follows.
\end{proof}

Roughly speaking, if the path $\pi_i$ of a robot $i$ in an $(\bs,\bt)$-plan $\pth$ contains a $\Delta$-far breakpoint,
robot $i$ must reach the $\Delta$-frontier at some time $\lambda$. We must produce an alternate plan
in which robot $i$ becomes $\Delta$-far in a controlled manner, to avoid plans with
arbitrarily far breakpoints. To do so, we park $i$ at a point $q$, contained in a WSRA, near the
$\Delta$-frontier at time $\lambda$. We also park all other robots that pass
through a neighborhood of $i$,
specifically $K(q,24k)$, in individual WSRAs. We keep each robot $j$ that has parked in a WSRA ``frozen'' there until we reach a time
$\lambda'$ such that $\pi_j(\lambda')$ becomes $\Delta/k$-close again.
Note that, since $t_i \in \Lambda$ for all $i \in \BR$, there always exists a time $\lambda$
where robot $i$ is $\Delta/k$-close.
At time instance $\lambda$, we freeze all other robots at their current positions
and move $j$ from its current position to $\pi_j(\lambda)$. We refer to this overall scheme of freezing and unfreezing
robots as the \emph{global surgery} and the modification in the plan that moves robots from their current positions
to WSRAs or vice versa as the \emph{local surgery}. We first describe the local surgery procedure and then the global
surgery that uses the local surgery as a subroutine.

We need the following lemma, giving a plan for two robots to move freely within revolving areas,
which we will use repeatedly to perform the local surgery.

\begin{figure}
    \centering
    \includegraphics[width=.45\textwidth]{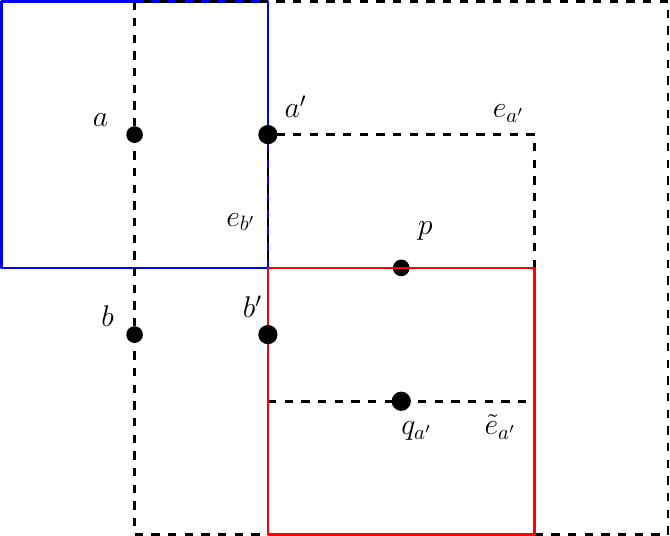} \hspace{1cm} \includegraphics[width=0.45\textwidth]{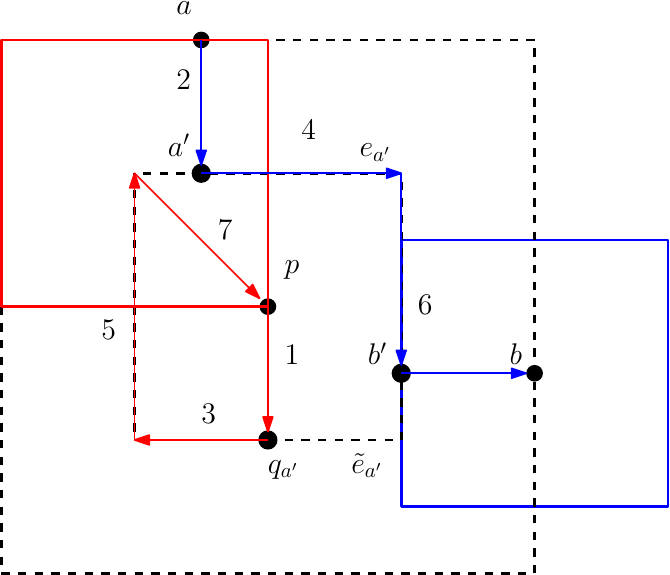}
    \caption{The decoupled $((p,a), (p,b))$-plan $\pth$ used in \lemref{jiggle}.
    (left) The points of interest in the procedure in \lemref{jiggle}.
    (right) The described (2-robot) plan $\pth$ with each move numbered in order for some example $a,b \in \partial B(p,2)$.}
    \label{fig:jiggle}
\end{figure}

\begin{lemma}\label{lem:jiggle}
    Let $p \in \F$ be a point such that $\RA(p) \subseteq \F$, and let $a,b \in \partial(B(p,2))$. Then there exists a feasible
    $((p,a),(p,b))$-plan $\tilde{\pth} = (\pi_1,\pi_2)$ for two robots (i.e., one robot moves from $a$ to $b$ while the other
    begins and ends at $p$) such that $\pi_i \subset \RA(p), \pi_2 \subset B(p,2)$, and $\cost(\tilde{\pth}) < 20$.
\end{lemma}
\begin{proof}
    \ben{updated for squares but needs proofread}
    We describe a decoupled $\left((a,p),(b,p)\right)$-plan $\pth$, whose 2D projection lies exclusively in $B(p,2)$,
    for arbitrary $a,b \in \partial B(p,2)$. Suppose robot $1$ (resp. robot $2)$ is the robot whose motion is
    defined by the first (resp. second) coordinate, and suppose
    $a'$ (resp $b'$) is the nearest point on $\partial(\RA(p))$ to $a$ (resp. $b$). Note that either
    $a',b'$ are corner squares of $\RA(p)$, or
    $a'$ (resp. $b'$) shift $a$ (resp. $b$) one unit in the $x$ direction if $a$ (resp. $b$) lies on
    a vertical edge of $B(p,2)$ and one unit in the $y$ direction otherwise.
    Suppose $e_{a'}$ (resp. $e_{b'}$) is the edge of $\RA(p)$
    for which $a' \in e_{s'}$ (resp. $b' \in e_{t'}$). Let $\opps$ (resp. $\oppt$) be the edge of
    $\partial(\RA(p))$ opposite to $e_{a'}$ (resp. $e_{b'}$). See \figref{jiggle}.
    We define $\tilde{\pth}$ by revolving the two robots around $\RA(p)$ in, say, clockwise orientation, as follows:
    \begin{enumerate}[1.]
        \item Move robot $1$ from $p$ to the nearest point $q_{a'}$ on $\opps$, via the straight segment $pq$.
        \item Move robot $2$ to $a'$, by the segment $aa'$ of length at most $\sqrt{2}$ perpendicular to $e_s$.
        \item Move robot $1$ from $q$ to the next clockwise vertex of $\partial(\RA(p))$.
        \item If robot $2$ is parked on the same edge as $b'$, move robot $2$ to $b'$, then to $b$ by the line segment
        $b'b$ of length at most $\sqrt{2}$ perpendicular to $e_{b'}$.
        Otherwise, move robot $2$ from its current position to the next clockwise vertex of $\partial(\RA(p))$.
        \item If robot $2$ is located at $b$, move robot $1$ from its parking position $q$ back to $p$ along the segment $qp$
        and terminate the process.
        Otherwise, repeat from step 3.
    \end{enumerate}
    We now argue the feasibility of $\tilde{\pth}$. Definitionally, the initial configuration $(p,a)$ is feasible;
    suppose WLOG that it is $x$-separated.
    Since $\opps$ is the opposing edge of $\RA(p)$ to $e_{s'}$, moving robot 1 to any point in $\opps$
    maintains that the robots are $x$-separated, as is the case when moving robot 2 to $a'$.
    Thereafter, on the $j$th repetition of the procedure beginning from step 3, clearly the robots will be $x$-separated
    if $j$ is even and $y$ separated if $j$ is odd, and so it follows that $\tilde{\pth}$ is feasible.

    Additionally, both robots trace a path whose length is bounded by the circumference of $\RA(p)$, along
    with the initial and final segments of length $\leq \sqrt{2}$ followed by robot 2. Since the length of $\partial \RA(p)$ is $8$,
    and the (at most two) segments taken by robot $2$ along $aa'$ and $b'b$ each have
    length at most $\sqrt{2}$, and $pq_{a'}$, $qp$ both have length at most one,
    $\cost(\pi_i) \leq 8 + 2\sqrt{2} < 11$ for each $i \in \{1,2\}$, and thus $\cost(\pth) < 22 = O(1)$.
    Moreover, definitionally $\pi_i \subset B(p,2)$, as desired.
\end{proof}

Having shown the existence of at least $j$ WSRAs inside the ball $K(q,24j)$ centered at a $72j$-far point $q$
which is in a wide area, we now argue that there exists a plan moving each robot $i$ from a
point on $\partial K(q,26j)$ into a WSRA inside the ball $K(q,24j)$.
(We need some buffer region between $K(q,26j)$ and $K(q,24j)$ to ensure that assigning WSRAs in this way is feasible.)
The interesting case for this is
when $k(q,25j)$ is almost a corridor, but wide enough to contain revolving areas, and uses
a similar proof to \lemref{portal-cases}.

\begin{lemma}\label{lem:boundary-k}
    Let $q$ be an $90k$-far point in a wide area, let $R = \{r_1, \ldots, r_k\}$ be the centers of $k$ WSRAs
    inside $K(q,24k)$, and let $i \in \BR$.
    Let $\bp = (p_1, \ldots, p_k) \in \BF$ be a free configuration such that 
    $p_i \in \partial K(q,26k)$, and $p_j = r_j$ for $j \not= i$, if $p_j \in \interior K(q,26k)$.
    Let $\bq = (q_1, \ldots, q_k) \in \BF$
    be a free configuration such that $q_i = r_i$ and $q_j = p_j$ for all $i \not= j$.
    Then there exists a $(\bp,\bq)$-plan $\bphi$, with $\cost(\bphi) \leq C_0 k^2$ for a constant $C_0 >0$.
\end{lemma}
\begin{proof}
    We prove the claim by moving the robot $i$ with $p_i \in \partial K(q,26k)$ to its corresponding point $q_i = r_i$
    inside a revolving area. Moving $i$ to $r_i$
    may move the other robots each only $O(1)$ distance away from and then back to their initial positions,
    by a plan $\pth$ with $\cost(\pth) = O(k)$.

    Let $i$ be the robot satisfying $p_i \in \partial K(q,26k)$, and
    let $\bar{e}$ be the edge of $\partial K(q,24k)$ such that $p_i \in \bar{e}$.
    We consider two cases:
    {
     \renewcommand{\labelenumi}{{(Case \arabic{enumi})}:}
     \setlength{\leftmargini}{5em}  
    \begin{enumerate}
        \item The length of $\bar{e}$ is at least $k+2$.
        \item The length of $\bar{e}$ is less than $k+2$.
    \end{enumerate}
    }
    
    We first handle the simplest
    case when $\bar{e}$ has length $||\bar{e}||_\infty \geq k+2$.
    We move $i$ by a straight segment in $\F$ from $p_i$
    onto the closest point $p_i'$ on $\partial B(q,26k-2)$,
    and then move it directly to $r_i$.
    If $p_i$ lies within distance 1 of a corner of $B(q,26k)$, it is possible
    that doing so creates a conflict with another robot $j$, as there could be an edge of $\partial \F$ nearly intersecting
    $\bar{e}$ and pushing robot $i$ into $j$. Suppose $p_i, p_j$ are $x$-separated, and $x(p_i) \leq x(p_j) - 2$.
    If such a conflict occurs, simply shift robot $j$ two units in the
    $+x$-direction to avoid conflicting with $i$. If this causes any cascading conflicts with other robots,
    push them identically. This process is clearly feasible, since $||\bar{e}|| \geq k+2$.
    We then move robot $i$ to $q_i$, and shift any robots that were pushed in the process two units in the $-x$ direction.
    The cost of the plan to move $i$ from $p_i$ to $p_i'$ does not exceed $2\sqrt{2}\cdot k + 4k$.
    We can finally move $i$ from $p_i'$ to $r_i$ by the segment $p_i'r_i$,
    and for all sections of the segment contained in an occupied revolving area in $R$,
    we use the surgery in \lemref{jiggle}. The cost to do so does not exceed $20$ per revolving
    area, of which there are $O(k)$. Hence, $\cost(\pth) \leq 20k + 2\sqrt{2}\cdot 52k + 4k < 172k$ in the first case.

    We now concentrate on the more complicated case when $\bar{e}$ has length less than $k+2$.
    By \lemref{fat-square} at most two edges of $\F$ can intersect the boundary of $K(q,26k)$.
    If there are two free space edges intersecting $\partial K(q,26k)$, denote them by $e_1$ and $e_2$.
    Alternatively, suppose there is only one edge $e_1$ intersecting $K(q,26k)$. Since $q \in K(q,26k)$,
    at least one edge $e_2$ of $\partial K(q,26k)$ is also an axis-aligned boundary edge of
    $B(q,26k)$. Note that $e_2\not= \bar{e}$, as otherwise on assumption it would've been
    handled in Case 1. It is possible that there are multiple edges of the form specified for $e_2$,
    if so, assign one to be $e_2$ arbitrarily. 
    If there are no edges of $\F$ intersecting the boundary of $K(q,26k)$, set $e_1$ and $e_2$
    to be any pair of opposite boundary edges of $B(q,26k)$. If there are two edges of $\F$ intersecting the boundary
    of $K(q,26k)$, let $e_1$ and $e_2$ denote these edges.

    Observe that $e_1$ and $e_2$ define a trapezoid $\widecorr \subseteq K(q,26k)$, which has all of the same properties
    defining a corridor, except that $\width(\widecorr) \geq 2$ on assumption that $q$ does not lie in a corridor and is $90k$-far.
    It may be that (i) the portals of $\widecorr$ are axis-aligned, or we could have the trickier case (ii)
    where the portals of $\widecorr$ have slope $\pm 1$.
    Note, it is possible that $\bar{e}$ defines a portal of $\widecorr$, in which case the portal must be axis-aligned.

    Suppose first that $\bar{e}$ defines the portal $\sigma^-$ of $\widecorr$, a related approach will work more generally for case (i).
    Without loss of generality, since $\sigma^- = \bar{e}$ is axis-aligned,
    assume it is vertical. We can apply a similar plan to the one in \lemref{portal-cases} to move $i$ from $p_i$ to $q_i$,
    where robot $i$ can move from $p_i$ to the nearest point $p_i' \in \partial K(q,24k - 2)$ directly if $\widecorr$ is widening in the direction $u$
    oriented from $\sigma^-$ to $\sigma^+$, and robots need to be pushed away from the portal if $\widecorr$ is narrowing.
    That is, suppose $\widecorr$ is narrowing and robot $j$ conflicts with the segment $p_ip_i'$ for $i$.
    By a similar argument to \lemref{portal-cases},
    we can move robot $j$ in the widening direction horizontally two units, pushing any robots it conflicts with as well,
    and then move robot $i$ along $p_iq_i$ directly, and reverse the horizontal push of the first step.
    Since in the worst case $O(k)$ robots move two units in both directions of each respective step induced by pushing robot $j$, and $i$ moves along
    $p_ip_i'$, the cost of the $(\bp,\bq)$-plan is also $2\sqrt{2}\cdot k + 4k$ in this case.
    As before, it is straightforward to use \lemref{jiggle} to subsequently move $i$ from $p_i'$ to $q_i$,
    adding at most $2k$ to the cost. Hence, $\cost(\pth) \leq 20k + 2\sqrt{2}\cdot 52k + 4k < 172k$ in this case
    as well.

    Suppose, finally, that $\bar{e}$ does not define a portal of $\widecorr$, and suppose $\sigma^-$ is the nearest portal.
    Again suppose $\bar{e}$ is vertical.
    If $\sigma^-$ is axis-aligned, then it is also vertical, and an identical plan to the above suffices to move robot $i$ from
    $\bar{e}$ to the point $r_i$ via the line segment $p_ir_i$.
    On the other hand, suppose the portal were diagonal.
     We consider the portal $\sigma^-(-r)$, shifting $\sigma^-$ outward
    in the $-u$ direction of $\corr$
    for a value $0 < r \leq k+2$ such that $p_i \in \sigma^-(-r)$. \todo{See \figref{} for a depiction.} Note that doing so still yields a well-defined portal since $q$ is $90k$-far
    and thus there are no vertices of $\F$ on the edge extensions of $e_1$ or $e_2$ in $K(q,90k)$.
    If, for every robot $j$ with $p_j \not \in \interior K(q,26k)$, $\langle p_j, u \rangle \leq \langle p_i, u \rangle$, then
    there is a straightforward motion plan from the portal to $r_i$, as follows.

    Let $m$ denote the midpoint of $\sigma^-(-r)$, and let $\alpha = ||\sigma^-(-r)||_\infty$.
    Then by definition of portals, and since there are no corridors intersecting $K(q,26k)$,
    $B(m, \alpha/2) \subset \F$, and there is a square of radius $\alpha/2$, centered at $m$,
    containing $p_i$. Hence, there is at least one direction $d \in \{x,y\}$ so that the segment $\bar{s}$
    of length two in direction $d$ is free, that is, not only $\bar{s} \subset \F$, but $\bar{s} \subset \F[\{p_j \not= p_j\}]$,
    since $\bp$ is a feasible configuration, every $p_j \in \interior K(q,26k)$ is inside a revolving area
    within $K(q,24k)$, and every $p_j \not \in \interior K(q,26k)$ is thus either $x$ or $y$-separated from $p_i$ in the $-u$ direction.
    So, we move robot $i$ along $\bar{s}$ to a point on $\sigma^-$, and then from $\sigma^-$ to $r_i$ by the shortest path directly.
    The cost of this plan $\pth$ is simply $\cost(\pth) \leq 2 + 2\sqrt{2} \cdot 52k < 150k$.

    However, it could be the case that
    $\langle p_j, u \rangle > \langle p_i, u \rangle$ for some $j$. In this case, we choose the index $j \not=i$, such
    that $p_j \not \in \interior K(q, 26k)$
    maximizing $\langle p_j, u \rangle$, and by the argument above, move it directly to its corresponding WSRA at the point $r_j$. We repeat the process for
    any other remaining robots with this property, until moving robot $i$. Finally,
    we move every robot $j \not= i$ that had a maximal index and was moved to a WSRA back to its original position.
    The cost of the plan $\pth$ in this case is the same as the previous case,
    but for each robot, which is at most

    \[\cost(\pth) \leq (2 + 4\sqrt{2} \cdot 52k + 20k)k \leq 317k^2.\]

    All together, we conclude that there exists a $(\bp,\bq)$-plan as specified, with cost $C_0 \cdot k^2$ for a constant $C_0 = 317$.
\end{proof}

An identical statement moving robots from the boundary of $K(q,\alpha k)$
into $K(q,24k)$ holds more generally for a value $\alpha$ if $\alpha \geq 26$.

\begin{corollary}\label{cor:boundary-k}
    Let $\alpha \geq 26$, let $q$ be an $3\alpha k$-far point in a wide area, let $R = \{r_1, \ldots, r_k\}$ be the centers of $k$ WSRAs
    inside $K(q,24k)$, and let $i \in \BR$.
    Let $\bp = (p_1, \ldots, p_k) \in \BF$ be a free configuration such that 
    $p_i \in \partial K(q,26k)$, and $p_j = r_j$ for $j \not= i$, if $p_j \in \interior K(q,26k)$.
    Let $\bq = (q_1, \ldots, q_k) \in \BF$
    be a free configuration such that $q_i = r_i$ and $q_j = p_j$ for all $i \not= j$.
    Then there exists a $(\bp,\bq)$-plan $\bphi$, with $\cost(\bphi) \leq C_0 k^2$ for a constant $C_0 >0$.
\end{corollary}

Note that the only reason that the bound in \lemref{boundary-k} is quadratic in $k$ is that
the proof of \lemref{boundary-k}, if we are unlucky with the orientation of portals in the trapezoid $\widecorr$
contained in $K(q,26k)$, may drag additional robots into WSRAs within $K(q, 24k)$ before moving them out again.
If we call \lemref{boundary-k} iteratively, we can be a little more careful and
avoid moving too many robots into WSRAs and also produce a plan still with cost $O(k^2)$.

\begin{lemma}\label{lem:boundary-extra}
    Let $q$ be a $90k$-far point that lies in a wide area, let $R = \{r_1, \ldots, r_k\}$ be the centers of $k$ WSRAs
    inside $K(q,24k)$, and let $Z \subseteq \BR$.
    Let $\bp = (p_1, \ldots, p_k) \in \BF$ be a free configuration such that
    $p_i \in \partial K(q,26k)$ for every $i \in Z$, $p_j = r_j$ if $p_j \in \interior K(q,26k)$, and $p_j \in K(q,30k)$ otherwise.
    Let $\bq = (q_1, \ldots, q_k) \in \BF$
    be a free configuration such that $q_i = r_i$ for all $i \in Z$ and $q_j = p_j$ for all $j \not \in Z$.
    Then there exists a $(\bp,\bq)$-plan $\bphi$, with $\cost(\bphi) \leq C_0' k^2$ for a constant $C_0' >0$.
\end{lemma}
\begin{proof}
    We proceed by checking the cases in \lemref{boundary-k}, and avoiding moving any robots occupying a WSRA
    out of $\interior K(q,26k)$ until the end of the plan. Fix a robot $i$ for which $i \in Z$, and thus $p_i \in \partial K(q,26k)$.
    Analogously to \lemref{boundary-k}, let $\bar{e}_i$ denote the edge of $\partial K(q,26k)$ containing $p_i$.

    In all cases of \lemref{boundary-k} except for when $||\bar{e}_i||_\infty \leq k+2$ and $\widecorr$ has diagonal portals,
    the plan $\bphi$ produced had $\cost(\varphi_i) \leq C_0k$ to move robot $i$ to $r_i$. So, if for every $i \in Z$ either
    $||\bar{e}_i||_\infty \leq k+2$ or $\widecorr$ has axis-aligned portals, then applying the same constructions as in
    \lemref{boundary-k} for every $i \in Z$ yields a plan $\bphi$ with $\cost(\bphi) = C_0 k^2$. If this is the case then we are done,
    so suppose $||\bar{e}_i||_\infty \leq k+2$ for at least one robot $i$ and that the resulting trapezoid $\widecorr$
    has diagonal portals. Let $u$ be the direction of $\widecorr$.

    As mentioned above, again we apply the construction in \lemref{boundary-k}, but leave any robots $j \not= i$ assigned a WSRA
    in the procedure at their position $r_j$ until the very end.
    In particular, we choose a robot $j\in \BR$ so that
    $\langle p_j, u \rangle > \langle p_i, u \rangle$ and the segment $p_ir_i$ conflicts with $p_j$.
    That is, we choose the robot $j \not=i$ conflicting with $i$, such
    that $p_j \not \in \interior K(q, 26k)$ and $\langle p_j, u \rangle$ is maximized. We move it to $r_j$ by the feasible plan
    described in \lemref{boundary-k}. We repeat the process for all robots. The produced plan, as in \lemref{boundary-k}, has
    cost $C_0k^2$. Now, for every robot $j \in \BR \setminus Z$ in reverse order of $\langle p_j, u \rangle$, we then move $j$ from
    $r_j$ back to $p_j$ by the segment $r_jp_j$, using \lemref{jiggle} any time $j$ intersects a revolving area of another parked robot.
    The cost of this step is at most $\sqrt{2} \cdot 52k + 20k$ per robot, and thus at most $94k^2$ overall.
    Hence, there exists a plan $\bphi$ with $\cost(\bphi) \leq C_0k^2 + 94k^2 = C_0'k^2$ for a constant $C_0' > 0$.
\end{proof}

As with \corref{boundary-k}, we can immediately extend \lemref{boundary-extra} to work when the set $Z$ of robots which we would like
to place at a revolving area inside $K(q,24k)$ lies within a larger square.

\begin{corollary}\label{cor:boundary-extra}
    Let $q$ be a $90k$-far point that lies in a wide area, let $R = \{r_1, \ldots, r_k\}$ be the centers of $k$ WSRAs
    inside $K(q,24k)$, and let $Z \subseteq \BR$.
    Let $\bp = (p_1, \ldots, p_k) \in \BF$ be a free configuration and $\alpha \geq 1$ be a value such that
    $p_i \in \partial K(q,\alpha 26k)$ for every $i \in Z$, $p_j = r_j$ if $p_j \in \interior K(q,26k)$, and $p_j \in K(q,30k)$ otherwise.
    Let $\bq = (q_1, \ldots, q_k) \in \BF$
    be a free configuration such that $q_i = r_i$ for all $i \in Z$ and $q_j = p_j$ for all $j \not \in Z$.
    Then there exists a $(\bp,\bq)$-plan $\bphi$, with $\cost(\bphi) \leq C'_0 k^2$ for a constant $C'_0 >0$.
\end{corollary}

\paragraph{Local surgery: moving robots to and from WSRAs.}

We are now ready to describe the full local surgery. Let $q \in \F$ be a $90k$-far point that lies in a wide area,
let $R = \{r_1, \ldots, r_k\}$ be the set of centers of WSRAs in $K(q,24k)$ (see \lemref{wsras-count}),
let $Z \subseteq \BR$ be a subset of $j \leq k$ robots, and let 
$\ba = (a_1, \ldots, a_k)$ and $\bb = (b_1, \ldots, b_k)$ be two feasible configurations such that $a_j \in K(q,30k)$
and $b_j \in R$ for all $i \in Z$ and $a_j = b_j \not\in K(q,26k)$ for all $j \not \in Z$. We describe a feasible $(\ba,\bb)$-plan
$\bphi$ with $\cost(\bphi) = O(k^2)$.

Since there are at most two edges of $\F$ intersecting $K(q,30k)$ by \lemref{fat-square},
we consider the following two cases, and for either case, we define a direction $\dir$ dictating the surgery:

\paragraph{Case 1:} $K(q,30k)$ intersects two edges of $\partial\F$:
    
    If two edges of $\F$ intersect $K(q,30k)$, then analogously to the definition for
    corridors, $\F \cap K(q,30k)$ must contain a trapezoid $\widecorr$ with portals $\sigma^-, \sigma^+$ and bisecting line segment $\ell$;
    note that the $\ell_\infty$-length of $\sigma^-$ and $\sigma^+$ here is at least two.
    We choose $\dir$ to be the direction parallel to $\ell$, oriented in the widening direction of $\widecorr$.

\paragraph{Case 2:} $K(q,30k)$ intersects fewer than two edges of $\partial\F$:
    
    If at most one edge $e_1$ of $\F$ intersects $K(q, 30k)$, then we define
    $\dir$ to be the direction parallel to $e_1$. Since $q \in K(q,30k)$, in this case
    there is at least one edge $e_2$ of $\partial K(q,30k)$ contained entirely in $\F$.
    Analogously to Case 1 we choose this edge $e_2$ to be a blocker, define a trapezoid $\widecorr$
    with blockers $e_1$ and $e_2$ identically, and let $\dir$ be the
    the widening direction with respect to $\ell$, the portal bisector of
    $\widecorr$. If there is no edge $e_1$ intersecting $K(q,30k)$
    then we simply choose $e_1$ and $e_2$ to be any opposing pair of edges of $B(q,30k)$,
    giving $\widecorr = B(q,30k)$, and let $\dir$ be the axis-aligned direction perpendicular to $e_1$ and $e_2$,
    oriented in either direction.

\begin{figure}
    \includegraphics[width=.9\textwidth]{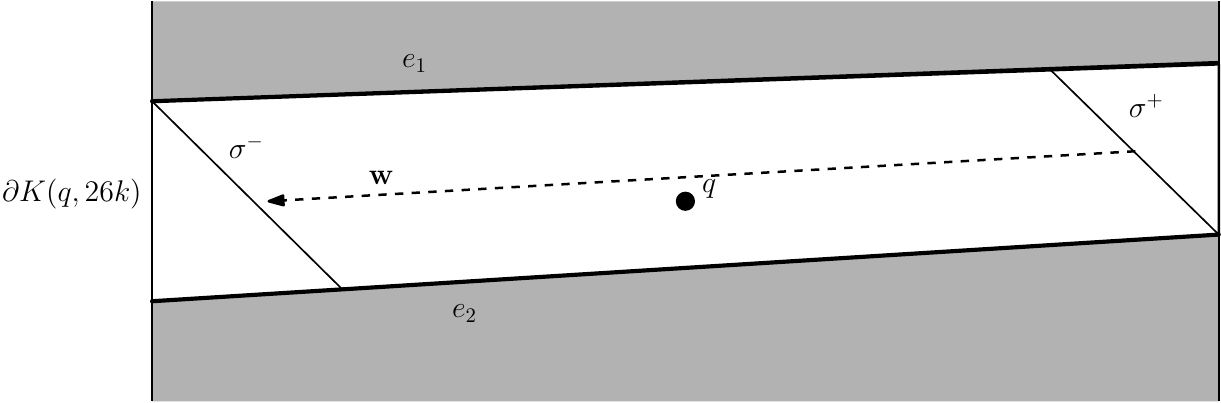}
    \caption{The direction $\dir$ as defined by the trapezoid $\widecorr$,
    which fits the criteria of a corridor except that $\width(\widecorr) \geq 2$, contained within $K(q,26k)$.}\label{fig:q-direction}
\end{figure}

\medskip
See \figref{q-direction} for an example.
Suppose, without loss of generality, the values of $a_1, \ldots, a_k$
are sorted in the order with respect to direction $\dir$,
that is, $\langle a_i, \dir \rangle \leq \langle a_{i+1}, \dir \rangle$
for all $i \in [k-1]$.
Thus, the local surgery proceeds in two phases
in both Case 1 and Case 2 by (i) sweeping over $K(q,30k)$ in direction $\dir$
and pushing the robots in direction $\dir$ along a segment parallel to $\dir$
in order to guarantee that no robots lie in the interior of $K(q,24k)$,
thus ensuring by \lemref{wsras-count} there are $k$ avaliable WSRAs within $K(q,24k)$.
After ensuring this property, then (ii) we assign the robots a WSRA one at a time.

At a high level, the idea of phase (i) is to push robot $i$, in order as defined by $\dir$
from $i=1$ to $k$, in the direction $\dir$ from $a_i$ until its position $p_i$ lies outside $\interior(K(q,26k))$ and
$\langle p_i, \dir \rangle >  \langle q, \dir \rangle$.
If doing so causes $i$ to conflict with another robot $j$, we push $j$ in direction $\dir$ as well.
(We do not stop pushing when $j$ reaches $\partial K(q,26k)$, only do this for $i$.
If $i$ reaches the boundary of $K(q,26k)$ then robot $j$ will go at least distance 2,
and not more than distance $2k$, beyond the boundary of $K(q,26k)$.)
If any robot $h$ still has its current position $p_h \in K(q,26k)$, or
$\langle p_h, \dir \rangle <  \langle q, \dir \rangle$ we repeat the pushing process above.
If there is no remaining robot $h$ with either property that $p_h \in K(q,26k)$ or $\langle p_h, \dir \rangle <  \langle q, \dir \rangle$,
we continue to Phase (ii).

At a high level, the idea of phase (ii) then is to move robots into $K(q,24k)$, in the order that they appear near the boundary of $K(q,26k)$.
We argue that there is a plan to do this, moving the robots into a WSRA inside $K(q,24k)$ one at a time.
So, initializing the procedure by setting $i = 1$ to be the robot furthest in the $-\dir$ direction,
the two phases proceed as follows:

\paragraph{Phase (i):}
    Let $p_i \in \partial K(q,26k)$ be the point in direction $\dir$
    intersecting $\partial K(q,26k)$ originating from $a_i$, or, if there is no such point, since $\langle a_i, \dir \rangle < \langle q, \dir \rangle$,
    let $p_i$ be the first point in direction $\dir$ from $a_i$ so that $\langle a_i, \dir \rangle \geq \langle q, \dir \rangle$.
    Let $\varphi_i^{(i)}$, the 2D projection of $\bphi^{(i)}:[0,1] \to \BF$ followed by robot $i$, trace the segment $a_i p_i$,
    where $\varphi_i^{(i)}(0) = a_i$, and $\varphi_i(1) = p_i$.
    Moving robot $i$ along $\varphi_i^{(i)}$ may conflict with other robots, so suppose $j$ is the first such robot where $\varphi_i^{(i)}$
    conflicts with $a_i$, i.e., $a_i\in \varphi_i^{(i)} \oplus \poly$.
    Let $\ell_j$ be the ray, originating from $a_j$, with direction $\dir$. Let $p_j$ be the first point on $\ell_j$ such that
    $||p_i - p_j||_\infty \geq 2$, $\langle p_j, \dir \rangle \geq \langle q, \dir \rangle$, and $p_j \not \in \interior K(q,26k)$.
    Define $\varphi^{(i)}_j$ to trace the segment $a_ip_j$. For the first robot $l$ either subsequently conflicting with $\varphi^{(i)}_i$, or conflicting with
    $\varphi^{(i)}_j$, define $\varphi^{(i)}_l$ similarly, along the segment $a_lp_l$. \todo{See figure.}
    Then move the robots along their paths as described earlier, whenever a conflict could occur.

\paragraph{Phase (ii):}
    Now every robot $i$ lies at a position $p_i$ (possibly $p_i = a_i$)
    such that $p_i \not \in \interior(K(q,26k))$ and $\langle a_i, \dir \rangle \geq \langle q, \dir \rangle$.
    Maintain a set $A$ of active robots not yet assigned a WSRA, and initially set $A = \BR$.
    Let $b = \min_{i \in A}||p_i - b||_\infty$, and suppose $i$ is a robot
    with $||p_i - q||_\infty = b$.
    First, assign $i$ to a WSRA $r_i$ defined from \lemref{wsras-count}, and then
    use the plan given in \lemref{boundary-extra} (or \corref{boundary-extra})
    to move $i$ to $r_i$. Remove robot $i$ from $A$, and repeat until every robot has been assigned a WSRA.

We will denote by $\bphi^{(i)}$ and $\bphi^{(ii)}$ the plans resulting
from phase (i) and phase (ii) of the above procedure, respectively.
\todo{See figure.}
We let $\bphi = \bphi^{(i)} \circ \bphi^{(ii)}$ be the $(\ba,\bb)$-plan as desired by the local surgery. 
We verify the feasibility of $\bphi^{(i)}$ in Case 1 in Lemma \ref{lem:dispersion-1-1}, of $\bphi^{(i)}$ in Case 2
in \lemref{dispersion-2-1}, of $\bphi^{(i)}$ in either Case 1 or Case 2 in \lemref{dispersion-2-2}.
In all cases, we can bound the cost of the plan we produce to be
$\cost(\bphi) = O(k^2)$, which we verify in
\lemref{dispersion-1-and-2-cost}. All together, this proves
\lemref{dispersion}.

\begin{lemma}\label{lem:dispersion-1-1}
    The plan $\bphi^{(i)}$ described in Phase (i) of Case 1 of the local surgery is feasible.
\end{lemma}
\begin{proof}
    Since, on assumption of Case 1, there are two edges of $\F$ intersecting
    $K(q,30k)$, denote these edges by $e_1$ and $e_2$.
    Since $q$ is not in any corridor, and $q$ is $90k$-far,
    there is also no corridor $\corr$ such that $\corr \cap K(q,30k) \not= \emptyset$,
    otherwise $K(q,30k)$ must contain a vertex of $\F^{(2)}$.
    This implies that $e_1$ and $e_2$, taken as blockers analogously to the definition
    for corridors, define a maximal trapezoid $\widecorr$, contained in $K(q,26k)$,
    with portals defined by squares supported in $\F$ of radius at least two.
    Let $\sigma^-, \sigma^+$ denote the portals of $\widecorr$, and let $\ell$ denote
    the portal bisector. Further, for every point $p \in \ell$, there is a maximum-radius
    square $B(p,R) \subset \F$, with radius $c \geq 2$, and the value of $R$ increases monitonically
    in the widening direction $\dir$ of $\widecorr$.
    Then, for every $a_i$, the segment $\bar{w}_i$ parallel to $\ell$ in direction $\dir$
    with $a_i$ as one endpoint and a point of $\partial K(q, 26k)$ as the other,
    satisfies that $\bar{w}_i \subset \F$.

    Thus, since every $\bar{w}_i \subset \F$, for every robot $i$
    with $a_i \in K(q,26k)$ or $\langle a_i, \dir \rangle <  \langle q, \dir \rangle$,
    pushing the robots in direction $\dir$ as described in Phase (i) of the construction
    is feasible, and the position $p_i$ of a robot after the procedure will be on the boundary
    of $K(q,26k)$ if the robot was never pushed by some other robot and had $a_i \in K(q,26k)$,
    and $p_i \in K(q,30k)$ otherwise. If robot $i$ originally had placement $a_i \in K(q,30k)\setminus K(q,28k)$,
    it can never be pushed by another robot in the described construction, and remains in $K(q,30k)$
    with $p_i = a_i$. The feasibility of $\bphi^{(i)}$ follows.
\end{proof}


\begin{lemma}\label{lem:dispersion-2-1}
    The plan $\bphi^{(i)}$ described in Phase (i) of Case 2 of the local surgery is feasible.
\end{lemma}
\begin{proof}
    Let $\ba \in K(q,30k)$ be as described in the construction above,
    and again suppose without
    loss of generality that the coordinates of $\ba$ are sorted
    in the direction $u$.
    Similar to \lemref{dispersion-1-1}, any robot $i$ placed initially
    so that $a_i \in K(q,28k)$ or $\langle a_i, u \rangle <  \langle q, u \rangle$ may need to move,
    either directly by the construction or it could be
    pushed by other robots; however, again if $a_i \in K(q,30k) \setminus K(q,28k)$
    and $\langle a_i, u \rangle \geq  \langle q, u \rangle$, we have that $a_i = p_i$. 
    
    In the former case,
    since there is at most one boundary edge of $\F$ intersecting
    $K(q,30k)$, for any $a_i$ observe that there is a segment $\bar{w}_i \subset K(q,30k)$
    parallel to $u$ with $a_i$ as one of its endpoints and a point of
    $\partial K(q,30k)$ as the other endpoint. Thus, as before, any robot $i$ called originally
    (as opposed to being pushed by another robot)
    to move in the construction that has $a_i \in K(q,26k)$ has also that $\bar{w}_i \subset K(q,26k)$
    and thus since $\bar{w}_i \subset K(q,30k)$ its motion is feasible, and any robots it pushes
    remain inside $K(q,28k)$ as in \lemref{dispersion-1-1}. 
    Alternatively, if $a_i \not \in K(q,26k)$, but $i$ was called originally to move,
    since $u$ was chosen in the widening direction, any robot $j$ pushed by $i$
    also remains in $K(q,30k)$; if $j$ enters $\interior(K(q,26k))$ its position
    will re-define $a_j$, and $j$ will
    subsequently be pushed to leave $\interior(K(q,26k))$.

    Hence, in all cases, the position after Phase (i) leaves every robot $i$ in $K(q,26k)$,
    and on some point of $\bar{w}_i$.
    Since $\bar{w}_i \subset K(q,26k) \subset \F$, the feasibility follows.
\end{proof}

Iteratively applying \lemref{boundary-extra} and \corref{boundary-extra} immediately gives
the feasibility of Phase (ii) in either case.

\begin{lemma}\label{lem:dispersion-2-2}
    The plan $\bphi^{(ii)}$ described in Phase (ii) of both Case 1 and Case 2 of the local surgery is feasible.
\end{lemma}

Similarly, we can check the total cost of the plan produced in either case throughout Phase (i) and Phase (ii) and
verify that it is $O(k^2)$.

\begin{lemma}\label{lem:dispersion-1-and-2-cost}
    The plan $\bphi = \bphi^{(i)} \circ \bphi^{(ii)}$ for either Case 1 or Case 2 has $\cost(\bphi) \leq C_1 k^2$
    for a constant $C_1 > 0$
\end{lemma}
\begin{proof}
    When executing Phase (i), the pushing motion in direction $\dir$ moves at most all $k$
    robots in a straight line in direction $\dir$ within $K(q,30k)$. Hence,
    the cost of the path for each robot $j$ is
    $\cost(\varphi^{(i)}_j) \leq 2\sqrt{2} \cdot 30k$, and so $\cost(\bphi^{(i)}) \leq 2\sqrt{2} \cdot 30k^2$.

    After Phase (i), it is straightforward to see that every $i \in \BR$, that originally was located at some $a_i \in K(q,30k)$
    still lies in $K(q,30k)$, and we can initially apply \lemref{boundary-extra} with $R = \emptyset$.
    Thereafter, supposing $i$ is the robot for which $d_i = ||p_i - q||_\infty$ is minimized,
    we can apply \corref{boundary-extra} on the connected component $K(q,d_i)$. Doing so successively
    for each possible definition of $d_i$ while $A \not= \emptyset$, gives a plan $\bphi$ with $\cost(\bphi) \leq C_0' k^2$,
    after using the same trick from \lemref{boundary-extra}
    to keep every robot that parks in a WSRA inside $K(q,24k)$ during the procedure at this parking place until the very end of the procedure,
    when $d_i$ is maximized. By \corref{boundary-extra},
    we obtain that the total cost charged to $\bphi$ is at most $O(k^2)$ with constant given by $2\sqrt{2} \cdot 30 \cdot 26/24C_0'$,
    and hence, that $\cost(\bphi^{(ii)}) = O(k^2)$

    So, we obtain that the plan $\bphi$ produced in Case 1 has 
    \[\cost(\bphi) \leq \cost(\bphi^{(i)}) + \cost(\bphi^{(ii)}) \leq 2\sqrt{2} \cdot 30k^2 + 2\sqrt{2} \cdot 30 \cdot 26/24C_0' = C_1k^2\]
    where $C_1 > 0$ is a constant.
\end{proof}


Putting everything together, we obtain the following.

\begin{lemma}\label{lem:dispersion}
    Let $q \in \F$ be a $90k$-far point that lies in a wide area, let $R = \{r_1, \ldots, r_k\}$ be a set of centers of WSRAs
    in $K(q,24k)$, let $\BS \subseteq \BR$ be a subset of $j \leq k$ robots, and let $\ba = (a_1, \ldots, a_k)$ and $\bb = (b_1, \ldots, b_k)$
    be two feasible configurations such that $a_j \in K(q,24k)$ and $b_j \in R$ for all $j \in Z$. Then there
    exists a feasible $(\ba,\bb)$-plan of cost at most $C_1k^2$ for some constant $C_1 > 0$.
\end{lemma}

\paragraph{Global surgery: eliminating far breakpoints}

We fix three parameters: $\delta := Ck$ where $C$ is a constant defined in \todo{lemma},
$\Delta = \delta k = Ck^2$, $\Delta^- = \Delta - 90k$ and $\Delta^+ = \Delta + 90k$. In this section, because
we use it heavily, for a point $q \in \F$ we write $K(q)$ as a shorthand for $K(q,24k)$.
We now describe the global surgery
to remove all $\Delta^+$-far breakpoints from a plan that lie in a wide area at a small increase
in the cost of the plan. As mentioned above, we perform the surgery by freezing and unfreezing
robots, parking frozen robots in WSRAs, and rerouting a robot when it unfreezes so that its motion does
does not have a $\Delta^+$-far breakpoint in a wide area. The global surgery, in detail, is as follows.

Let $\pth:[0,1] \to \BF$ be a feasible (piecewise-linear) $(\bs,\bt)$-plan. We maintain the following information:
\begin{enumerate}[(i)]
  \item A set $Z \subseteq \BR$ of frozen robots,
  \item a set $Q = \{q_1, \ldots, q_s\}$
  where $s \leq k$, of $\Delta$-frontier points such that the robots of
  $Z$ are parked in WSRAs of $K(q_i) := K(q,24k)$ for some $q_i \in Q$,
  \item $R_i = \{r_{i1}, \ldots, r_{ik}\} \subset K(q_i)$, the set of centers of $k$ WSRAs according to \lemref{wsras-count},
  \item $R_i^{O} \subseteq R_i \subseteq R_i^{U} \subseteq R_i$: the set \emph{occupied} and \emph{unoccupied} (centers of) WSRAs of $R_i$,
  \item $Z_i \subset Z$, the set of robots parked in the occupied WSRAs of $K(q_i)$; note that
  $|Z_i| = |R_i^O|$ and $Z_1, \ldots, Z_s$ is a partition of $Z$ (a frozen robot may lie in the neighborhood
  of multiple points in $Q$, but it is assigned to only one of them), and
  \item a matching $M_i \subseteq Z_i \times R_i^O$ where $(j,p)$ implies that robot $j$ is parked at $p \in R_i^O$
  (only one robot is parked at $p$).
\end{enumerate}

Initially, $Z = \emptyset$, $Q = \emptyset$, and thus the rest of the sets are undefined.

By performing a surgery on $\pth$, we construct an $(\bs,\bt)$-plan $\pth':[0,1] \to BF$
in which at any time $\lambda \in [0,1]$, any frozen robots in $Z$ are parked at a $\left(\Delta^-\right)$-far
(but $\Delta$-close) WSRA,
and for any unfrozen robot $j \in \BR \setminus Z$, $\pi_j'(\lambda) = \pi_j(\lambda)$. Any robot $i \in Z$ remains frozen while $\pi_i(\lambda)$
is $\delta$-far, and any robot $j \in \BR \setminus Z$ remains thawed while $\pi_j(\lambda)$ is both $\Delta$-close and outside of any neighborhoods $K(q_i, 80k)$, where
$q_i \in Q$.
Thus, we have one of three types of \emph{critical events} at a time
$\lambda \in [0,1]$ that cause a robot to either become frozen or thawed:

{
  \renewcommand{\labelenumi}{\textbf{(T\arabic{enumi})}:}
  \setlength{\leftmargini}{3em}  
\begin{enumerate}
    \item A thawed robot $i \in \BR \setminus Z$ reaches the $\Delta$-frontier at time $\lambda$, i.e.,
    $\pi_i(\lambda)$ lies at the $\Delta$ frontier.
    \item A thawed robot $j \in \BR \setminus Z$ enters $K(q_i, 80k)$ for some $q_i \in Q$ at time $\lambda$, i.e., $\pi_j(\lambda)\in \partial K(q_i, 80k)$.
    \item A frozen robot $i \in Z$ becomes $\delta$-close at time $\lambda$, i.e., $\pi_i(\lambda)$ lies at the $\delta$-frontier.
\end{enumerate}
}



By traversing $\pth$ in increasing order of time and maintaining $Z$ and $Q$, we can determine all critical events in
a straightforward manner. Let $\lambda_1 = 0 < \lambda_2 < \ldots < \lambda_m = 1$ be the sequence
of critical events. For simplicity, we assume that only one event occurs at each critical event,
but it is easy to adapt the procedure to handle multiple simultaneous events.
Processing $\lambda_1$ is easy: we set $\pth'[0,\lambda_1) = \pth[0,\lambda_1)$, $Z = \emptyset$, and $Q = \emptyset$.
Suppose we have processed $\lambda_1, \ldots, \lambda_{i-1}$ and constructed $\pth'[0,\lambda_i)$. We process the event $\lambda_i$
and construct $\pth'[\lambda_i, \lambda_{i+1})$.
The surgery proceeds by case analysis
depending on the type of critical event for $\lambda_i$:

    \paragraph{(T1)} Suppose robot $j \in \BR \setminus Z$ reaches the $\Delta$-frontier. Add $q := \pi_j(\lambda_i)$
    to $Q$. Compute WSRAs $R_{s+1}$ of $K(q)$ using \lemref{wsras-count}. Let $Z_{s+1} \subseteq \BR \setminus Z$
    be the set of unfrozen robots including $j$, that lie in $K(q_{s+1})$, i.e., the set of all $j \in \BR \setminus Z$
    such that $\pi_j(\lambda_i) \in K(q_{s+1})$. Assign WSRAs of $R_{s+1}$ to $Z_{s+1}$, set them to be
    $R^{O}_{s+1}$ and set $R^{U}_{s+1} = R_{s+1} \setminus R_{s+1}^{O}$.
    Let $M_{s+1}$ be the resulting assignment. Using the local surgery procedure described above, move all robots in $Z_{s+1}$
    to their assigned parking placements. Let $\bphi$ be the plan computed by the local surgery.
    We define $\bpsi:(\lambda_i, \lambda_{i+1}) \to \BF$ as follows:
    For $j \in Z$, $\psi_{j}(\lambda) = \varphi_j$ for all $\lambda \in (\lambda_i, \lambda_{i+1})$
    where $\varphi_j$ is the assigned parking place of $j$. For $j \in \BR \setminus Z$,
    $\psi_{j}(\lambda) = \pi_j(\lambda)$ for all $\lambda \in (\lambda_i, \lambda_{i+1})$. We set $\pth'[\lambda_i, \lambda_{i+1}] = \bphi \circ \bpsi$.

    \paragraph{(T2)} Suppose robot $j \in \BR \setminus Z$ enters $\partial K(q_u, 80k)$ for some $q_u \in Q$. We choose a point $\xi \in R_u^U$
    and assign $\RA(\psi)$ to $j$. Add $j$ to $Z_u$ and $Z$, move $\xi$ from $R_u^U$ to $R_u^O$, and add $(j, \xi)$ to $M_u$.
    Using the local surgery we compute a feasible plan $Q$ that moves $j$ from $\pi_j(\lambda_i)$ to $\psi$ while the other robots
    begin and end at their current position. Define $\bpsi$ as in (T1) event and set $\pth'[\lambda_i, \lambda_{i+1}] = \bphi \circ \bpsi$.

  \paragraph{(T3)} Suppose $j \in Z_u$ for some $u \leq s$. 
    Remove $j$ from $Z$ and $Z_u$, update the sets $R_u^O, R_u^U, M_u$, and remove $q_u$
    from $Q$ if $R_u^O = \emptyset$ (all the sets associated with $q_u$ are also removed).
    Let $P_j$ be the shortest $\delta$-far path from the WSRA of $j$ to its next $\delta$-close
    position, $\pi_j(\lambda_i)$ in $\pth$. If there exists a point $\xi \in P_i$
    and an unfrozen robot $h \in \BR \setminus Z$ such that $\pi_h(\lambda_i) \in B(\xi,2)$,
    i.e., $h$ conflicts with $P_i$, apply the local surgery for the unfrozen robots in $K(\xi, 80k)$ and freeze them.
    Recurse on any remaining conflicts with robots not in a WSRA.
    Let $Y_i \subseteq \BR \setminus Z \cup \{j\}$ be the set of robots that are moved to WSRAs in this step.
    Move $j$ to $\pi_j(\lambda_i)$ along $P_i$ using \lemref{jiggle} if $j$ ever enters a WSRA of 
    some robot. Once $R_i$ reaches $\pi_i(\lambda)$, undo the surgery performed above to move all robots $h \in Y_i$
    back to their original positions at $\pi_h(\lambda_i)$. Let $\bphi = \bphi_1 \circ \bphi_2 \circ \bphi_3$ be the plan
    describing the above motion, where $\bphi_1$ is the local surgery that moves the robots interfering with $P_i$
    to WSRAs, $\bphi_2$ is the motion of $j$ along $P_j$ including the local motion of other robots if $j$ passes through
    their revolving areas, and $\bphi_3$ is the motion that brings the robots in $Y_i$ to their original positions.
    Then we set $\pth'[\lambda_i, \lambda_{i+1}) = \bphi \circ \pth(\lambda_i, \lambda_{i+1})$.

Let $\pth'$ be the $(\bs,\bt)$-plan obtained after processing all events. We now prove that $\pth'$ is feasible
and does not contain any $\Delta^+$-far breakpoints. We then bound $\cost(\pth')$.
First, to prove feasibility of $\pth$, we need the following lemma.

\begin{lemma}\label{lem:far-not-in-corridor}
  Let $\ba,\bb \in \BF$ be two points, such that $\ba$ is $\Delta$-far and $\bb$ is on the $\delta$-frontier.
  Let $\bphi$ be a piecewise-linear $(\ba,\bb)$-plan such that $\bphi(\lambda)$ is $\delta$-far for every $\lambda \in [\lambda^-,\lambda^+)$.
  Then, for every $\lambda \in [\lambda^-,\lambda^+)$, there is no robot $i \in \BR$ for which $\varphi_i(\lambda)$ is in a corridor.
\end{lemma}
\begin{proof}
  Suppose for contradiction that there existed a $\lambda^* \in [\lambda^-, \lambda^+)$ where $\varphi_i(\lambda^*) \in \sigma$
  for a portal $\sigma$ of some corridor $\corr$. Then by definition there is a vertex $v$ of $\F^{(2)}$ for which
  $||\psi_i(\lambda^*) - v||_\infty < 2$. Yet, recall that the vertices of $\F^{(2)}$ are included as landmarks,
  so $v \in \Lambda$, and $\varphi_i(\lambda^*)$ is $\delta$-close. This contradicts the assumption that $\bphi(\lambda)$
  is not $\delta$-close for all $\lambda \in [\lambda^-, \lambda^+)$.
\end{proof}

\begin{lemma}\label{lem:feasible-global}
  The $(\bs,\bt)$-plan $\pth'$ as constructed by the global surgery is feasible and does not contain any $\Delta^+$-far breakpoints where
  $\Delta^+ \leq C_2k^2$ for a constant $C_2 > 0$.
\end{lemma}
\begin{proof}
  \todo{this mostly just needs an update with newest notation, and will call above lemma.}
Suppose $\pth$ is a min-sum $(\bs,\bt)$-plan and $\pth'$ is the plan produced from the global surgery.
By construction, the procedure above if $\lambda$ is a (T1) event immediately guarantees that $\pth'$ is feasible
on the feasibility of the plans described in \lemref{dispersion} and \lemref{jiggle}, as long as the frozen $24k$-radius neighborhoods
containing WSRAs,
centered at the points of $Q$, are disjoint. Indeed, we can guarantee this property, by the fact that
the (T2) events occur when robots enter an $80k$-radius neighborhood $K(q_i, 80k)$ of some $q_i \in Q$.
Thus, the only way for robots to lie in $K(q_i, 80k)$ for some $q_i \in Q$ is if they are parked at a center of a WSRA
of $R_i$,  contained in $K(q_i)$.
So, for any $q_i, q_j \in Q$, their corresponding sets $R_i$ and $R_j$ of WSRAs lie inside $K(q_i)$ and $K(q_j)$
respectively, and are all pairwise disjoint; 
$K(q_i) \cap K(q_j) = \emptyset$. By \lemref{wsras-count}, there are sufficiently many WSRAs inside each 
$K(q_i)$ to park every robot. So, the (T1) events as specified are all feasible.

Similarly, the (T2) events by construction are feasible, by \corref{boundary-k} and the fact that there exists a WSRA
inside $K(q_i)$ for robot $j$, if $\pi_j(\lambda) \in \partial K(q_i)$ spurred the (T2) event at time $\lambda$.

Finally, to check that the (T3) events are feasible we examine each of the subplans $\bphi_1, \bphi_2,$
and $\bphi_3$ comprising the resulting plan $\bphi$ that arises when robot $i$ causes such an event (when it next becomes
$\delta$-close in the original plan). To see that $\bphi_1$ is feasible, we note from \lemref{far-not-in-corridor}
that any points along $P_i$ conflicting with other robots do not do so in a corridor. Hence, at the conflict point,
we may apply \lemref{dispersion}. To see that $\bphi_2$ is feasible follows immediately from the feasibility of $P_i$
and \lemref{jiggle}. Finally, $\bphi_3$ is feasible after noting that it is feasible to use the surgery of
\lemref{dispersion} in the reverse direction, moving the robots from WSRAs to an initial configuration rather than the
usual direction.

So, $\pth'$ during all of the critical events is feasible. During the time intervals where no critical events are occuring,
$\pth'$ is feasible on the assumption that the original plan $\pth$ was feasible, and since the parking places of any robots in $Z$
are feasible.

Finally, by construction, our global surgery has maintained the invariant that no more than one robot can be $\Delta^+$-far at a time,
which occurs during the handling of (T3) events. So, the shortest path $P_i$ described will simply be a line segment while $P_i$ is $\Delta^+$-far,
and $P_i \subset \F[\{\pi_j(\lambda) \mid j \not= i\}]$, because of this property. We conclude that the plan, while $P_i$ is $\Delta^+$-far,
does not have any breakpoints. The claim follows.
\end{proof}

\begin{lemma}
  $\cost(\pth') \leq (1+C_3 k^2/\Delta) \cost(\pth)$ where $C_3 > 0$ is a constant.
\end{lemma}

\begin{proof}
As in \lemref{feasible-global}, we check each of the cases to bound $\cost(\pth')$ during critical events.
We charge the added error incurred from each critical event caused by a critical event due to robot $i$
to the cost of its plan; we denote the charge for robot $i$ by $\charge_i$.
That is, $\charge_i$ is defined to be $\cost(\pi_i)$ plus the total cost of the
plan in the global surgery that was called as a result of a critical event spurred by robot $i$.

Observe that if the critical event $\lambda_j$ is either a (T1) or (T2) event then it requires that
there must be at least one subsequent critical event $\lambda_{j+1} > \lambda_j$,
since $t_i$ is $0$-close. We note that a (T3) event $\lambda^*$ from robot $i$ incurs cost at most
$\cost(P_i)$ plus $C_1h^2$ for every conflict at a point $p$, which induces a call to \lemref{dispersion}
and moves each of the $h$ robots located in $K(p,80k)$ inside a WSRA in $K(p)$, plus $80k\sqrt{2} + 20h$ for $i$ to traverse
$K(p,80k)$ and revolve around any frozen robots with \lemref{jiggle}, and one more call to \lemref{dispersion} in reverse, adding
$C_1h^2$ once more at the very end. Since each conflict along $P_i$ involves disjoint sets of robots,
the total charge incurred by the $(T3)$ event is thus not exceeding
\[\charge_i(\lambda^*) \leq \cost(P_i) + 2C_1 k^2 + 180k.\]

Thus, let
$\Upsilon_i = \{[\zeta_1^-, \zeta_1^+], [\zeta_2^-, \zeta_2^+], \ldots [\zeta_l^-, \zeta_l^+]\}$ denote the \emph{frozen intervals},
$[\zeta_j^-, \zeta_j^+] \subset [0,1]$, in which $\zeta_j^-$ is either a (T1) or (T2) critical event for $i$ in which it is frozen,
and $\zeta_j^+$ is a (T3) event in which it is thawed. We now charge the cost of any local surgeries
to the robot $i$ causing the critical event, by summing the costs over all of the frozen intervals.
In particular, using the preceding local surgeries,
for a frozen interval $[\zeta^-, \zeta^+] \in \Upsilon_i$, if $\zeta^-$ is a (T1) event we have
\[\charge[\zeta^-, \zeta^+] \leq C_1 k^2 + \left(\charge_i(\zeta^+)\right) \leq C_1k^2 + \left(\cost(P_i) + 2C_1 k^2 + 180k\right).\] 
where the first term is from the call to \lemref{dispersion} by $\zeta^-$ and the second set of terms is from $\zeta^+$ as discussed above.
Similarly, if $\zeta^-$ is a (T2) event we have
\[\charge[\zeta^-, \zeta^+] \leq C_1k^2 + \left(\charge_i(\zeta^+)\right) \leq C_1k^2 + \left(\cost(P_i) + 2C_1 k^2 + 180k\right)\]
by \lemref{dispersion} (and more specifically, \corref{boundary-k}). So, let
\[\tilde{C}_3 = C_1k^2 + \left(\cost(P_i) + 2C_1 k^2 + 180k\right)\]
Hence, since $\cost(\pi_i[\zeta^-, \zeta^+]) \geq \Delta - \delta$ for every $[\zeta^-, \zeta^+] \in \Upsilon_i$,
\[\charge_i \leq \left(1 + \frac{\tilde{C}_3}{\Delta - \delta}\right)\cost(\pi_i).\]
An easy algebraic manipulation, recalling the formula for geometric series and since $0 < \delta/\Delta < 1$, gives
\begin{align*}
  1 + \frac{1}{\Delta - \delta} &= 1 + \frac{1}{\Delta(1-\delta/\Delta)}\\
  &= 1 + \frac{1}{\Delta}\left(1 + \frac{\delta}{\Delta} + \left(\frac{\delta}{\Delta}\right)^2 + \ldots\right) \\
  &\leq 1 + \frac{1}{\Delta}\left(1 + \frac{\delta}{\Delta}\right) \\
  &= 1 + \frac{1 + \delta/\Delta}{\Delta}.
\end{align*}

Using just that $\delta/\Delta < 1$, and so $1 + \frac{1 + \delta/\Delta}{\Delta} < 1 + \frac{2}{\Delta}$,
we obtain
\[\charge_i \leq \left(1 + \frac{2\tilde{C}_3}{\Delta}\right)\cost(\pi_i).\]

Finally, putting everything together, and letting $C_3 = 2\tilde{C}_3$,

\[\cost(\pth') \leq \sum_{i \in [k]}\charge_i \leq \sum_{i \in [k]}\left(1 + \frac{C_3}{\Delta}\right)\cost(\pi_i) = \left(1 + \frac{C_3}{\Delta}\right)\cost(\pth)\]
\end{proof}

To conclude, we obtain the following result, allowing us to approximate an optimal plan with a near-optimal one that does not have any $\Delta^+$-far breakpoints.

\begin{lemma}\label{lem:far-explicit}
  Let $\pth$ be a min-sum $(\bs,\bt)$-plan, and let $\Delta \geq C_1k^2$.
  Then there exists a feasible, piecewise-linear $(\bs,\bt)$-plan $\pth'$,
  that does not contain any $(\Delta^+)$-far breakpoints,
  with $\cost(\pth') \leq \left(1 + \frac{C_3k^2}{\Delta}\right)\cost(\pth)$.
\end{lemma}

\section{Omitted proofs}\label{sec:append}

For completeness, we include omitted proofs from claims earlier in the paper. Specifically,
we give the details proving the correctness of our algorithm.

\subsection{Analysis of the algorithm}\label{sec:correct}

\begin{figure}
    \centering
    \includegraphics[width=0.6\textwidth]{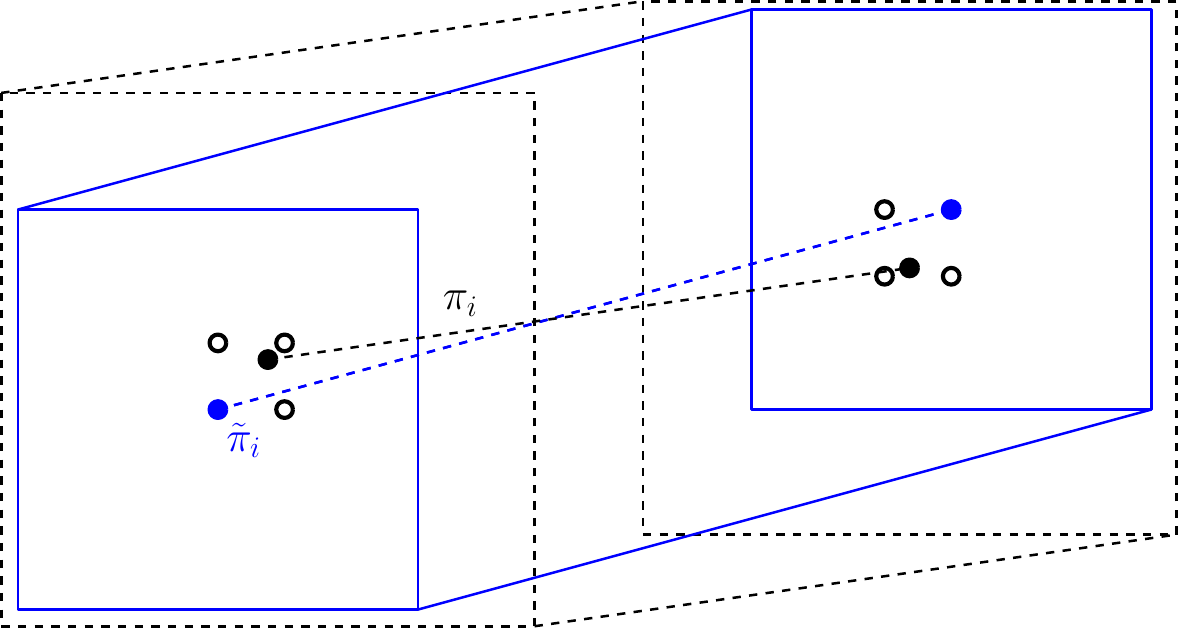}
    \caption{Snapping the breakpoints of a path $\pi_i$ (black points) to $\epsilon$-width grid vertices (empty points, with chosen vertices in blue)
    will perturb the endpoints of each segment in $\pi_i$ by $\ell_\infty$ distance at most $\epsilon$
    to produce the path $\tilde{\pi}_i$. If $\pi_i$ is a path in an $\epsilon$-robust plan,
    then $\tilde{\pi}_i \oplus \protect \poly \subset \pi_i \oplus (1+\epsilon) \protect \poly$,
    and thus $\tilde{\pi}_i \subset \F$.}
    \label{fig:perturb}
\end{figure}

Given a polygonal environment $\P$, $k$ unit squares $\BR$ that can translate in $\P$,
and a pair of start/target configurations $(\bs,\bt) \in \C^2$ let $\tilde{\pth}$
be the $(\bs,\bt)$-plan output by the algorithm in \secref{algo}.
We proceed by bounding the cost of $\tilde{\pth}$ and then analyze the runtime of the algorithm.
First, we note our robustness assumption that $\rho \geq \epsilon$
enables the following elementary lemma, whose proof is omitted. See \figref{perturb}.

\begin{lemma}\label{lem:perturb}
    Let $p, q \in \mathbb{R}^2$, and let $\rho, \epsilon \in (0,1)$ be two parameters. Let $\poly$ be a unit square
    as defined above. Let $p', q' \in \mathbb{R}^2$,
    so that $\|p - p'\|_\infty < \epsilon$ and $\|q - q'\|_\infty < \epsilon$.  If $\rho \geq \epsilon$,
    then $p'q' \oplus \poly \subset pq \oplus (1 + \rho)\poly$.
\end{lemma}

We can thus bound $\cost(\tilde{\pth})$.

\begin{lemma}\label{lem:cost-bound}
    Assuming there exists an $\epsilon$-robust $(\bs,\bt)$-plan, the $(\bs,\bt)$-plan $\tilde{\pth}$ we
    compute is feasible, and satisfies $\cost(\tilde{\pth}) \leq (1+\epsilon)\opt(\bs,\bt,1+\epsilon)$.
\end{lemma}
\begin{proof}
    Given the existence of a $\epsilon$-robust $(\bs,\bt)$-plan,
    by \lemref{bound} there exists an $(\bs,\bt)$-plan
    $\pth$ so that the number of breakpoints $\beta$ of $\pth$
    is proportional to the cost of an optimal $\epsilon$-robust plan;
    $\beta \leq n \cdot f(k, \epsilon^{-1})\left(\opt(\bs,\bt,1+\epsilon) + 1\right)$,
    where $f(k,\epsilon) = \left(k/\epsilon\right)^{k^{O(1)}}$.
    From \lemref{far} we can modify $\pth$ into a $\Delta$-tame plan $\pth'$ for which
    $\cost(\pth') \leq (1+\epsilon) \cost(\pth)$, where the modifications to ensure that $\pth'$ is
    $\Delta$-tame contribute only a constant factor to the function $f(k,\epsilon^{-1})$; \todo{see Lemma in section 3}.
    Let $\beta'$ denote the number of breakpoints of $\pth'$.
    Let $\tilde{\pth}$ be the $(\bs,\bt)$-plan that results from snapping the placement of every robot at a breakpoint of
    $\pth'$ to its nearest vertex in $\X$. By \lemref{perturb}, $\tilde{\pth}$ is feasible. Moreover,
    since every breakpoint contributes at most $\sqrt{2} \cdot \bar{\epsilon}$ additive error to $\cost(\pi_i)$ for every
    $i \in \BR$,
    \[\cost(\tilde{\pth}) \leq \cost(\pth') + \beta'\left(k \cdot \sqrt{2}\bar{\epsilon}\right).\]
    Using \lemref{far} and \lemref{bound},
    \[\cost(\tilde{\pth}) \leq (1+\epsilon)\cost(\pth) + n \cdot f(k,\epsilon^{-1})\opt(\bs,\bt,1+\epsilon)\left(k \cdot \sqrt{2} \bar{\epsilon}\right).\]
    For reasons that will soon become clear, reapply \lemref{far} with $\epsilon/3$ in place of $\epsilon$,
    adding a factor of $3$ to the involved constant $C$ defining $\Delta$, which gives
    \[\cost(\tilde{\pth}) \leq (1+\epsilon/3)\cost(\pth) + n \cdot f(k,\epsilon^{-1})\opt(\bs,\bt,1+\epsilon)\left(k \cdot \sqrt{2} \bar{\epsilon}\right).\]

    Recalling the choice of $\bar{\epsilon}$,
    \begin{align*}
    \cost(\tilde{\pth}) &\leq (1+\epsilon/3)\cost(\pth) + \opt(\bs,\bt,1+\epsilon)\cdot \epsilon/3 \\
    &\leq (1+\epsilon/3) \opt(\bs,\bt,1+\epsilon)+ \opt(\bs,\bt,1+\epsilon)\cdot \epsilon/3 \\
    &\leq (1 + \epsilon/3)\opt(\bs,\bt,1+\epsilon) + (1+\epsilon/3)\opt(\bs,\bt,1+\epsilon)\cdot \epsilon/3\\
    &= (1+\epsilon/3)^2\opt(\bs,\bt,1+\epsilon)\\
    &\leq (1+\epsilon)\opt(\bs,\bt,1+\epsilon).
    \end{align*}
    We thus obtain that $\cost(\pth) \leq (1+\epsilon)\opt(\bs,\bt,1+\epsilon)$.
\end{proof}

We complete the proof of \thmref{result} by proving the runtime of the algorithm.

\begin{lemma}\label{lem:runtime}
    Given $(\bs,\bt) \in \BF^2$, our algorithm computes an $(\bs,\bt)$-plan $\tilde{\pth}$
    as specified above, when one exists, in
    $O(f(k,\epsilon)n^{2k})$ time.
\end{lemma}
\begin{proof}
We examine each step of the algorithm.
First, we compute $\F^{(1)},\F^{(2)}$, and $\F^{(24k)}$, and thus all of the points in $\Lambda$,
via standard techniques \cite[Chapter 10]{debergbook} in $O(n\log^2{n})$ time.
We next compute the placements in $\X$ directly by centering a radius $\Delta$ square about each vertex in $\Lambda$
and overlaying $\G$, noting that

\begin{align*}
    |\X| &= O(n) \cdot \left(\frac{\Delta}{\bar{\epsilon}}\right)^2 = n\left(\frac{3Ck^2\epsilon^{-1}}{\epsilon/(f(k, \epsilon^{-1}) \sqrt{2} k)}\right)^2 \\
    &= 3\sqrt{2}C\left(n\left(\frac{k^3\epsilon^{-2}}{2^{-k}\epsilon^k}\right)^2\right) \leq 3\sqrt{2}C n k^{2(k+3)}\epsilon^{-2(k+2)} = n \cdot \epsilon^{-O(k)}k^{O(k)}.
\end{align*}

To compute the configuration graph, we can then test for every free configuration in $\X^k$ and add each to
$\PV$ in $O(|\X|^k) = n^k\epsilon^{-O(k^2)}k^{O(k^2)}$ time. For $\bv,\bu \in \PV$, add an edge $(\bv,\bu)$ to
$\PE$ if there is a feasible $(\bv,\bu)$-plan $\pth_{\bv\bu}$ in which every robot moves along a straight line.
The feasibility of $\pth_{\bv \bu}$ can be tested in $O(k^2)$ time
by testing the feasibility of linearly interpolating each robot as
in \lemref{geodesic-plan} in constant time for each of the $O(k^2)$ pairs. 
If $\pth_{\bv \bu}$ is feasible, assign the weight $w(\bv, \bu) = \cost(\pth_{\bv \bu})$ to the edge.
Altogether, constructing $\PG$ can thus be done in time $n^{2k}\epsilon^{-O(k^2)}k^{O(k^2)}$.

Finally, we compute a shortest $(\bs,\bt)$ path in $\PG$ using Dijkstra's algorithm,
which takes $O(|\PV \log{\PV} + \PE) = n^{2k}\cdot(k/\epsilon)^{O(k^2)}$ time.
\end{proof}



\bibliography{references}


\end{document}